\documentclass[usenatbib,usegraphicx,useAMS]{mn2e}

\usepackage{times}
\usepackage{amsmath}
\usepackage{amssymb}

\newfont{\itbf}{cmbxti10 at 9pt}
\newfont{\slbf}{cmbxsl10 at 9pt}

\newcommand{\mbi}[1]{\mbox{\boldmath$#1$}}

\newcommand{\mat}[1]{\mbox{\rm\bf #1}}
\newcommand{\lsim}[1]{\mbox{${\,\hbox{\hbox{$ < $}\kern -0.8em \lower 1.0ex\hbox{$\sim$}}\,}$}}
\newcommand{\gsim}[1]{\mbox{${\,\hbox{\hbox{$ > $}\kern -0.8em \lower 1.0ex\hbox{$\sim$}}\,}$}}

\def\etal{{\it et al.\ }}

\def\beqn{\vspace{2mm}
\begin{eqnarray}} 
\def\eeqn{\vspace{2mm} 
\end{eqnarray}}

 \def\la{\langle}

\begin{document}
\title[Cosmic Cartography of the LSS with SDSS DR6]{Cosmic Cartography of the Large-Scale Structure with Sloan Digital Sky Survey Data Release 6}

\author[Kitaura \etal]{Francisco S.~Kitaura,$^{1,2}$\thanks{E-mail:
    kitaura@sissa.it, kitaura@mpa-garching.mpg.de} Jens Jasche,$^2$  Cheng Li,$^{2,3}$ Torsten A.~En$\ss$lin,$^2$ R.~Benton Metcalf,$^2$   \and
Benjamin D.~Wandelt,$^4$ Gerard Lemson,$^{5,6}$  and Simon D.~M.~White,$^2$
   \\
 $^1$ SISSA, Scuola Internazionale Superiore di Studi Avanzati, via Beirut 2-4 34014 Trieste, Italy \\
 $^2$ MPA, Max-Planck Institut f\"ur Astrophysik, Karl-Schwarzschildstr.~1, D-85748 Garching, Germany \\
 $^3$ MPA/SHAO Joint Center for Astrophysical Cosmology at Shanghai Astronomical Observatory, Nandan Road 80, Shanghai 200030, China \\
 $^4$ Department of Physics University of Illinois at Urbana-Champaign, 1110 West Green Street Urbana, IL 61801-3080, USA\\
 $^5$ Astronomisches Rechen-Institut, Zentrum f\"ur Astronomie der Universit\"at
Heidelberg, M\"onchhofstr. 12-14, 69120 Heidelberg, Germany\\
 $^6$ MPE, Max-Planck Institut f\"ur Extraterrestrische Physik, Giessenbachstraße, D-85748 Garching, Germany \\
} 

\maketitle

\begin{abstract}
We present the largest Wiener reconstruction of the cosmic density field made to date. The reconstruction is based on the Sloan Digital Sky Survey data release 6  covering the northern Galactic cap.
We use a novel supersampling algorithm to suppress aliasing effects and a Krylov-space inversion method to enable high performance  with high resolution.  These techniques are implemented in the \textsc{argo} computer code.
 We reconstruct the field over a 500 Mpc cube  with Mpc grid-resolution while accounting both for the angular and radial selection functions of
the SDSS, and the shot noise giving an effective resolution of the order of $\sim$10 Mpc.
 In addition, we correct for the redshift distortions in the linear and nonlinear regimes in an approximate way.  We show that the commonly used method of inverse weighting the galaxies by the corresponding selection function heads to excess noise in regions where the density of the observed galaxies is small. It is more accurate and conservative to adopt a Bayesian framework in which we model the galaxy selection/detection process to be Poisson-binomial. This results in heavier smoothing in regions of reduced sampling density.   Our results show a complex cosmic web structure with huge void regions indicating that the recovered matter distribution is highly non-Gaussian. Filamentary structures are clearly visible on scales up to  $\sim$20 Mpc. We also calculate the statistical distribution of  density after smoothing the reconstruction with  Gaussian kernels of different radii $r_{\rm S}$ and  find good agreement with a log-normal distribution for  $10\,{\rm Mpc}\lsim1{r_{\rm S}}\lsim130\,{\rm Mpc}$.  
\end{abstract}


\section{Introduction}
Measuring the Large-Scale Structure (LSS) of the Universe has become a major task in cosmology in recent years. The relics of the seed fluctuations, originating from the inflationary phase of the early Universe, are mainly encoded in the linear regime of the LSS in which structure formation has not significantly degraded the primordial phase information. In particular there has recently been a focus on measuring the baryon acoustic signal imprinted in the galaxy distribution which has been suggested as a powerful standard ruler for our Universe \citep[see for example][]{2005NewAR..49..360E}. 

Upcoming and ongoing galaxy redshift surveys such as DEEP2 or Baryon Oscillation Spectroscopic Survey (BOSS) will cover higher and higher redshifts \citep[see for example][]{2005ASPC..339..128D,2007AAS...21113229S}. They are designed  to trace complex structures in the Universe and to study the environment of galaxies and their evolution.

We carry a reconstruction of the density field dealing with statistical and systematic errors of the galaxy distributions with the \textsc{argo}\footnote{\textsc{\bf A}lgorithm for the \textsc{\bf R}econstruction of the  \textsc{\bf G}alaxy-traced \textsc{\bf O}verdensity field} computer code described in \citet[][]{kitaura}. \textsc{argo} is a high-performance implementation of a three-dimensional Wiener-filter, permitting treatments of an inhomogeneous and incomplete window function acting on the galaxy distribution. It exploits the power of fast Fourier transforms (FFTs) and iterative Krylov-space based inversion schemes for the otherwise intractable data inversion step.

Reconstructions permit us  to characterize the
large-scale structure, helping to deepen our understanding of structure
formation, to gain insight into the physical processes involved,
 to construct signal templates for the detection of weak physical effects. These can be used to study
 the cosmic microwave background and to
reveal signals ranging from the Integrated Sachs-Wolfe effect \citep[see for example][]{2008MNRAS.391.1315F},
over the Sunyaev-Zel'dovich effect in the diffuse gas, to metal
absorption lines. 
An interesting further application would be to constrain the bias
between luminous and dark matter using reconstructions made by
\textsc{argo} and correlating them with simulations and
reconstructions of the matter distribution coming from other
observables like weak lensing, Lyman alpha forest, etc. 
Topological studies could be made from the reconstructed data, leading
to a geometrical characterization of the actual large-scale structure \citep[see for example][]{2005astro.ph..2105S}.
It  is  also interesting  to  study  how  the physical  properties  of
galaxies  depend  on their  large-scale  environment \citet{Li-06a,Lee-Lee-08}  and  \citet{Lee-Li-08}. 
The reconstructed structures of a galaxy catalogue can be traced
back in time with various methods, like those based on the
\citet{1970A&A.....5...84Z} approximation \citep[see for example][]{1992ApJ...391..443N}. These early matter density
fluctuations can be used as initial conditions for N-body
simulations. The results of such a constrained simulation have a wide
application in structure formation theory \citep[see for
  example][]{2002MNRAS.333..739M}.
A joint estimation of the matter field and its power-spectrum would also be a natural next step given the technology we develop below \citep[for similar work in CMB-analysis see, for example,][]{2004PhRvD..70h3511W,2004ApJ...609....1J,2007ApJ...656..641E}. 

We present the first application of \textsc{argo} to
observational data.  In particular we have applied our method to recover  the galaxy density field based on data from {\tt Sample dr6fix} of the New York University Value
Added Catalogue (NYU-VAGC), which was constructed from the Sloan Digital Sky Survey
\citep[SDSS;][]{York-00} Data Release 6 \citep[DR6;][]{Adelman-McCarthy-08}. This leads to the largest Wiener-reconstruction of
the Large-Scale Structure made so far effectively requiring the inversion of a matrix with about $10^8\times 10^8$ entries. The use of optimized iterative
inversion schemes within an operator formalism \citep[see][]{kitaura}
together with a careful treatment of aliasing effects
\citep[see][]{dsp} permits us to recover the overdensity field on Mpc scales 
\citep[for previous Wiener reconstructions see][]{WienerFSL,1994ASPC...67..185H,1994ApJ...423L..93L,1994ASPC...67..171L,1995ApJ...449..446Z,1995MNRAS.272..885F,1997MNRAS.287..425W,1999ApJ...520..413Z,1999AJ....118.1146S,2004MNRAS.352..939E,2006MNRAS.373...45E}.
Note, that alternative density reconstruction techniques like Voronoi and Delaunay tesselations \citep[see
e.g.][]{1991QJRAS..32...85I,1993PhRvE..47..704E,1995A&AS..109...71Z,1996MNRAS.279..693B,1997A&AS..123..495D,1999ASPC..176..333M,2000ASPC..200..422K,2000A&A...363L..29S,2001misk.conf..268V,2001A&A...368..776R,2004ogci.conf..245P,2006ChJAA...6..387Z} are tuned to optimally represent the density field from a geometrical point of view, but are not explicit in the statistical assumptions made on the galaxy or matter distribution, which is an important aspect of our analysis here. 

 We investigate in detail the statistical problem of  finding an expression for a noise covariance which includes the survey angular and radial selection functions. The expression we find assumes a binomial model for the galaxy selection/detection process. 

We show that including our proposed noise covariance matrix in the
Wiener-filter leads to a more conservative reconstruction of matter structures than using the inverse weighting scheme. We also compare the linear WF expression which is derived from a least squares approach and the non-linear WF which uses a signal dependent noise covariance \citep[see appendix A in][]{kitaura}. The latter shows to be even more conservative than the linear WF since it strongly suppresses the cells with higher number counts.   

Due to the fine mesh of the reconstruction ($\sim1$ Mpc) a
treatment of the redshift distortion in the linear and non-linear
regime is required.  We choose a redshift distortion deconvolution
method, as presented by \cite{2004MNRAS.352..939E}, which aims to
correct in both regimes. This treatment only corrects the power and neglects any phase information. For this reason, the effective resolution of the reconstruction is lower than the resolution of the grid ($\sim10$ Mpc).

Our paper is structured as follows. 
We start by describing the input galaxy sample of the Sloan Digital Sky Survey (SDSS) Data Release 6 (DR6) in section \ref{sec:IGS}.
Then we present the methodology used to perform an estimation of the matter field (section \ref{sec:methodology}). 
In detail, the galaxy distribution is first
transformed into the comoving frame (section \ref{sec:trafo})  and then assigned to a grid using our newly developed {\it supersampling} method \citep[described
  in][]{dsp} to correct for aliasing effects, ensuring a correct
spectral representation of the galaxy distribution even up to the
highest modes contained in the grid (section \ref{sec:supersampling}).
Completeness on the sky and radial galaxy
selection function are then translated into a three dimensional
mask, which will be part of the response operator used in the
filtering step (section \ref{sec:mask}).
 Then, an observed galaxy
overdensity field is calculated which fulfils the
statistical requirements we want to impose on the matter field (section \ref{sec:DM}).
Taking the observed galaxy field as the data vector we finally apply a
Wiener-filtering step with the \textsc{argo} computer code (section \ref{sec:WF}) followed by a deconvolution step, effectively correcting for the redshift distortion (section \ref{sec:deconvolution}).  Here, we distinguish between a linear WF expression which is derived from a least squares approach and a non-linear WF which uses a signal-dependent noise covariance.
Both WF formulations are tested with mock data and quantitatively compared to a simple  procedure in which the galaxies are inverse weighted with the completeness, then gridded and finally smoothed to give a matter field estimate.  

We present a reconstruction of the density field for the DR6 main sample in section \ref{sec:results}. First, we analyze the survey sky mask (section \ref{sec:maskres}).
Results for the Sloan Great Wall are then presented in
detail. Some other prominent structures, for example, the Coma, the Leo, and the Hercules clusters
are also discussed (section \ref{sec:GW}) together with the detection of a large void region (section \ref{sec:VOID}).
The proper implementation of the filter  enables us to deal with complex masks which include
unobserved regions. We demonstrate the improved detection of
overdensity regions close to edges of the mask and the prediction of
structures  in gaps, as demonstrated by comparing with data from the
Data Release 7 (DR7) where those gaps are filled (section \ref{sec:GAP}).
In section \ref{sec:statistics} we analyze the statistical distribution of the density field and find  good agreement with a log-normal distribution for smoothing radii in a Gaussian filter in the range $10\,{\rm Mpc}\lsim1{r_{\rm S}}\lsim130\,{\rm Mpc}$.
Finally, we make a summary of the work, and present our conclusions  and future outlook.

\section{Input galaxy sample}
\label{sec:IGS}

In   this  study   we   use   data  from   the   sixth  data   release
\citep[DR6;][]{Adelman-McCarthy-08}  of the  Sloan Digital  Sky Survey
\citep[SDSS;][]{York-00}.  The survey contains  images of a quarter of
the sky obtained using a drift-scan camera \citep{Gunn-98} in the {\em
  u,  g,   r,  i,  z}   bands  \citep{Fukugita-96,Smith-02,Ivezic-04},
together  with spectra  of almost  a million  objects obtained  with a
fibre-fed double  spectrograph \citep{Gunn-06}. Both  instruments were
mounted  on a special-purpose  2.5~meter telescope  \citep{Gunn-06} at
Apache  Point  Observatory.   The  imaging  data  are  photometrically
\citep{Hogg-01,Tucker-06}    and    astrometrically    \citep{Pier-03}
calibrated, and were used to select spectroscopic targets for the main
galaxy  sample  \citep{Strauss-02},  the  luminous red  galaxy  sample
\citep{Eisenstein-01},  and  the  quasar  sample  \citep{Richards-02}.
Spectroscopic fibres  are assigned to  the targets using  an efficient
tiling     algorithm     designed     to     optimize     completeness
\citep{Blanton-03c}.  The details of  the survey strategy can be found
in \citet{York-00} and an overview  of the data pipelines and products
is provided in the Early Data Release paper \citep{Stoughton-02}. More
details on the photometric  pipeline can be found in \citet{Lupton-01}
and on the spectroscopic pipeline in \citet{Subbarao-02}.

We take data from {\tt Sample dr6fix} of the New York University Value
Added Catalogue (NYU-VAGC).  This is an update of the catalogue
constructed by \citet{Blanton-05b} and is based on the SDSS DR6 data
and publicly available selection
masks\footnote{http://sdss.physics.nyu.edu/vagc/}.  Starting from {\tt
  Sample dr6fix}, we construct a magnitude-limited sample of galaxies
with spectroscopically measured redshifts in the range $0.001<z<0.4$,
$r$-band Petrosian apparent magnitudes $14.5< m \le 17.6$, and
$r$-band absolute magnitudes $-23<M_{^{0.1}r}<-17$.  Here $m$ is
corrected for Galactic extinction, and the apparent magnitude limits
are chosen in order to get a sample that is uniform and complete over
the entire area of the survey.  The absolute magnitude $M_{^{0.1}r}$
is corrected to its $z=0.1$ value using the K-correction code of
\citet{Blanton-03b} and the luminosity evolution model of
\citet{Blanton-03a}.  We also restrict ourselves to galaxies located
in the main area of the survey in the northern Galactic cap, excluding
the three survey strips in the southern cap, i.e. we include galaxies with right
ascension ($\alpha$) and declination ($\delta$) in the following
ranges: $105^{\rm o}<\alpha<270^{\rm o}$ and $-5^{\rm
  o}<\delta<70^{\rm o}$.  In addition, we considered only galaxies
which are inside a comoving cube of side 500 Mpc  (with equal side lengths: $Lx\times Ly\times Lz$), as we describe below.
These restrictions result in a final sample of 255,818 galaxies.

In order to correct for incompleteness in our spectroscopic sample, we
need to have complete knowledge  of its selection effects.  A detailed
account  of  the   observational  selection  effects  accompanies  the
NYU-VAGC release.  These include two  parts: a mask  on the sky  and a
radial  selection function  along the  line-of-sight.  The  mask shows
which areas of the sky have  been targeted, and which have not, either
because they are  outside the survey boundary, because  they contain a
bright confusing source, or  because observing conditions were too poor to
obtain all the required data.  The effective area of the survey on the
sky defined by this mask  is 5314 square degrees for the sample
we  use here. It  is divided  into a  large number  of  smaller subareas,
called  {\em  polygons},  for  each  of which  the  NYU-VAGC  lists  a
spectroscopic  completeness. This is  defined as  the fraction  of the
photometrically  defined  target galaxies  in  the  polygon for  which
usable spectra were obtained. The average completeness over our sample
galaxies is 0.86. The radial  selection function gives the fraction of
galaxies   in   the   absolute   magnitude  range   being   considered
($-23<M_{^{0.1}r}<-17$  in  our case)  that  are  within the  apparent
magnitude  range of the  sample ($14.5<m\le17.6$ in  our case)  at a
given redshift.

In certain cases we also work with a sample of galaxies drawn from
SDSS data release 7 \citep[DR7;][]{Abazajian-08} for which the galaxy
positions, redshifts and fluxes are publicly available from the SDSS
website\footnote{http://www.sdss.org/dr7} but the survey completeness
as described above was not released at the moment this work started.  With this sample we apply
only a gridding scheme and a subsequent Gaussian smoothing, without
accounting for any selection effects, in order to {\em qualitatively}
check for overdense regions present in the gap in the SDSS DR6.

\section{Methodology}
\label{sec:methodology}

In this section, we describe the main algorithms required to perform a
Wiener-filter reconstruction of the matter field as described in
\cite{kitaura} \citep[see also the pioneering works][]{1949wiener,1992ApJ...398..169R,1995ApJ...449..446Z}.  We start with the preparation of the data followed by
a filtering step and a final deconvolution.  Detailed descriptions of
the methodology used for each step are described in the following
subsections.

\subsection{Preparation of the data}

Reconstructing a signal like the matter density field from the
observed galaxy sample requires a model which relates the
underlying matter field to the galaxy distribution. This model will define
the inverse problem, which can be  solved with a reconstruction
algorithm. In this subsection, we describe how to prepare the input
data in such a way that it is consistent with the data model underlying the
\textsc{argo}-code.

\subsubsection{Transformation of the data into comoving coordinates}
\label{sec:trafo}

To apply a reconstruction algorithm which uses the correlation
function in comoving space, we first have to transform the redshift
distances into comoving distances for each galaxy by performing the
integral\footnote{Not to be confused with the $r$-band.}:
\begin{equation}
r\equiv \int{\rm d}z\frac{1}{cH(z)}{,}
\end{equation}
with $H(z)$ being the Hubble parameter given by:
\begin{equation}
H(z)={H_0}{\sqrt{\Omega_{\rm m}(1+z)^3+\Omega_{\rm K}(1+z)^2+\Omega_{\Lambda}}}{,}
\end{equation}
where we chose the concordance $\Lambda$CDM-cosmology with
$\Omega_{\rm m}=0.24$, $\Omega_{\rm K}=0$ and $\Omega_{\Lambda}=0.76$
\citep[][]{spergel}.  In addition, we assumed a Hubble constant: $H_0=h
      {{\rm km/s}}/{{\rm Mpc}}$ with $h=73$.

With this definition the three-dimensional galaxy positions (X,Y,Z) in
comoving space are calculated as follows:
\begin{eqnarray}
{\rm X}&=&r\cdot\cos(\delta)\cdot\cos(\alpha)\nonumber\\
{\rm Y}&=&r\cdot\cos(\delta)\cdot\sin(\alpha)\nonumber\\
{\rm Z}&=&r\cdot\sin(\delta){.}
\label{eq:transf}
\end{eqnarray}

\subsubsection{Supersampling step}
\label{sec:supersampling}

Now, we can sort the galaxies onto a grid with a {\it supersampling}
scheme, which will permit us to apply a reconstruction scheme based on
FFTs. The much lower computational costs of FFTs permits us to tackle
much more ambitious matter reconstructions than have been attempted
previously with Wiener-filtering techniques.  The main difficulty in
signal processing via FFT techniques arises from the need to
represent a continuous signal which extends to infinity on a finite discrete
grid. Various methods to approximate the real continuous
signal by a discrete representation have been proposed in literature,
e.g. Nearest Grid Point (NGP), Cloud In Cell (CIC) or Triangular
Shaped Clouds (TSC) \citep[see e.g][]{1981csup.book.....H}.
However, all of these methods are only approximations to the ideal
low-pass filter, and introduce discretisation artifacts such as
aliasing. For a detailed discussion see
e.g.~\citet[][]{1981csup.book.....H,2005ApJ...620..559J, Cui,dsp}.
In recent years a number of methods have been proposed to correct for
these artifacts, especially for the purpose of power-spectrum estimation
\citep[][]{2005ApJ...620..559J, Cui}. However, common methods to
suppress these artifacts in the discretised signals, tend to be
numerically expensive.  

To circumvent this problem, \citet{dsp}
proposed a supersampling technique, which is able to provide discrete
signal representations with strongly suppressed aliasing contributions at reasonable computational cost.
This method relies on a two-step filtering process, where in the first
step the signal is pre-filtered by sampling the signal via the TSC
method to a grid with twice the target resolution. In our case we use
a $1024^3$ grid.
In a second step the ideal discrete low-pass filter is applied to the
pre-filtered signal, allowing us to sample the low-pass filtered field at
the lower target resolution.
In this fashion we obtain an aliasing free
signal sampled at a target resolution of $512^3$ cells (with equal number of cells in each axis: $Nx\times Ny\times Nz$).

Let us define the observed galaxy sample as a point source distribution $n^{\rm o}_{\rm p}({\mbi s})$ with coordinate  $\mbi s$
\begin{equation}
n^{\rm o}_{\rm p}({\mbi s})\equiv\sum_{i=1}^{N^{\rm o}_{\rm c}}\delta_{\rm D}(\mbi s-\mbi s_i){,}
\end{equation}
with $N^{\rm o}_{\rm c}$ being the total observed galaxy number count and $\delta_{\rm D}$ the Dirac-delta function.
The process of putting the galaxies on a regular grid is equivalent to a
convolution in real-space followed by a grid-point selection step according to \cite{1981csup.book.....H}
\begin{equation}
{n}^{\rm o}(\mbi s)\equiv\Pi\big(\frac{\mbi s}{H}\big)\int {\rm d}{\mbi s'}\,K_{\rm S}(\mbi s-\mbi s') n^{\rm o}_{\rm p}(\mbi s'){,}
\label{def:ngrid}
\end{equation}
with $\Pi({\mbi r})=\sum_{\mbi n\in \mathbb{Z}}\delta_{\rm
  D}(\mbi s-\mbi n)$, $H$ being the grid-spacing and $K_{\rm S}$ the supersampling
  kernel.
We define the resulting field as the observed galaxy number density $n^{\rm o}(\mbi s)$.
The observed galaxy number density is a function of the Cartesian position in
comoving space, but includes redshift distortion. For this
reason, we say that the distribution is in redshift-space denoted
by the coordinate $\mbi s$.

\subsubsection{Calculation of the three-dimensional mask: completeness on the sky and selection function}
\label{sec:mask}

To define the data vector we need to model the three dimensional
mask. We do this by processing the two-dimensional sky mask in several
steps.  First, the sky mask or completeness on the sky $w_{\rm
  SKY}(\alpha,\delta)$ is evaluated using the survey mask provided in
{\tt Sample dr6fix} of the NYU-VAGC (see Section \ref{sec:IGS}) on an
equidistant $\alpha\times \delta$-grid with $165000\times75000$ cells
having a resolution of $36''$ both in right ascension and declination
(see panel {(a)} in Fig.~\ref{fig:SKY}). Then, we project the sky
mask on a comoving Cartesian ${\rm X}\times {\rm Y}\times {\rm
  Z}$-grid containing $512^3$ cells.

This is done with the transformation given by Eqs.~\ref{eq:transf} taking projected values of the mask every 0.25
Mpc in the radial direction which are then assigned on the grid using the
Nearest Grid Point (NGP) method and normalized by the number of mask counts
at that grid cell. The analogous procedure is done with the radial
completeness $w_r(z)$, i.e.~the selection function which is available
as a function of redshift.  

Finally, we obtain the three dimensional
mask $w(\mbi s)$ as a product of the projected two dimensional mask,
i.e.~the completeness on the sky $w_{\rm SKY}(\alpha,\delta)$ and the
projected selection function $w_r(\mbi s)$ (see Fig.~\ref{fig:sel} and panels a in figures 6, 8, 9 and 10 in section
\ref{sec:results}). We define $w(\mbi s)\leq1$.

\subsection{Definition of the data model}
\label{sec:DM}

Let us define the observed galaxy overdensity field as\footnote{Not
  to be confused with the declination $\delta$.}:
\begin{equation}
{\delta}^{\rm o}_{\rm g}(\mbi s)\equiv \frac{n^{\rm o}(\mbi
  s)}{\overline{n}}-w(\mbi s){,}
\label{eq:datamodel} 
\end{equation}
with $\overline{n}$ being the mean galaxy number density.  

 The mean galaxy number density on the grid $\overline{n}$ is defined
 by the quotient of the total number of observed galaxies $N^{\rm o}_{\rm c}$
 and the observed volume $V^{\rm o}$. Note, that this assumes that the
 observed volume is a {\bf fair sample} of the Universe. We can then
 write:
\begin{equation}
\label{eq:nmean}
\overline{n}\equiv \frac{N^{\rm o}_{\rm c}}{V^{\rm o}}\equiv\frac{\sum_{i=1}^{N_{\rm cells}}
  N^{\rm o}_{{\rm c}i}}{\int{\rm d}\mbi r\, w(\mbi r)}{,}
\end{equation}
with $N^{\rm o}_{{\rm c}i}$ being the number of observed galaxies at cell $i$:
$N^{\rm o}_{\rm c}\equiv\sum_{i=1}^N N^{\rm o}_{{\rm c}i}$, $N_{\rm cells}$ being the total number of cells  and the observed volume
being defined by the integral: $V^{\rm o}\equiv\int{\rm d}\mbi r\,
w(\mbi r)$.  The relation between the expected galaxy number density
in a small volume $\Delta V$ around position $\mbi r$ $\rho_{\rm g}(\mbi r)$ and the mean
galaxy number density  in the whole volume under consideration $V$ is
given by:
\begin{equation}
\rho_{\rm g}(\mbi r)\equiv\overline{n}\,(1+\delta_{\rm
  g}(\mbi r)){,}
\label{eq:galden}
\end{equation}
where $\delta_{\rm g}(\mbi r)$ is the galaxy overdensity field, which
describes the spatial density distribution of
galaxies. Here we assume that effects due to galaxy evolution are
negligible in the observed region, and, especially, that the mean
number density is redshift independent.

The observed quantity $\delta_{\rm g}^{\rm o}(\mbi s)$ defined in
Eq.~\ref{eq:datamodel} has to be related to the signal, we seek to
recover, via a data model. This relation is to be inverted
by the reconstruction algorithm.

\subsubsection{Physical model}

In this section we describe the physical model which will enable us to apply linear reconstruction methods and obtain an estimate of the matter field  valid on large-scales ($>$1 Mpc).  
Let us assume a continuous matter field $\delta_{\rm m}(\mbi r)$ in
comoving space $\mbi r$ as well as a continuous galaxy field
${\delta}_{\rm g}$. We model the actual galaxies as being Poisson
distributed according to this field with an expectation density of
$\overline{n}\,(1+\delta_{\rm g}(\mbi r))$.  In general, the relation
between the galaxy overdensity field ${\delta}_{\rm g}$ and the
underlying matter field $ \delta_{\rm m}$ will be given by a non-local
and nonlinear bias operator.
However, the formalism we present here, without any further
development, allows us to account only for a non-local linear
translation-invariant bias operator $B(\mbi r-\mbi r')$ of the form:
\begin{equation}
\label{eq:bias}
{\delta}_{\rm g}(\mbi r)\equiv\int {\rm d}{\mbi r'} \,B(\mbi r-\mbi
r')\delta_{\rm m}(\mbi r'){.}
\end{equation}
Note, that this linear operator is known to fail at least at sub-Mpc
scales.  Several non-local biasing models are described in the literature,
which are mainly used to correct for the shape of the power-spectrum
on large-scales \citep[][]{Tegmark-04,2008JCAP...07..017H}. We will
carry this general bias through the algebraic calculations. However,
in this work we consider the galaxy field to be
a fair sample of the matter field. Thus, we assume the special case of
a linear constant bias equal to unity: $B(\mbi r,\mbi r')=\delta_{\rm
  D}(\mbi r-\mbi r')$, so that $\delta_{\rm g}=\delta_{\rm m}$.  Nevertheless, any non-local bias scheme of the
form of Eq.~\ref{eq:bias} can be adapted without the need to repeat
the filtering. We show that one can easily deal with non-local bias
models in a final deconvolution step (see Eq.~\ref{eq:WF}). As a
result, various posterior biasing assumptions can be applied based on
this reconstruction to test different biasing models.

We will also assume the existence of a redshift distortion operator\footnote{Not to be confused with the Z-axis in our
  Cartesian grid.}
$Z(\mbi s,\mbi r)$, which transforms the density field from real-space
into redshift-space.  Note, that the redshift distortion operator
cannot be a linear operator, since it depends on the matter field
$\delta_{\rm m}(\mbi r)$. However, we will approximate it with a
linear redshift distortion  operator $Z(\mbi s,\mbi r)$ here:
\begin{equation}
{\delta}_{\rm g}(\mbi s)\equiv\int {\rm d}{\mbi r} \,Z(\mbi s,\mbi
r)\delta_{\rm g}(\mbi r){,}
\end{equation}
 and postpone a matter field dependent treatment, sampling the
 peculiar velocity field as proposed in \citep[][]{kitaura}, for
 later work.

Let us further assume an additive noise term resulting in a data model for the
observed galaxy overdensity as:
\begin{equation}
{\delta}^{\rm o, th}_{\rm g}(\mbi s)\equiv w(\mbi s)\int {\rm
  d}{\mbi r} \,Z(\mbi s,\mbi r)\int {\rm
  d}{\mbi r'} \,B(\mbi r-\mbi r')\delta_{\rm m}(\mbi r')+{\epsilon}(\mbi s){,}
\label{eq:datamodele}
\end{equation}
with $\epsilon$ being the noise term.  
The corresponding vector representation of the data model can be approximated as:
\begin{equation}
{\mbi \delta}^{{\rm o, th}}_{{\rm g},s}\equiv \mat W_s\mat Z_{s,r}\mat
B_r \mbi\delta_{{\rm m},r}+{\mbi\epsilon}_{s}{,}
\label{eq:datalinear}
\end{equation}
with the subscripts $r$ and $s$ denoting real-space and
redshift-space, respectively. The response operator can be defined by
\begin{equation}
\mat R_{s,r}\equiv \mat W_s\mat Z_{s,r}\mat B_r{,}
\label{eq:RO}
\end{equation}
with $\mat W_s$ being the three dimensional mask operator defined in
continuous space by: $W(\mbi s,\mbi s')=w(\mbi s)\delta_{\rm D}(\mbi
s-\mbi s')$, $\mat Z_{s,r}$ being the redshift distortion operator,
and $\mat B_r$ being the bias operator.
Now we need to specify a model for the noise term. 

\subsubsection{Statistical model}

Assuming that the galaxy distribution is generated by an inhomogeneous
Poissonian distribution, the number galaxy count $N_{\rm c}$ within a volume
$\Delta V$ around position $\mbi r$ is distributed as:
\begin{equation}
N_{\rm c}(\mbi r)\sim P_{\rm Pois}(N_{\rm c}(\mbi r)|\lambda(\mbi r)){.}
\label{eq:pois}
\end{equation}
with
\begin{equation}
P_{\rm Pois}(N_{\rm c}(\mbi r)|\lambda(\mbi r))=\frac{\lambda(\mbi r)^{N_{\rm c}(\mbi r)}}{N_{\rm c}(\mbi r)!}\exp({-\lambda(\mbi r)}){,}
\end{equation}
where the expected number of galaxy counts is given by the Poissonian
ensemble average: $\lambda(\mbi r)=\langle N_{\rm c}(\mbi r)\rangle_{\rm g}$
and is directly related to the expected galaxy density $\rho_{\rm g}$
at that position: $\rho_{\rm g}(\mbi r)\equiv \langle N_{\rm c}(\mbi r)\rangle_{\rm g}/\Delta V$. Here $\langle \{ \, \} \rangle_{\rm g} \equiv \langle \{ \, \} \rangle_{(N_{\rm c}\mid \lambda)} \equiv \sum^{\infty}_{N_{\rm c}=0} \, P_{\rm Pois}(N_{\rm c}\mid \lambda) \{ \, \}$ denotes an ensemble average over the
Poissonian distribution. We further model the observational selection of  $N^{\rm
  o}_{\rm c}(\mbi r)$ galaxies out of the $N_{\rm c}$ present within the small volume $\Delta V$ to be a binomial selection with an acceptance rate $w(\mbi r)$. We then can write:
\begin{equation}
N^{\rm o}_{\rm c}(\mbi r)\sim P_{\rm Bin}(N^{\rm o}_{\rm c}(\mbi r)|N_{\rm c}(\mbi r),w(\mbi r)){,}
\label{eq:bin}
\end{equation}
with
\begin{eqnarray}
\lefteqn{P_{\rm Bin}(N^{\rm o}_{\rm c}(\mbi r)\mid N_{\rm c}(\mbi r), w(\mbi r))}\nonumber\\
&&=
\begin{pmatrix}
N_{\rm c}(\mbi r)\nonumber\\
N^{\rm o}_{\rm c}(\mbi r)
\end{pmatrix}
(w(\mbi r))^{N^{\rm o}_{\rm c}(\mbi r)}(1-w(\mbi r))^{(N_{\rm c}(\mbi r)-N^{\rm o}_{\rm c}(\mbi r))}{.}
\end{eqnarray}
The expected mean observed number of galaxies in the volume 
$\Delta V$ is:
\begin{equation}
\langle N^{\rm o}_{\rm c}(\mbi r)\rangle_w=w(\mbi r)N_{\rm c}(\mbi r){,}
\label{eq:binmean}
\end{equation}
where $\langle \{ \, \} \rangle_{w} \equiv \langle \{ \, \} \rangle_{(N_{\rm c}^{\rm o}\mid N_{\rm c},w)}\equiv \sum^{\infty}_{N^{\rm o}_{\rm c}=0} \, P_{\rm Bin}(N^{\rm o}_{\rm c}\mid N_{\rm c},w) \{ \, \}$ represents the ensemble average over the
binomial distribution with a selection probability $w$.  Consequently,
one can model the observed number of galaxies, as a single Poissonian
process:
\begin{equation}
N^{\rm o}_{\rm c}(\mbi r)\sim P_{\rm Pois}(N^{\rm o}_{\rm c}(\mbi r)|\lambda^{\rm o}(\mbi r)){,}
\label{eq:poisbin}
\end{equation}
with mean  
\begin{equation}
\lambda^{\rm o}(\mbi r)\equiv w(\mbi r)\lambda(\mbi r)= w(\mbi r)\langle N_{\rm c}(\mbi r)\rangle_{\rm g}=\langle\langle N^{\rm o}_{\rm c}(\mbi r)\rangle_{\rm g}\rangle_w{.}
\label{eq:poisbinmean}
\end{equation}
Note, that the Poissonian and the binomial distributions commute with each other.

\subsubsection{Noise covariance and data autocorrelation matrix}

 Let us define the noise covariance matrix, according to the
 assumptions made in the previous section, as the shot noise resulting
 from an {\bf inhomogeneous Poisson distribution} for the galaxy
 distribution $n(\mbi s)$, and a {\bf binomial distribution} for
 describing the observation process which reduces the
 fraction of observed galaxies following the selection function. We
 then obtain an expression for the noise covariance\footnote{Not to be confused with the galaxy number counts $N_{\rm c}$.}:
\begin{eqnarray}
\label{eq:noisecov}
\lefteqn{N^{\rm SD}(\mbi s_1,\mbi s_2)\equiv\langle\epsilon(\mbi  s_1)\epsilon(\mbi s_2)\rangle_{(\epsilon\mid \delta_{\rm m},\mbi p_\epsilon)}\equiv\langle\langle\epsilon(\mbi  s_1)\epsilon(\mbi s_2)\rangle_{\rm g}\rangle_w}\nonumber\\ 
&\equiv&\frac{1}{\overline{n}^2}(\langle\langle n^{\rm o}(\mbi s_1)n^{\rm o}(\mbi s_2))\rangle_{\rm  g}\rangle_w-\langle\langle n^{\rm o}(\mbi s_1)\rangle_{\rm  g}\rangle_w\langle\langle n^{\rm o}(\mbi s_2)\rangle_{\rm  g}\rangle_w)\nonumber\\
&=&\frac{1}{\overline{n}^2}\langle \langle n^{\rm o}(\mbi s_1)\rangle_{\rm g}\rangle_w\delta_{\rm D}({\mbi  s_1-\mbi s_2})\nonumber\\ 
&=&\frac{1}{\overline{n}^2} w(\mbi s_1)\langle n(\mbi s_1)\rangle_{\rm g}\delta_{\rm D}({\mbi s_1-\mbi  s_2}){,}
\end{eqnarray}
where we have used the properties of the variance and mean of these
distribution functions and have added the superscript SD to denote that this covariance matrix is signal-dependent \citep[see section 2.5.3 and appendix A in][]{kitaura}. 
Note, that this noise covariance is defined as the ensemble average of the correlation matrix of the noise over all possible noise realizations denoted by the subscript $(\epsilon\mid \mbi \delta_{\rm m},p_\epsilon)$ with $\mbi p_\epsilon$ being a set of parameters which determine the noise.
 Here, we have neglected the cell to cell
correlation introduced by the gridding scheme we have used (TSP) as the first step in our {\it supersampling} scheme. 

Having defined the data model, together with the noise model, we can
calculate the expected data autocorrelation matrix, which is defined
as the ensemble average over all possible galaxy realizations and
density realizations (cosmic variance) leading to the following
expression:
\begin{eqnarray}
\label{eq:dataauto}
\lefteqn{\langle\langle\langle {\delta}_{\rm g}^{\rm o,th}(\mbi
  s_1){\delta}_{\rm g}^{\rm o,th}(\mbi s_2)\rangle_{w}\rangle_{\rm
    g}\rangle_{\rm m}}\\ &=&w(\mbi s_1)w(\mbi s_2)\int {\rm d}{\mbi
  r_1} \,Z(\mbi s_1,\mbi r_1)\int {\rm d}{\mbi r_2} \,Z(\mbi s_2,\mbi
r_2)\nonumber\\ &&\times\,\int {\rm d}{\mbi r_1'} \,B(\mbi r_1-\mbi
r_1')\int {\rm d}{\mbi r_2'} \,B(\mbi r_2-\mbi r_2')\,\langle
\delta_{\rm m}(\mbi r'_1)\delta_{\rm m}(\mbi r'_2)\rangle_{\rm
  m}\nonumber\\ && +\langle N(\mbi s_1,\mbi s_2)\rangle_{\rm
  m}\nonumber{,}
\end{eqnarray}
with $\langle\{ \, \} \rangle_{\rm m}\equiv\langle\{ \, \} \rangle_{(\delta_{\rm m}\mid \mbi p_{\rm m})}\equiv \int {\rm d}\delta_{\rm m}P(\delta_{\rm m}\mid \mbi p_{\rm m})$ being the ensemble average over all
possible matter density realizations with some prior distribution $P(\delta_{\rm m}\mid \mbi p_{\rm m})$ with $\mbi p_{\rm m}$ being a set of parameters which determine the matter field, say the cosmological parameters.
Note, that this equation is only valid in the approximation where the
bias and the redshift distortion operators are linear.

The noise term is the
in Eq.~\ref{eq:dataauto} has the following form:
\begin{eqnarray}
\lefteqn{N^{\rm LSQ}(\mbi s_1,\mbi s_2)\equiv\langle N^{\rm SD}(\mbi s_1,\mbi s_2)\rangle_{\rm  m}\equiv\langle{\epsilon(\mbi s_1)}{\epsilon(\mbi s_2)}\rangle_{(\delta_{\rm m},\epsilon\mid \mbi p)}}\nonumber\\
&&\equiv\frac{1}{\overline{n}}w(\mbi s_1)\delta_{\rm D}(\mbi s_1-\mbi s_2){,}
\label{eq:noiseav}
\end{eqnarray}
since $\langle\langle n(\mbi r)\rangle_{\rm g}\rangle_{\rm
  m}=\langle\overline{n}\,(1+\delta_{\rm g}(\mbi r')) \rangle_{\rm
  m}=\overline{n}$, assuming again, that the observed volume is a {\bf fair sample} of the Universe.  
The noise covariance has been denoted with the superscript LSQ because it corresponds to the expression which is obtained by performing the LSQ approach to derive the WF, i.e.~minimising the ensemble average of the squared difference between the real underlying density field $\delta_{\rm m}$ and the LSQ estimator $\delta^{\rm LSQ}_{\rm m}$ over all possible signal $\delta_{\rm m}$ and noise $\epsilon$ realizations: $\langle (\delta_{\rm m}-\delta^{\rm LSQ}_{\rm m})^2\rangle_{\rm (\delta_{\rm m},\epsilon\mid \mbi p)}$ with $\mbi p$ being the joint set of parameters: $\mbi p\equiv\{\mbi p_{\rm m},\mbi p_{\epsilon}\}$ \citep[for a derivation see appendix B in][]{kitaura}.
We have also assumed that the 
cross terms between the noise and the signal are negligible: $\langle \mbi
\delta_{\rm m}\mbi\epsilon^\dagger\rangle_{\rm m}=0$.
 This should be further analyzed in future work. Higher order correlations between noise and signal in fact exist, and
can be exploited using schemes like the Poissonian
scheme proposed in \citet[][]{kitaura}.
Note, however that we consider a signal-dependent noise for the WF Eq.~\ref{eq:noisecov} which requires a model for the expected observed galaxy number density $\langle \langle
n^{\rm o}(\mbi s_1)\rangle_{\rm g}\rangle_w$  \citep[for differences in the derivation see][]{kitaura}.
We restrict ourselves to the LSQ noise covariance model  $N^{\rm LSQ}$ given
by Eq.~\ref{eq:noiseav} in our application to the SDSS data (section \ref{sec:results}). 
 Note, that the LSQ representation of the Wiener-filter is a linear operator in contrast to the alternative formulation which depends on the signal and thus is a nonlinear filter.
We explore methods to deal with the signal-dependent noise formulation with mock galaxy catalogues and compare the results to the LSQ version of the Wiener-filter (see section \ref{sec:results0}). 

Note, that by construction the data autocorrelation
matrices for the observed galaxy overdensity field and the
theoretical overdensity field are identical given the noise model in
Eq.~\ref{eq:noisecov}:
\begin{equation}
\langle\langle\langle {\delta}_{\rm g}^{\rm o,th}(\mbi
s_1){\delta}_{\rm g}^{\rm o,th}(\mbi s_2)\rangle_{\rm
  g}\rangle_w\rangle_{\rm m}=\langle\langle\langle {\delta}_{\rm
  g}^{\rm o}(\mbi s_1){\delta}_{\rm g}^{\rm o}(\mbi s_2)\rangle_{\rm
  g}\rangle_w\rangle_{\rm m}{.}
\end{equation}

\subsection{Reconstruction algorithm}

In this section we propose a two step reconstruction process:
first a Wiener-filter step and second a deconvolution step.

\subsubsection{Wiener-filtering}
\label{sec:WF}

First, we recover the galaxy field in redshift-space
(${\mbi \delta}_{{\rm g},s}$) applying the Wiener-filter. The version of the Wiener-filter we use can be derived as follows.
Let us approximate the posterior distribution assuming a Gaussian prior and a Gaussian likelihood:
\begin{eqnarray}
\lefteqn{\label{eq:postdist}P({\mbi \delta_{{\rm g},s}}\mid{\mbi \delta^{\rm o}_{{\rm g},s}},\mbi p)\propto} \\
&&\hspace{-.5cm}{\rm exp}\left(-\frac{1}{2}\left[
    {\mbi \delta_{{\rm g},s}}^{\dagger}{\mat S_{{\rm g},s}}^{-1}{\mbi \delta_{{\rm g},s}}+({\mbi \delta^{\rm o}_{{\rm g},s}-\mat W_s\mbi \delta_{{\rm g},s}})^{\dagger}{\mat N_s}^{-1}({\mbi \delta^{\rm o}_{{\rm g},s}-\mat W_s\mbi \delta_{{\rm g},s}})\right]\right)\nonumber{,}
\end{eqnarray}
with the signal autocorrelation matrix $\mat S_{{\rm
    g},s}\equiv\langle\mbi\delta_{{\rm g},s} (\mbi\delta_{{\rm
    g},s})^\dagger\rangle$ being the inverse Fourier transform of the assumed model galaxy
power-spectrum in redshift-space: $\hat{\hat{S}}_{{\rm g},s}(\mbi
k,\mbi k')\equiv (2\pi)^3P_{{\rm g}}^s(\mbi k')\delta_{\rm D}(\mbi
k-\mbi k')$ and the hats denoting the Fourier transform of the signal
autocorrelation matrix. Note, that the posterior distribution depends also on a set of parameters $\mbi p$ which determine the power-spectrum $P_{{\rm g}}^s(\mbi k)$.   
The log-posterior distribution is then given by:
\begin{eqnarray}
\lefteqn{\log P({\mbi \delta_{{\rm g},s}}\mid{\mbi \delta^{\rm o}_{{\rm g},s}},\mbi p)\propto}\\ 
&&\hspace{-.5cm}{\mbi \delta_{{\rm g},s}}^{\dagger}{\mat S_{{\rm g},s}}^{-1}{\mbi \delta_{{\rm g},s}}+({\mbi\delta^{\rm o}_{{\rm g},s} -\mat W_s\mbi \delta_{{\rm g},s}})^{\dagger}{\mat N_s}^{-1}({\mbi\delta^{\rm o}_{{\rm g},s}-\mat W_s\mbi \delta_{{\rm g},s}}) \nonumber\\
&&\hspace{-.5cm}= {\mbi \delta_{{\rm g},s}}^{\dagger}{\mat S_{{\rm g},s}}^{-1}{\mbi \delta_{{\rm g},s}}+{\mbi \delta_{{\rm g},s}}^\dagger\mat W_s^\dagger{\mat N_s}^{-1}\mat W_s{\mbi \delta_{{\rm g},s}}-{\mbi \delta_{{\rm g},s}}^\dagger\mat W_s^\dagger{\mat N_s}^{-1}{\mbi \delta^{\rm o}_{{\rm g},s}}\nonumber\\
&&\hspace{-.5cm}-{\mbi \delta^{\rm o}_{{\rm g},s}}^\dagger{\mat N_s}^{-1}\mat W_s\mbi \delta_{{\rm g},s}+{\mbi \delta^{\rm o}_{{\rm g},s}}^\dagger{\mat N_s}^{-1}{\mbi \delta^{\rm o}_{{\rm g},s}}\nonumber{.}
\label{app:post1}
\end{eqnarray}
The first two terms can be combined to one term: ${\mbi \delta_{{\rm g},s}}^{\dagger}(\mbi\sigma_{\rm WF}^2)^{-1}{\mbi \delta_{{\rm g},s}}$, using the Wiener-variance: $\mbi\sigma_{\rm WF}^2\equiv (\mat S^{-1}+\mat W_s^\dagger \mat N_s^{-1}\mat W_s)^{-1}$.
To find the mean of the posterior distribution we seek an expression for the log-posterior of the form:
\begin{equation}
\log {P({\mbi \delta_{{\rm g},s}}\mid{\mbi \delta^{\rm o}_{{\rm g},s}},\mbi p)}\propto ({\mbi \delta_{{\rm g},s}-\langle{\mbi \delta_{{\rm g},s}}\rangle_{\rm WF}})^{\dagger}{(\mbi\sigma_{\rm WF}}^2)^{-1}({\mbi \delta_{{\rm g},s}-\langle{\mbi \delta_{{\rm g},s}}\rangle_{\rm WF}}){,}
\label{app:post2}
\end{equation}
with $\langle{\mbi \delta_{{\rm g},s}}\rangle_{\rm WF}=\mat F_{\rm WF}\mbi \delta^{\rm o}_{{\rm g},s}$ being the mean after applying the Wiener-filter $F_{\rm WF}$ to the data. 
Now the third and the fourth term of eq.~(\ref{app:post1}) can be identified with the terms in eq.~(\ref{app:post2}) as:
\begin{equation}
-{\mbi \delta_{{\rm g},s}}^{\dagger}\mat W_s^\dagger{\mat N_s}^{-1}{\mbi \delta^{\rm o}_{{\rm g},s}}=-{\mbi \delta_{{\rm g},s}}^{\dagger}(\mbi\sigma_{\rm WF}^2)^{-1}\mat F_{\rm WF}{\mbi \delta^{\rm o}_{{\rm g},s}}{,}
\label{app:WFCOV1}
\end{equation}
and
\begin{equation}
-{\mbi \delta^{\rm o}_{{\rm g},s}}^{\dagger}{\mat N_s}^{-1}\mat W_s{\mbi \delta_{{\rm g},s}}=-{\mbi \delta^{\rm o}_{{\rm g},s}}^{\dagger}\mat F_{\rm WF}^\dagger(\mbi\sigma_{\rm WF}^2)^{-1}{\mbi \delta_{{\rm g},s}}{,}
\label{app:WFCOV2}
\end{equation}
respectively.
The remaining term depends only on the data and is thus factorized in the posterior distribution function as part of the evidence.
From both eq.~(\ref{app:WFCOV1}) and eq.~(\ref{app:WFCOV2}) we conclude that the Wiener-filter has the form 
\begin{equation}
\mat F_{\rm WF}=\mbi\sigma_{\rm WF}^2\mat W_s^\dagger{\mat N_s}^{-1}=(\mat S^{-1}+\mat W_s^\dagger \mat N_s^{-1}\mat W_s)^{-1}\mat W_s^\dagger{\mat N_s}^{-1}{.}
\end{equation}
The mean $\langle{\mbi \delta}_{{\rm g},s}\rangle_{\rm WF}$ of the posterior distribution defined by Eq.~\ref{eq:postdist} can be obtained by: 
\begin{equation}
\langle{\mbi \delta}_{{\rm g},s}\rangle_{\rm WF}=\left(\mat S^{-1}_{{\rm
    g},s}+\mat W^{\dagger}_s\mat N^{-1}_s\mat W_s\right)^{-1}\mat
W^{\dagger}_s\mat N^{-1}_s{\mbi \delta}_{{\rm g},s}^{{\rm o}}{.}
\label{eq:WF}
\end{equation}
We favor this signal-space representation\footnote{We use here the
  terminology introduced in \citet[][]{kitaura}.} of the Wiener-filter
with respect to the equivalent and more frequently used data-space representation in LSS
reconstructions: $\langle{\mbi \delta}_{{\rm g},s}\rangle_{\rm WF}=\mat S_{{\rm g},s} \mat W_s^\dagger\left(\mat W_s\mat S_{{\rm g},s}\mat W_s^\dagger+\mat N\right)^{-1}{\mbi \delta}_{{\rm g},s}^{{\rm o}}$ \citep[see for example][]{1995ApJ...449..446Z},
because it avoids instabilities which otherwise arise in our rapid algorithm for evaluating the filter. 

Let us distinguish between the linear LSQ  and the nonlinear signal-dependent noise formulation of the Wiener-filter. The first takes the matter field averaged noise, covariance Eq.~\ref{eq:noiseav} $N=N^{\rm LSQ}$ and is used below when analyzing the SDSS data (see section \ref{sec:results}). In the  case of a signal-dependent noise: $N=N^{\rm SD}$ one needs an estimate of the expected observed galaxy number density  $w(\mbi s)\lambda(\mbi s)\equiv\langle \langle n^{\rm o}(\mbi s)\rangle_{\rm g}\rangle_w$ (see Eq.~\ref{eq:noisecov} and section \ref{sec:bayesWF}). Such an approach was done by \citet{2004MNRAS.352..939E}.

\subsubsection{Deconvolution step}
\label{sec:deconvolution}

In the second reconstruction step, we deconvolve the galaxy field
$\langle{\mbi \delta}_{{\rm g},s}\rangle_{\rm WF}$ from the assumed redshift distortion and
 galaxy bias operators, obtaining an
estimate for the underlying matter field in real-space:
\begin{equation}
\langle{\mbi \delta}_{{\rm m},r}\rangle_{\rm WF}=\mat B^{-1}_r \mat
Z^{-1}_{r,s}\langle{\mbi \delta}_{{\rm g},s}\rangle_{\rm WF}{.}
\label{eq:DEC}
\end{equation}
In this approximation, we can easily transform the reconstructed
galaxy field into the matter field by just performing a final
deconvolution with some scale-dependent bias of the form:
$\hat{\hat{B}}(\mbi k,\mbi k')\equiv b(\mbi k)\delta_{\rm D}(\mbi
k-\mbi k')$.  As already mentioned above, our result should not be
restricted to a single arbitrary chosen bias model. We therefore
choose to recover the galaxy field by assuming a bias equal to unity
from which matter reconstructions for all possible linear (and
invertible) bias schemes can easily be constructed via
Eq.~\ref{eq:DEC}.  Note, that an alternative representation of the
Wiener-filter which regularizes the bias and the redshift distortion
operator when they are not be invertible, consists of including them
in the response operator (Eq.~\ref{eq:RO}) when calculating the
Wiener-filter, leading to: $\langle{\mbi \delta}_{{\rm
    m},r}\rangle_{\rm WF}=\left(\mat S^{-1}_{{\rm m},r}+\mat
R^{\dagger}_{r,s}\mat N^{-1}_s\mat R_{s,r}\right)^{-1}\mat
R^{\dagger}_{r,s}\mat N^{-1}_s{\mbi \delta}_{{\rm g},s}^{{\rm o}}$.

\begin{figure*}
\includegraphics[width=10cm]{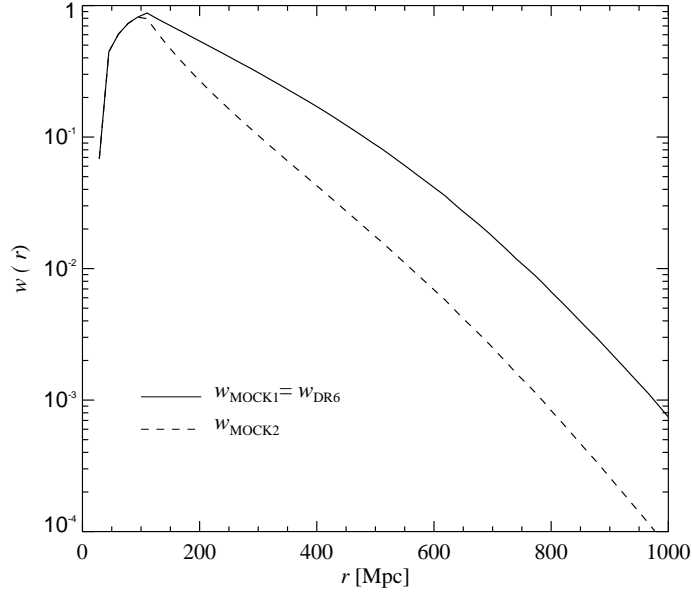}
\caption{\label{fig:sel} Radial selection functions used for the mock tests. Note, that the selection function used for the first mock test $w_{\rm MOCK1}$ is identical to the radial completeness of the DR6 catalogue $w_{\rm DR6}$. The second selection function $w_{\rm MOCK2}$ is calculated by weighting $w_{\rm DR6}(r)$ with the factor 100 Mpc/$r$ for $r\geq100$ Mpc. }  
\end{figure*}

\subsubsection{Redshift distortion operator}
\label{sec:RDO}

Following \citet{2004MNRAS.352..939E} we define the power-spectrum in
redshift-space as the product of the power-spectrum in real-space and
an effective redshift distortion factor given by the angle-averaged
Kaiser factor\footnote{Not to be confused with the supersampling kernel $K_{\rm S}$.} $K(\mbi k,\mu)$ times the damping Lorentzian factor
$D(\mbi k,\mu)$:
\begin{equation}
P^s_{\rm m}(\mbi k)\equiv\langle K(\mbi k,\mu)D(\mbi k,\mu)\rangle_{\mu}P^r_{\rm
  m}(\mbi k){,}
\end{equation}
with $\mu=\mbi k\cdot\mbi r/(|\mbi k| |\mbi r|)$.
 The Kaiser factor is given by \citep[see][]{Kaiser-87}:
\begin{equation}
K(\mbi k,\mu)\equiv (1+\beta\mu^2)^2{,}
\end{equation} 
with $\beta$
being the redshift distortion parameter which can be approximated by:
$\beta\simeq\Omega_{\rm m}^{0.6}$ assuming a constant bias equal to unity
and neglecting  dark energy dependences
\citep[see][]{1991MNRAS.251..128L}.  The Lorentzian damping factor is 
based on an exponential distribution in real-space for the pairwise
peculiar velocity field and is given by:
\begin{equation}
D(\mbi k,\mu)\equiv \frac{1}{1+(k^2\sigma_v^2\mu^2)/2}{,}
\end{equation}  
with $k\equiv|\mbi k|$ and $\sigma_v$ being
the average dispersion velocity of the galaxies, which we assume to be
$\sigma_v=500$ km s$^{-1}H_0^{-1}$ \citep[see for
  example][]{1996MNRAS.282..877B,Jing-Mo-Boerner-98,
Jing-Borner-04,Li-06b}.  

We refer to \citet{2004MNRAS.352..939E} for the angle-average
expression of the product of the Kaiser factor and the damping factor.
Consequently, we introduce the angular averaged redshift distortion
operator defined as the square root of the factor in the previous
Eqs.:
\begin{equation}
\hat{\hat{\mat Z}} (\mbi k,\mbi k')\equiv \sqrt{\langle K(\mbi k',\mu)D(\mbi
  k',\mu)\rangle_{\mu}}\delta_{\rm D}(\mbi k-\mbi k'){.}
\end{equation}
By construction, this operator yields the correct power-spectrum
modification for the translation from real- to redshift
space\footnote{Note, that we deviate here from
  \citet{2004MNRAS.352..939E} in the order of the angular averaging
  and square root. An inspection of the power-spectrum corresponding
  to the reconstructions shows, however that only the 
  prescription as implemented here leads to agreement with the nonlinear
  \citet[][]{2003MNRAS.341.1311S} power-spectrum.}.

 Note, that this approximation is valid up to second order statistics,
 and gives only an effective solution to the redshift distortion due to the angular averaging. A
 proper solution would require a phase and direction dependent redshift distortion
 operator.  
If we assume that
 the galaxy bias is unity, we then can write the galaxy power-spectrum
 in redshift space as: $P_{{\rm g}}^s(\mbi k')=\langle K(\mbi k,\mu)D(\mbi
 k,\mu)\rangle_{\mu}P^r_{\rm m}(\mbi k)$. Note, that this reduces the validity of our reconstruction to scales larger than the mesh resolution which is of about 1 Mpc to scales of about 10 Mpc.  The power
 spectrum in real-space $P^r_{\rm m}$ is given by a nonlinear
 power-spectrum that also describes  the effects of virialised
 structures with a halo term as given by
 \citet[][]{2003MNRAS.341.1311S} at redshift $z=0$. 
In addition to the cosmological parameters presented in section \ref{sec:trafo}, we assume a spectral index $n_s=1$.
With each of the required operators defined, we can now apply our
reconstruction algorithm as we demonstrate in the next section.

\subsection{Signal-dependent noise formulation of the Wiener-filter}
\label{sec:bayesWF}

To apply the signal-dependent noise formulation of the Wiener-filter one needs to find estimators for the expected density field in the signal-dependent noise covariance (Eq.~\ref{eq:noisecov}). We require either a good estimator for $\lambda^{\rm o}(\mbi r)\equiv\langle\langle N_{\rm c}^{\rm o}(\mbi r)\rangle_{\rm g}\rangle_w$ or for $\lambda(\mbi r)\equiv\langle N_{\rm c}(\mbi r)\rangle_{\rm g}$ since $\lambda^{\rm o}(\mbi r)\equiv w(\mbi r)\lambda(\mbi r)$.

\subsubsection{Flat prior assumption}
\label{sec:FPA}

The inverse weighting estimator used in previous works to estimate the noise covariance \citep[see for example][]{2004MNRAS.352..939E} can be derived from the frequentist
approach by assuming a flat prior for the overdensity distribution or
equivalently infinite cosmic variance.

Let us start with Bayes theorem:
\begin{equation}
P(\lambda^{\rm o}|N_{\rm c}^{\rm o})=\frac{P(N_{\rm c}^{\rm o}|\lambda^{\rm o})P(\lambda^{\rm o})}{P(N_{\rm c}^{\rm o})}{.}
\end{equation}
The flat prior is defined as: $P(\lambda^{\rm o})={c}$, with $c$ being a constant.
The evidence is then given by
\begin{equation}
P(N_{\rm c}^{\rm o})=\int_0^\infty {\rm d}\lambda^{\rm o}\, P(N_{\rm c}^{\rm o}|\lambda^{\rm o}) c=c{,}
\end{equation}
since
\begin{equation}
\int_0^\infty {\rm d}\lambda^{\rm o}\, P(N_{\rm c}^{\rm o}|\lambda^{\rm o}) = \int_0^\infty {\rm d}\lambda^{\rm o}\, \frac{(\lambda^{\rm o})^{N_{\rm c}^{\rm o}}{\rm e}^{-\lambda^{\rm o}}}{N_{\rm c}^{\rm o}!}=\frac{\Gamma(N_{\rm c}^{\rm o}+1)}{N_{\rm c}^{\rm o}!}=1{.}
\end{equation}
Consequently, we obtain that the posterior distribution is equal to the likelihood
\begin{equation}
P(\lambda^{\rm o}|N_{\rm c}^{\rm o})={P(N_{\rm c}^{\rm o}|\lambda^{\rm o})}{.}
\label{app:flatprior}
\end{equation}
The maximum likelihood estimator $\lambda_{\rm max}$ is obtained by looking at the extrema:
\begin{eqnarray}
0&=&\frac{\partial P(\lambda_{\rm max}|N^{\rm o}_{\rm c})}{\partial \lambda_{\rm max}}\\
&=& (N^{\rm o}_{\rm c}(w\lambda_{\rm max})^{-1}w-w)\frac{(w\lambda_{\rm max})^{N^{\rm o}_{\rm c}}{\rm e}^{-w\lambda_{\rm max}}}{N^{\rm o}_{\rm c}!}\nonumber\\
&=& N^{\rm o}_{\rm c}\lambda_{\rm max}^{-1}-w\nonumber{,}
\end{eqnarray}
leading to:
\begin{equation}
\lambda_{\rm max}=\frac{N^{\rm o}_{\rm c}}{w}{.}
\end{equation}
 Note, that the maximum estimator $\lambda_{\rm max}$ is not a valid estimator for the noise covariance matrix, since it can become zero at cells in which no galaxy count is present even if the cell belongs to the observed region.  
The mean estimator $\lambda^{\rm o}_{\rm mean}$ can be found by performing the following integral:
\begin{eqnarray}
\lambda^{\rm o}_{\rm mean}&\equiv&\int_0^\infty {\rm d}\lambda^{\rm o}\, \lambda^{\rm o}\, P(\lambda^{\rm o}|N_{\rm c}^{\rm o})\\
&=&\int_0^\infty {\rm d}\lambda^{\rm o} \,\frac{(\lambda^{\rm o})^{N_{\rm c}^{\rm o}+1}{\rm e}^{-\lambda^{\rm o}}}{(N_{\rm c}^{\rm o}+1)!}(N_{\rm c}^{\rm o}+1)\nonumber{.}
\label{app:pois3}
\end{eqnarray}
Thus, we have:
\begin{equation}
\lambda_{\rm mean}\equiv\frac{\lambda^{\rm o}_{\rm mean}}{w}=\frac{1}{w}(N_{\rm c}^{\rm o}+1){.}
\end{equation}
The mean estimator $\lambda_{\rm mean}$ gives a regularized solution with respect to the maximum estimator $\lambda_{\rm max}$ overcoming the problem of having zero noise at cells with zero observed number counts. 
Both estimators however, rely on the flat prior assumption which can be dominated by the shot-noise for low completeness. This can be a problem when the reconstruction is performed on a fine mesh with extremely low completeness.  For this reason, we test the SD Wiener-filter with an alternative scheme presented in the next section. 

\subsubsection{Statistically unbiased Jackknife-like scheme}
\label{sec:bayesWF}

The Jackknife-like scheme we present here and test in the next section produces subsamples from a galaxy distribution with selection function effects which are statistically unbiased with the underlying mean number density having a noise term with a structure function depending only on $\lambda(\mbi r)$. 
The first step of the scheme consists of generating a subsample using the binomial distribution given the observed number counts and the selection probability $\alpha/w(\mbi r)$ with
 a tunable parameter $\alpha<{\rm min}(w(\mbi r))$:
\begin{equation}
\label{eq:selfunc}
\hspace{0.cm}N'_{\rm c}(\mbi r)\sim 
\begin{cases}
P_{\rm Bin}\left(N'_{\rm c}(\mbi r)\mid N_{\rm c}^{\rm o}(\mbi r),\frac{\alpha}{w(\mbi r)}\right)\nonumber\\
P_{\rm Pois}(N'_{\rm c}(\mbi r)\mid \alpha\lambda(\mbi r)){.}
\end{cases}
\end{equation}
In the second step the subsample $N'_{\rm c}(\mbi r)$ is inverse weighted with $\alpha$:
\begin{equation}
N''_{\rm c}(\mbi r)\equiv \frac{1}{\alpha} N'_{\rm c}(\mbi r){.}
\end{equation}
One can notice, that the ensemble average over all possible $\alpha$ realizations leads to the mean number density $\lambda(\mbi r)$:
\begin{equation}
\langle \langle N''_{\rm c}(\mbi r)\rangle_{(N_{\rm c}\mid \lambda)}\rangle_\alpha = \frac{1}{\alpha} \langle \langle N'_{\rm c}(\mbi r)\rangle_{(N_{\rm c}\mid \lambda)}\rangle_\alpha=\langle N_{\rm c}(\mbi r)\rangle_{(N_{\rm c}\mid \lambda)}=\lambda(\mbi r){.}
\end{equation}
Here, $\langle\{\,\}\rangle_\alpha$ is a binomial average with acceptance frequency $\alpha$.
The estimator for $\langle\lambda^{\rm o}(\mbi r)\rangle_{JK}\equiv w(\mbi r)N''_{\rm c}(\mbi r)$, with the subscript $JK$ standing for the Jackknife estimator.
We test the estimator proposed here to sample the noise covariance (see section \ref{sec:results0}).

\section{Quality validation of the radial selection function treatment}
\label{sec:results0}

\begin{figure*}
\begin{tabular}{cc}
\includegraphics[width=8.5cm]{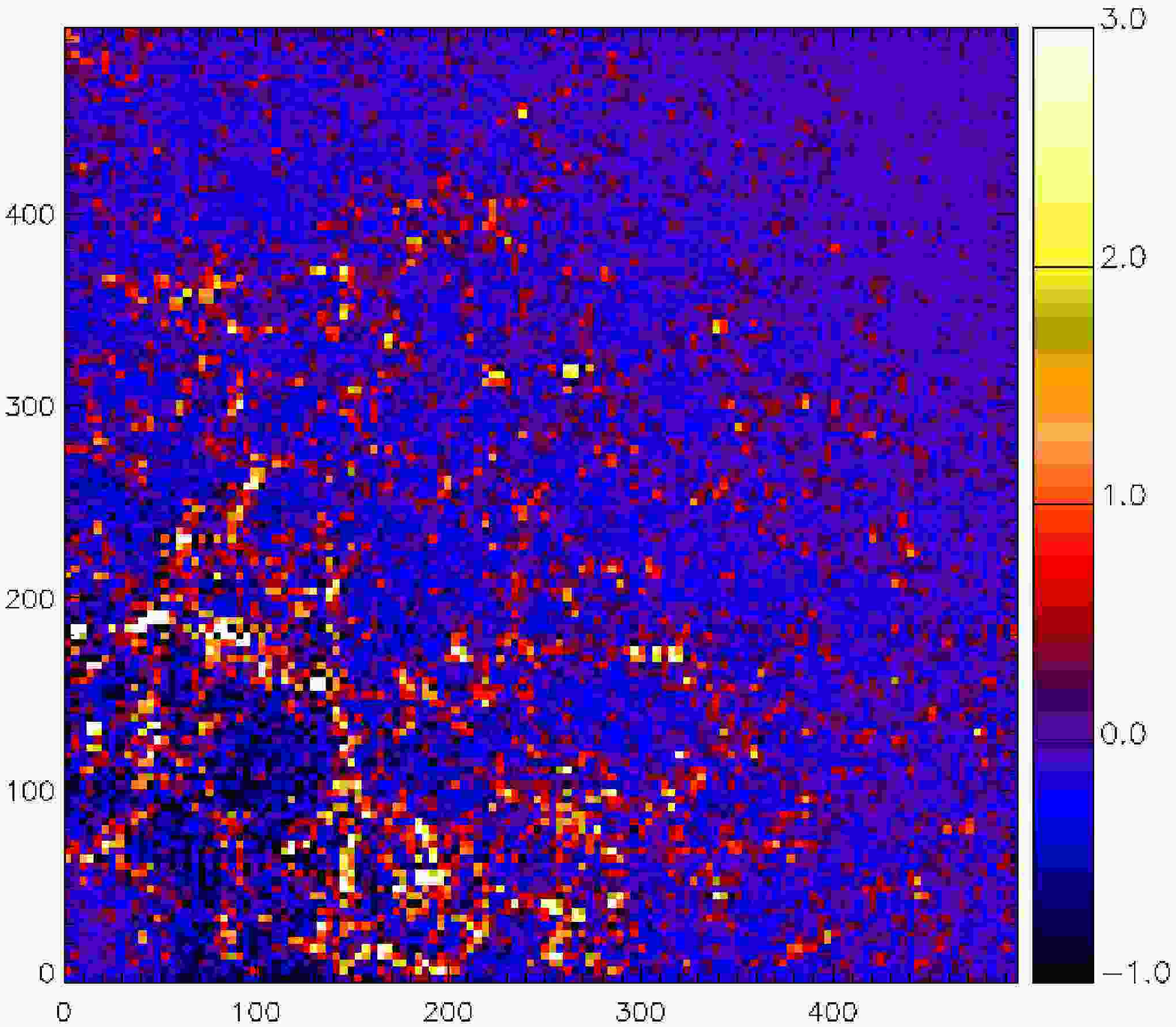}
\put(-252,0.5){{\huge (a)}}
\put(-138,-10){{X [Mpc]}}
\put(-249,93){\rotatebox[]{90}{Z [Mpc]}}
\hspace{0.5cm}
\includegraphics[width=8.5cm]{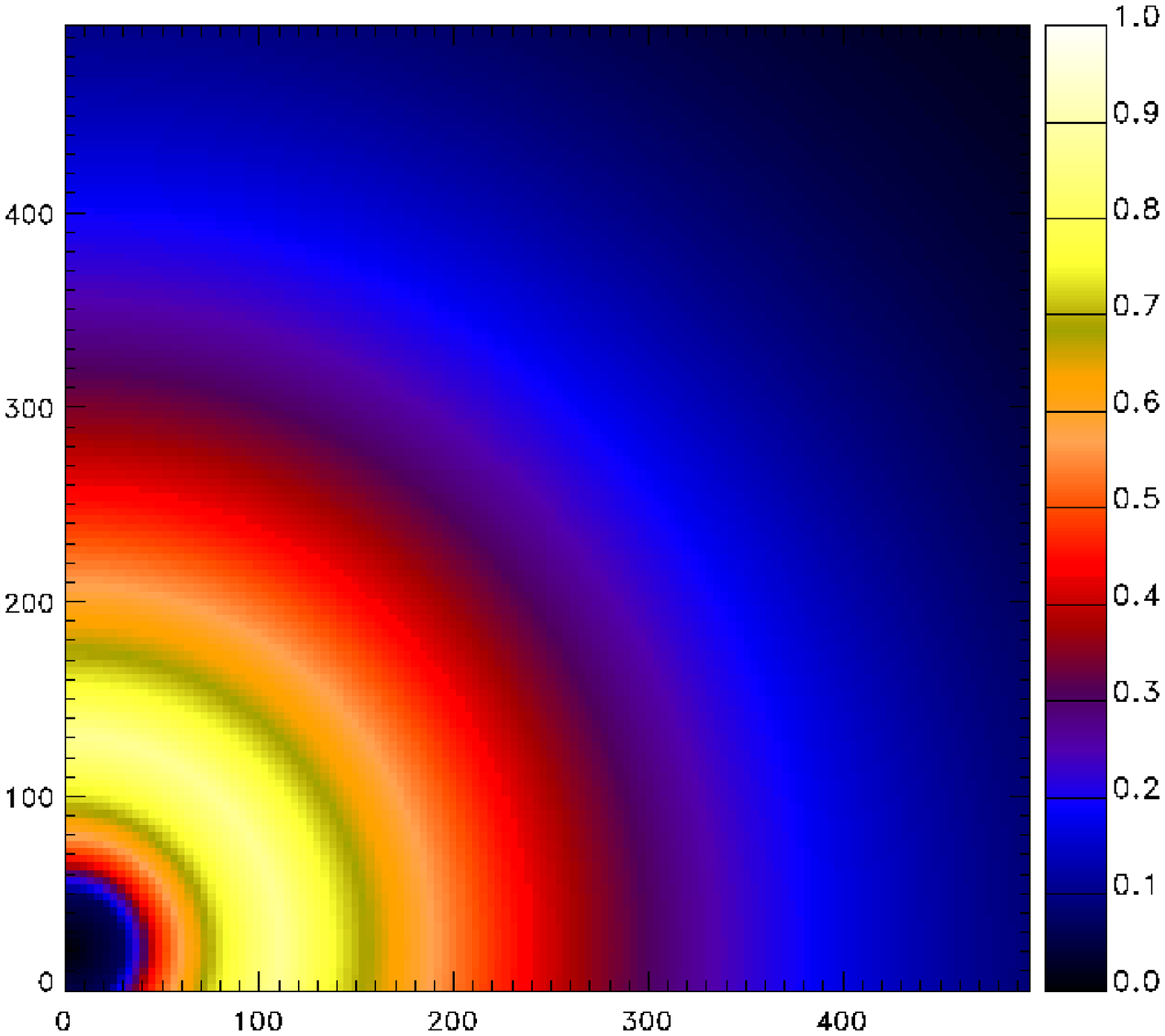}
\put(-252,1.5){{\huge (b)}}
\put(-138,-10){{X [Mpc]}}
\put(-249,93){\rotatebox[]{90}{Z [Mpc]}}
\\\\
\includegraphics[width=8.5cm]{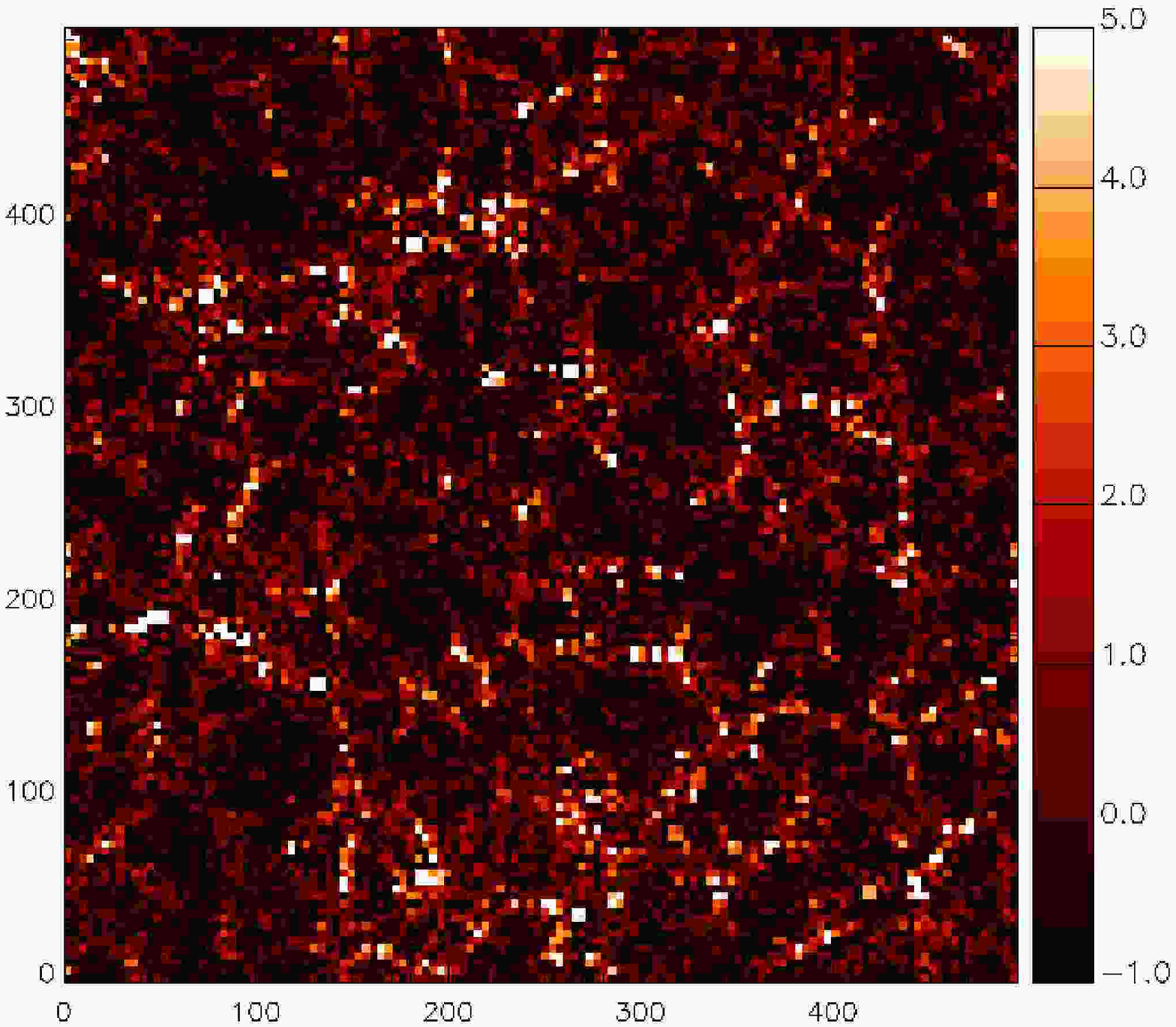}
\put(-252,0.5){{\huge (c)}}
\put(-138,-10){{X [Mpc]}}
\put(-249,93){\rotatebox[]{90}{Z [Mpc]}}
\hspace{0.5cm}
\includegraphics[width=8.5cm]{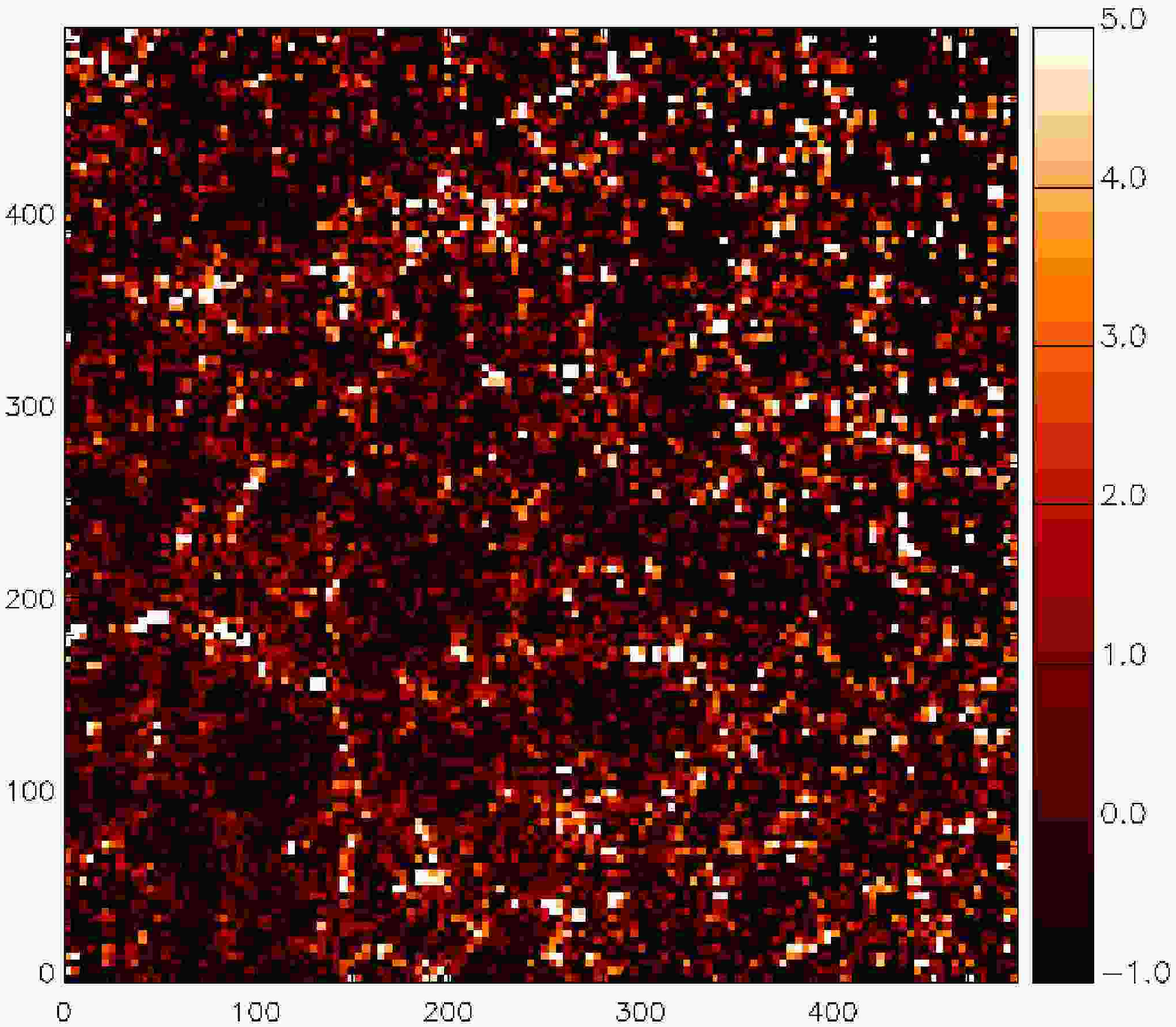}
\put(-252,1.5){{\huge (d)}}
\put(-138,-10){{X [Mpc]}}
\put(-249,93){\rotatebox[]{90}{Z [Mpc]}}
\\\\
\end{tabular}
\caption{Mock test 1 using $w_{\rm MOCK1}$. Input galaxy sample $\sim20$\% of the complete galaxy sample. Slices around Y$\sim270$ Mpc through a 500 Mpc cube box with a $128^3$ grid for different quantities without smoothing. Panel {(a)}: observed  mock galaxy overdensity field before correcting for the incompleteness. Panel {(b)}: DR6 radial completeness corresponding to this test.  Panel {(c)}: underlying complete mock  galaxy field. Panel {(d)}: inverse weighting scheme applied to the sample represented in (a). Note, that panels (a), (c) and (d) were created taking the mean over 10 neighboring slices around the slice at  Y$\sim270$ Mpc, corresponding to a thickness of 40 Mpc. } 
\label{fig:MOCKS}
\end{figure*}

\begin{figure*}
\begin{tabular}{cc}
\includegraphics[width=8.5cm]{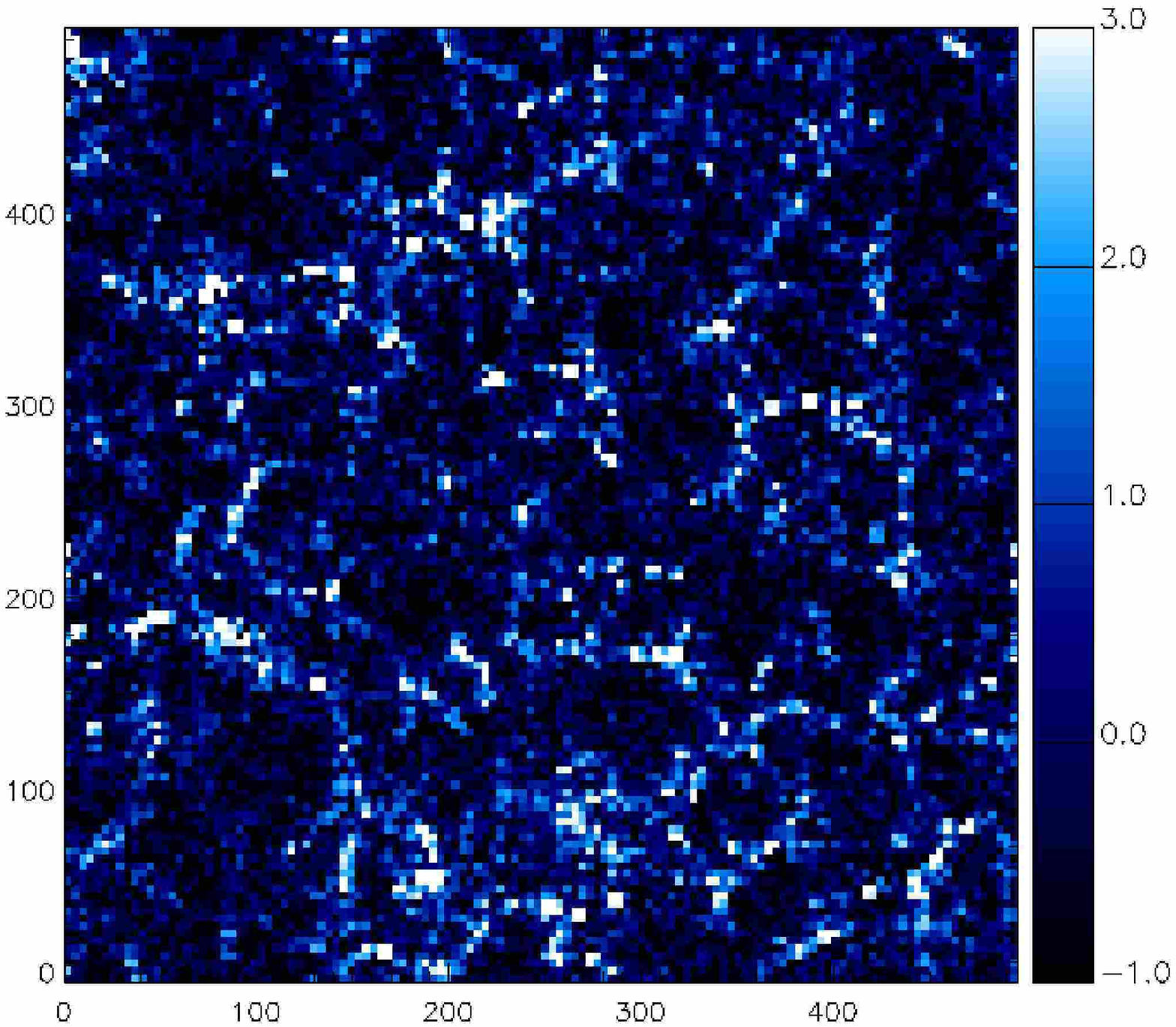}
\put(-252,0.5){{\huge (a)}}
\put(-138,-10){{X [Mpc]}}
\put(-249,93){\rotatebox[]{90}{Z [Mpc]}}
\hspace{0.5cm}
\includegraphics[width=8.5cm]{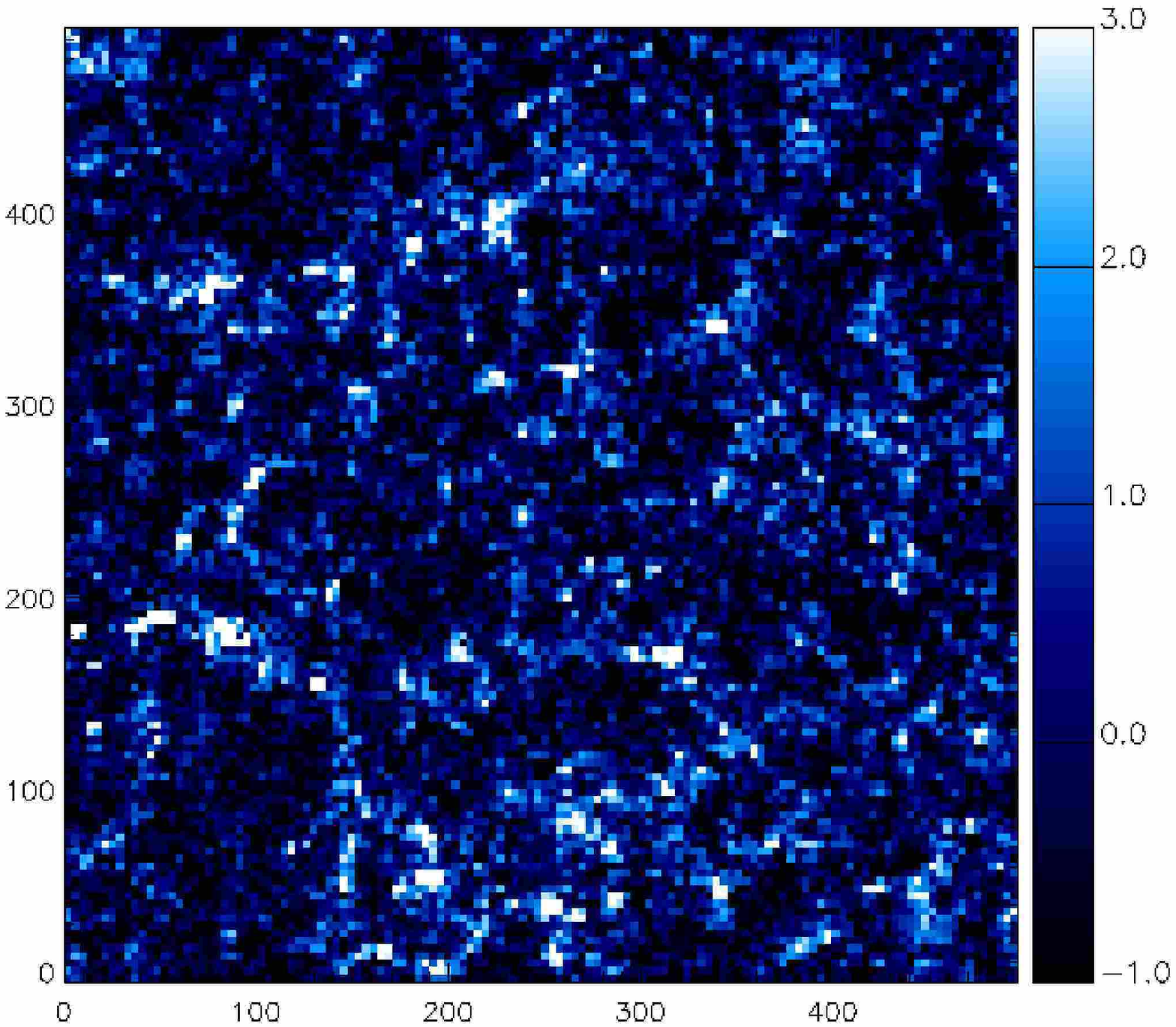}
\put(-252,1.5){{\huge (b)}}
\put(-138,-10){{X [Mpc]}}
\put(-249,93){\rotatebox[]{90}{Z [Mpc]}}
\\\\
\includegraphics[width=8.5cm]{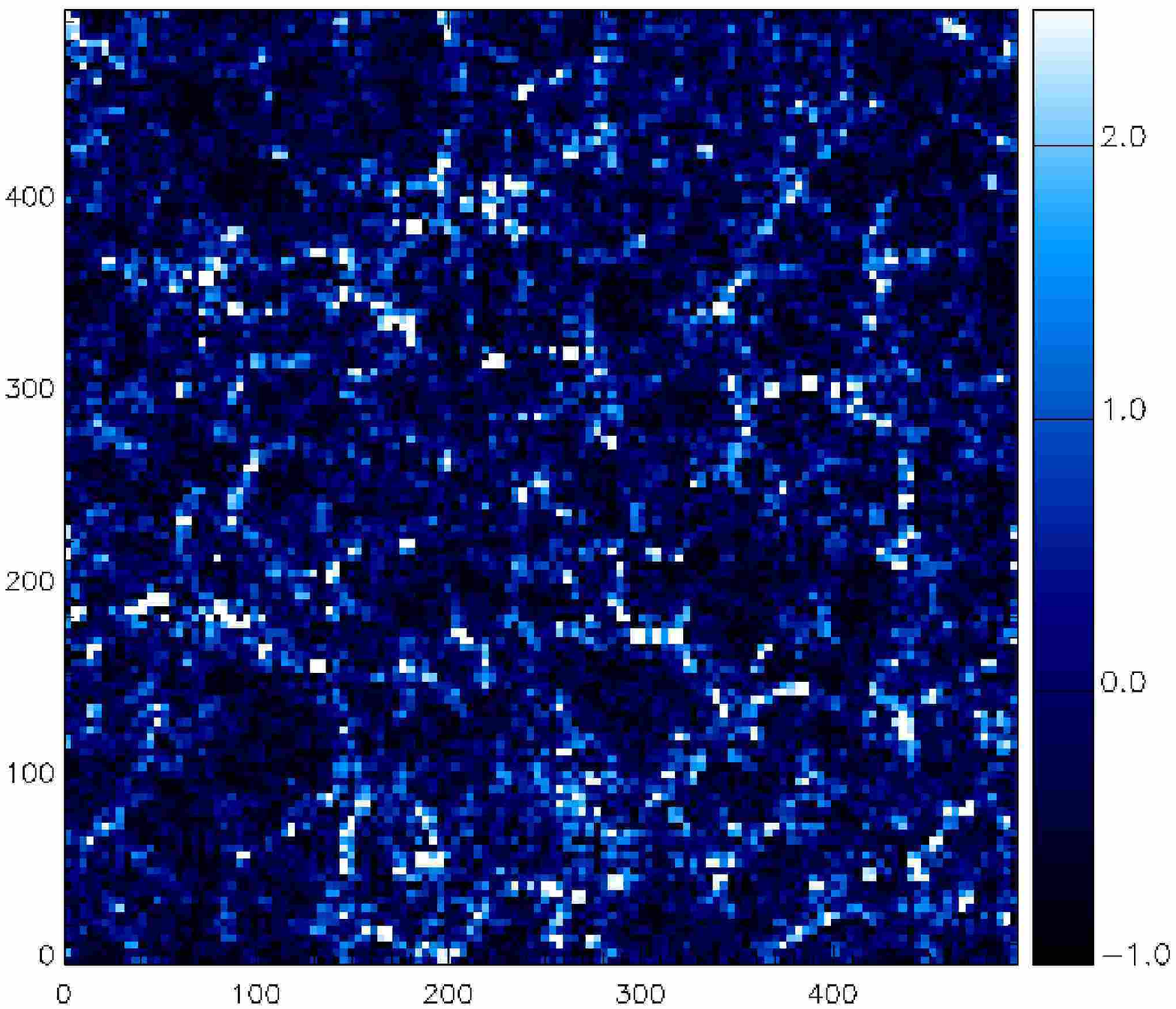}
\put(-252,0.5){{\huge (c)}}
\put(-138,-10){{X [Mpc]}}
\put(-249,93){\rotatebox[]{90}{Z [Mpc]}}
\hspace{0.5cm}
\includegraphics[width=8.5cm]{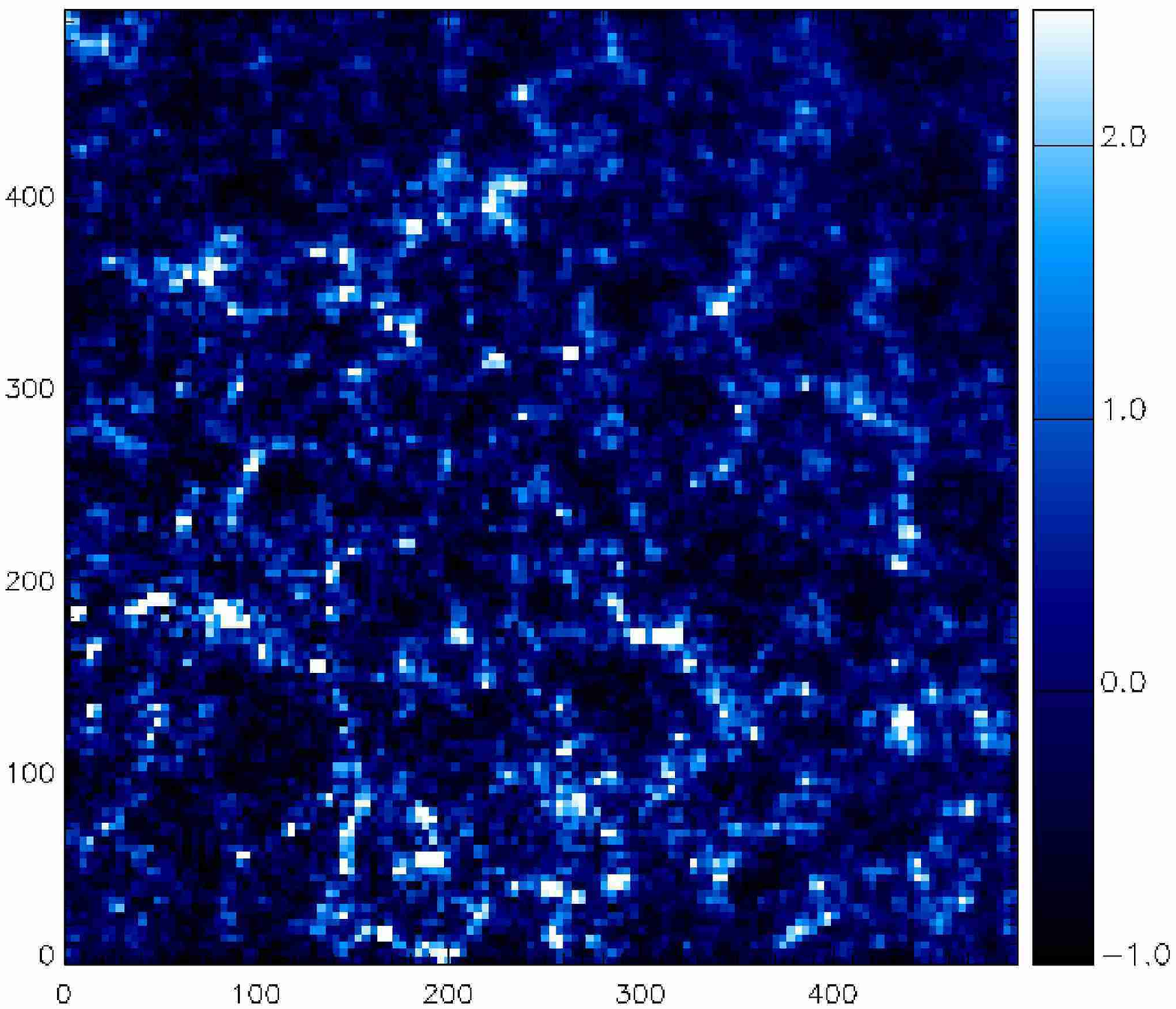}
\put(-252,1.5){{\huge (d)}}
\put(-138,-10){{X [Mpc]}}
\put(-249,93){\rotatebox[]{90}{Z [Mpc]}}
\\\\
\end{tabular}
\caption{Mock test 1 using $w_{\rm MOCK1}$. Input galaxy sample $\sim20$\% of the complete galaxy sample. Slices around Y$\sim270$ Mpc through a 500 Mpc cube box with a $128^3$ grid for different quantities without smoothing.  Panel (a): LSQ Wiener reconstruction to correct for the shot noise of the  mock  galaxy field taking the complete sample. Panel (b): LSQ Wiener reconstruction of the incomplete  mock  galaxy field taking into account the averaged shot noise and the radial selection function. Panel (c): mean over 200 Bayesian Wiener reconstructions to correct for the shot noise of the  mock  galaxy field taking the complete sample. Panel (d): mean over 200 SD Wiener reconstruction of the incomplete  mock  galaxy field taking into account shot noise and the radial selection function.  Note, that  all the panels  were created taking the mean over 10 neighboring slices around the slice at  Y$\sim270$ Mpc, i.e. over a slice of thickness 40 Mpc. } 
\label{fig:MOCKS12}
\end{figure*}

\begin{figure*}
\includegraphics[width=16cm]{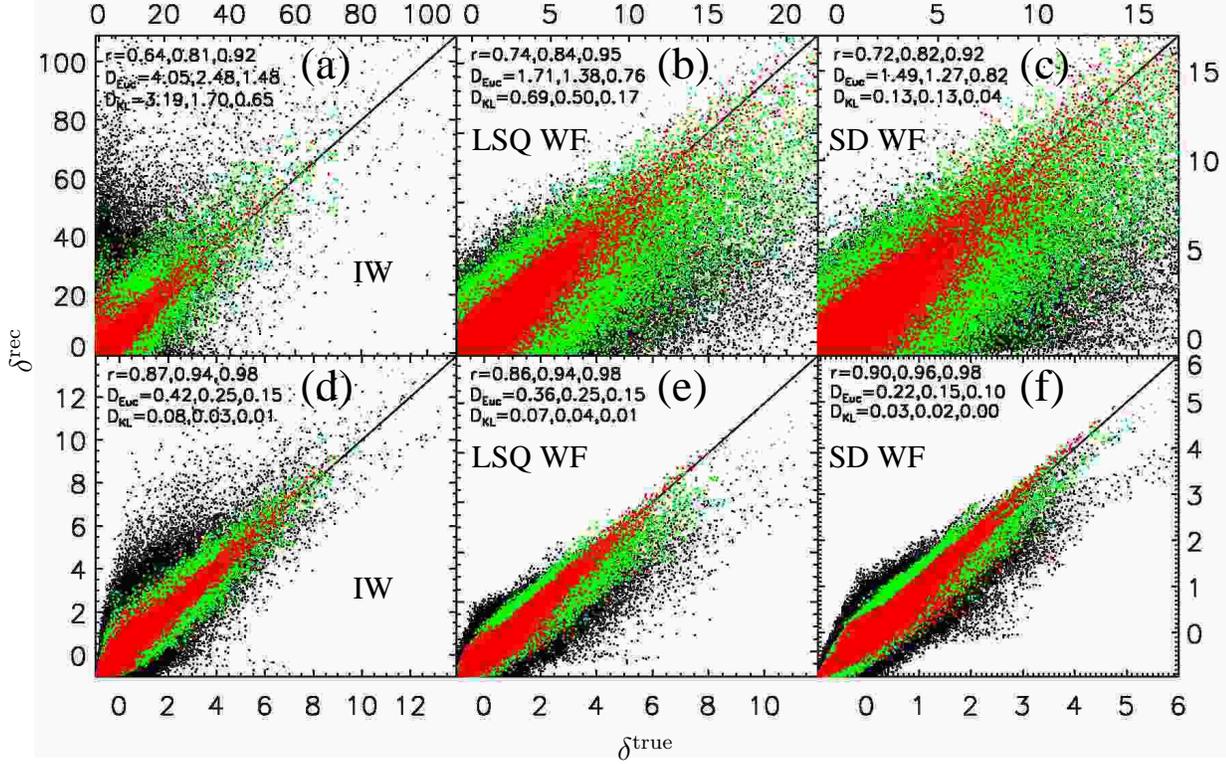}
\put(-355,257){{\huge (a)}}
\put(-335,180){{\Large IW}}
\put(-220,257){{\huge (b)}}
\put(-290,230){{\Large LSQ WF}}
\put(-83,257){{\huge (c)}}
\put(-155,230){{\Large SD WF}}
\put(-355,135){{\huge (d)}}
\put(-335,60){{\Large IW}}
\put(-220,135){{\huge (e)}}
\put(-290,110){{\Large LSQ WF}}
\put(-83,135){{\huge (f)}}
\put(-155,110){{\Large SD WF}}
\put(-235,-1.){\Large \bf $\delta^{\rm true}$}
\put(-465,150.){\rotatebox[]{90}{\Large\bf $\delta^{\rm rec}$}}\\
\caption{\label{fig:statscorr} Statistical cell to cell correlation between the mock {\it true} density field $\delta^{\rm true}$ and the reconstructed density field $\delta^{\rm rec}$ at different scales for our first test case using $w_{\rm MOCK1}$. Input galaxy sample $\sim20$\% of the complete galaxy sample. Also indicated: the statistical correlation coeficient r, the Euclidean distance D$_{\rm Euc}$ and the Kullback-Leibler distance D$_{\rm KL}$ first for all the sample (black dots), then for the sample in the radial comoving radius range between 200 and 400 Mpc (green dots), and finally in the range between 0 and 200 Mpc (red dots) away from the observer.  The upper panels correspond to the comparison without smoothing and the lower panels after smoothing with a smoothing radius of $r_{\rm S}=5$ Mpc.
Comparison between the complete  mock galaxy field (in this case: $\delta^{\rm true}$) and the inverse weighting scheme applied to the incomplete sample (in this case: $\delta^{\rm rec}$) without smoothing (a)  and after smoothing panel (d).  Panel (b) and  (e) represent the comparison between the average shot noise corrected complete mock  galaxy field (in this case: $\delta^{\rm true}$) and the LSQ Wiener reconstruction of the incomplete sample (in this case: $\delta^{\rm rec}$) with the corresponding scale at bottom or top. Panel (c) and  (f) represent the comparison between the local shot noise corrected complete mock  galaxy field (in this case: $\delta^{\rm true}$) and the SD Wiener reconstruction of the incomplete sample using the Jackknife estimator (in this case: $\delta^{\rm rec}$) with the corresponding scale at bottom or top.}  
\end{figure*}

\begin{figure*}
\includegraphics[width=16cm]{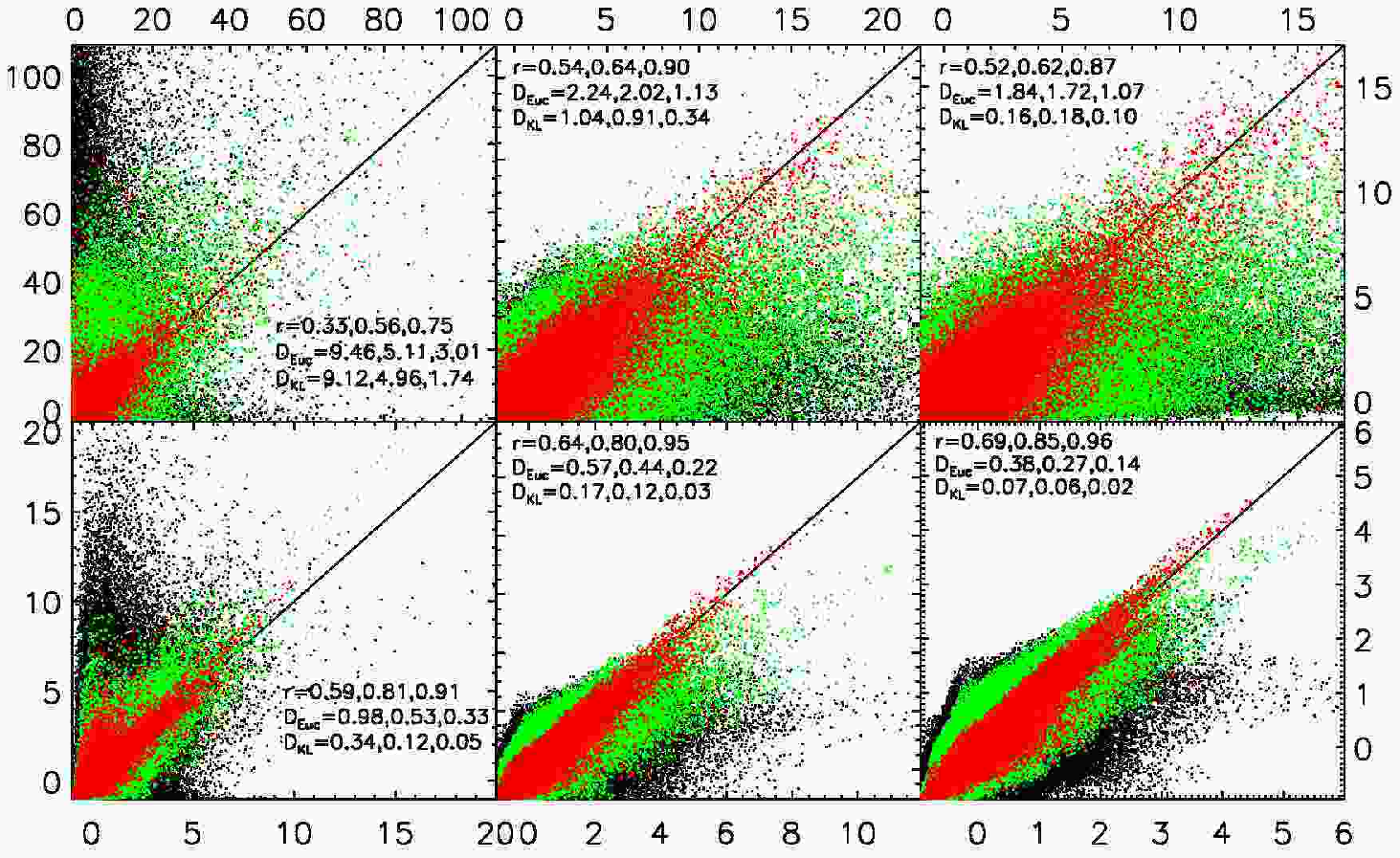}
\put(-355,257){{\huge (a)}}
\put(-335,200){{\Large IW}}
\put(-220,257){{\huge (b)}}
\put(-290,230){{\Large LSQ WF}}
\put(-83,257){{\huge (c)}}
\put(-155,230){{\Large SD WF}}
\put(-355,135){{\huge (d)}}
\put(-335,80){{\Large IW}}
\put(-220,135){{\huge (e)}}
\put(-290,110){{\Large LSQ WF}}
\put(-83,135){{\huge (f)}}
\put(-155,110){{\Large SD WF}}
\put(-235,-1.){\Large \bf $\delta^{\rm true}$}
\put(-465,150.){\rotatebox[]{90}{\Large\bf $\delta^{\rm rec}$}}\\
\caption{\label{fig:statscorr2}
Same as Fig.~\ref{fig:statscorr}, but using $w_{\rm MOCK2}$.
}  
\end{figure*}

In this section, we evaluate the quality of the reconstruction method under several incompleteness conditions. We restrict the study to a mesh of $128^3$ cells for a cube with 500 Mpc side length  and ignore bias and redshift distortion effects. The necessity of performing a reconstruction step to make further studies of the large-scale structure is addressed. More simple schemes in which the galaxies are just gridded and the resulting field smoothed are shown to lead to significantly worse estimates of the matter field. 

For this study, we consider a homogeneous subsample of $10^6$ galaxies in a 500 Mpc cube box from the mock galaxy catalogue by \citet[][]{DeLucia-Blaizot-07} selected at random based on the Millennium Simulation \citep[][]{2005Natur.435..629S}. We define the $10^6$ galaxy sample as our complete sample. 
Then, we generate two incomplete samples by radially selecting the galaxies according to two different radial completeness functions $w_{\rm MOCK1}$ and $w_{\rm MOCK2}$ (see Fig.~\ref{fig:sel}). This is done by drawing random uniform numbers between 0 and 1 for each mock galaxy  and selecting the galaxies depending on whether the drawn number is above or below the value of the completeness at the corresponding distance to the observer. Note, that this ensures a perfect binomial observation process treating all the galaxies independent of their luminosity and thus avoiding the problem of galaxy biasing.  The observer is defined in both cases at an equivalent position in the box to the real observer in the application to the observed DR6 data (section \ref{sec:results}), namely at X=0 Mpc, Y=250 Mpc, and Z=20 Mpc. Note, that the arbitrary coordinates of the mock data range from 0 to 500 Mpc in each direction X, Y, and Z.

We consider the LSQ formulation of the Wiener-filter, which is a linear filter with a homogeneous noise term multiplied with a structure function given by the selection function Eq.~\ref{eq:noiseav}, and the signal-dependent noise formulation, which is a nonlinear filter as it depends on the signal (see Eq.~\ref{eq:noisecov}),and the inverse weighting scheme.
In addition, to the Wiener-reconstruction methods, we define an inverse weighting scheme (IW) to estimate the underlying matter field as follows: first each galaxy is weighted with the inverse of the completeness at its location, then the galaxy sample is gridded according to the corresponding particle masses  (we use our supersampling scheme to suppress aliasing), and finally the resulting field is convolved with different smoothing kernels. The first part  of this scheme, leaving the smoothing for a later step, can be summarized by the following Eq.:  
\begin{equation}
\langle {n}(\mbi r)\rangle_{\rm IW}\equiv\Pi\big(\frac{\mbi r}{H}\big)\int {\rm d}{\mbi r'}\,K_{\rm S}(\mbi r-\mbi r')\frac{1}{w(\mbi r')} n^{\rm o}_{\rm p}(\mbi r'){,}
\label{def:ngrid}
\end{equation}
where we have denoted the corresponding estimator by the angles: $\langle \{ \, \}\rangle_{\rm IW}$.
Note, that the completeness cannot be zero at a position in which a galaxy was observed. 
In order to make  a quantitative comparison between the two Wiener-filtering methods and the inverse weighting method,  a {\it true} underlying field $\delta^{\rm true}$ needs to be defined. Since the inverse weighting scheme does not correct for the shot noise, we will compare  with the complete mock galaxy sample (see panel (c) in Fig.~\ref{fig:MOCKS}) after smoothing on different scales. Note, that a consistent comparison for this case is difficult, since the shot noise varies with the different galaxy samples and with the distance to the observer. For the Wiener reconstruction case study we define the {\it true} underlying matter field $\delta^{\rm true}$ as the resulting Wiener reconstruction taking the complete mock galaxy sample (see panel (e) in Fig.~\ref{fig:MOCKS}).   Note that the {\it true} field thus also differs between our two Wiener filtering schemes.
We will denote the reconstructed fields with each method as $\delta^{\rm rec}$.


\subsection{Statistical correlation measures}

To give a quantitative measurement of the quality of the reconstructions, we  define the correlation coefficient $\rm r$ between the reconstructed and the {\it true} density field by\footnote{Not to be confused with the comoving distance $r$.}
\begin{equation}
{\rm r}(\delta^{\rm rec},\delta^{\rm true})\equiv\frac{\sum^{N_{\rm cells}}_i\delta^{\rm true}_{ i}\delta^{\rm rec}_i}{\sqrt{\sum^{N_{\rm cells}}_i\left(\delta^{\rm true}_{i}\right)^2}\sqrt{\sum^{N_{\rm cells}}_j\left(\delta^{\rm rec}_j\right)^2}}{.}
\end{equation}
The cell to cell plot of the reconstruction against the {\it true} density field 
 is highly informative because the scatter in the alignment of the cells around the line of perfect correlation (45$^\circ$ slope) gives a qualitative goodness of the reconstruction.  
 In general, the quality of the recovered density map is better represented by the Euclidean distance between the {\it true} and the reconstructed signal \citep[see][]{kitaura}. The ensemble average of this quantity over all possible density realizations can also be regarded as an action or loss function that leads to the Wiener-filter through minimization \citep[see][]{kitaura}.
Here we introduce the Euclidean distance:
\begin{equation}
{\rm D}_{\rm Euc}(\delta^{\rm rec},\delta^{\rm true})\equiv\sqrt{\frac{1}{N_{\rm cells}} \sum^{N_{\rm cells}}_i\,\left(\delta^{\rm rec}_i-\delta^{\rm true}_i)\right)^2}{,}
\label{eq:enseres}
\end{equation}
 with $N_{\rm cells}=128^3$ for the mock tests).
Let us, in addition, define the normalized Kullback-Leibler distance\footnote{also called relative entropy in information theory} \citep[see][]{KL}  as 
\begin{equation}
{\rm D}_{\rm KL}(1+\delta^{\rm rec},1+\delta^{\rm true})\equiv\frac{1}{N_{\rm cells}}\sum^{N_{\rm cells}}_i\,(1+\delta^{\rm rec}_i)\log{\left(\frac{1+\delta^{\rm rec}_i}{1+\delta^{\rm true}_i}\right)}{.}
\label{eq:enseres}
\end{equation}

In our analysis we also compute smoothed versions of the density field convolving  it with a Gaussian kernel given by: 
\begin{equation}
G(\mbi r, r_{\rm S})\equiv\exp\left(\frac{|\mbi r|^2}{2r_{\rm S}^2}\right){,}
\label{eq:GAUSS}
\end{equation}
with $r_{\rm S}$ being the smoothing radius.


\subsection{First mock test}

In the first mock test we try to emulate the same completeness conditions as given in the observed DR6 sample. For that, we take the complete mock galaxy catalogue ($10^6$ galaxies) and select according to the DR6 radial selection function ($w_{\rm MOCK1}=w_{\rm DR6}$) a subsample  leaving about 20\% of the total number of galaxies (218020) (see Fig.~\ref{fig:MOCKS}). 
The DR6 radial selection function can be seen as the black line in Fig.~\ref{fig:sel}. A section through the box showing the completeness can be also seen in panel (b) of Fig.~\ref{fig:MOCKS}. The observer can be identified as being at the center of the spherical shells with equal completeness.
The resulting overdensity field after applying this selection function to the complete mock sample can be seen in panel (a) of Fig.~\ref{fig:MOCKS}.
Note, that we show here the mock observed galaxy field setting $w=1$ in Eq.~\ref{eq:datamodel} in order to clearly see the selection effects.
In the following, the discrete galaxy field (including Poisson noise) is represented with red color and the noise corrected field is represented in blue color.
 We will define the complete mock galaxy field including Poisson noise (panel (c) in Fig.~\ref{fig:MOCKS}) as the {\it true} galaxy density field for the inverse weighting scheme. The corresponding noise corrected fields using the LSQ WF (panel (a) in Fig.~\ref{fig:MOCKS12})  and the SD WF (panel (b) in Fig.~\ref{fig:MOCKS12}) are defined as the {\it true} galaxy density field for the Wiener reconstructions. The {\it true} dark matter field is approximately related to this via Eq.~\ref{eq:bias}, however, here we want to exclude the complication of galaxy biasing.  

Panel (d) in Fig.~\ref{fig:MOCKS} shows the result after applying the inverse weighting scheme.
Panels (b) and (d) of Fig.~\ref{fig:MOCKS12} show the respective reconstructions using the LSQ and the SD WF.
One can clearly see the noisy reconstruction produced by the inverse weighting scheme for structures located at large distances to the observer in contrast to the smoother estimation made by the Wiener-filtering schemes.
The SD WF was applied for the complete galaxy sample using our statistically unbiased Jackknife-like scheme with an $\alpha$ parameter of $10^{-3}$. The means after 200 reconstructions are shown in panels (c) and (d) for the complete and the selected samples respectively. 
 The corresponding statistical analysis can be seen in Fig.~\ref{fig:statscorr}. 
The cell to cell correlation plots show the tendency of the inverse weighting scheme to overestimate the density while the opposite is {true} in a significantly more moderate way when applying the Wiener-filter. 
In the case without smoothing (a mesh of size $\sim3.9$ Mpc) (panels (a) and (d) in Fig.~\ref{fig:statscorr})) the qualitative and quantitative difference between the methods is very large, showing significantly better correlation coefficient and lower Euclidean and Kullback-Leibler distances for the Wiener reconstructions than for the inverse weighting scheme. Only when the fields are  smoothed with a Gaussian of radius $r_{\rm S}=5$ Mpc does the difference between the matter field estimators drop. With this smoothing the statistical correlation coefficient are similar for the Wiener-filter and the inverse weighting scheme. However, the Euclidean and Kullback-Leibler distances remain being lower for the Wiener-filter (WF) reconstructions (see Fig.~\ref{fig:statscorr}).


\subsection{Second mock test}

For the second mock test results we modify the DR6 selection function to drop faster towards larger radii leaving less than 10\% of the galaxies (87220)  by weighting $w_{\rm DR6}(r)$ with the factor 100 Mpc/$r$ for $r\geq100$ Mpc.
The corresponding radial selection function ($w_{\rm MOCK2}$) can be seen as the dashed line in Fig.~\ref{fig:sel}. The dramatic difference from DR6 completeness can be seen. 
 using LSQ and SD formulations respectively.
The noisy reconstruction produced by the inverse weighting scheme for structures located at large distances to the observer is now even more visible than in the previous test.
Cells far away from the observed are excessively weighted. 
The Wiener-filter in contrast gives a smoother and more conservative estimation in regions in which the data are more incomplete. However, it remains sharp in regions where the information content is high (see structures close to the observer).

 The corresponding statistical analysis can be seen in Fig.~\ref{fig:statscorr2}. 
The tendency to overestimate the density of the inverse weighting scheme is now extreme. Smoothing helps to raise the correlation coefficient values and to decrease the Euclidean and Kullback-Leibler distances. They remain, however, clearly above the ones achieved with the Wiener-filter schemes. 


\begin{figure*}
\begin{tabular}{cc}
\includegraphics[width=9.0cm]{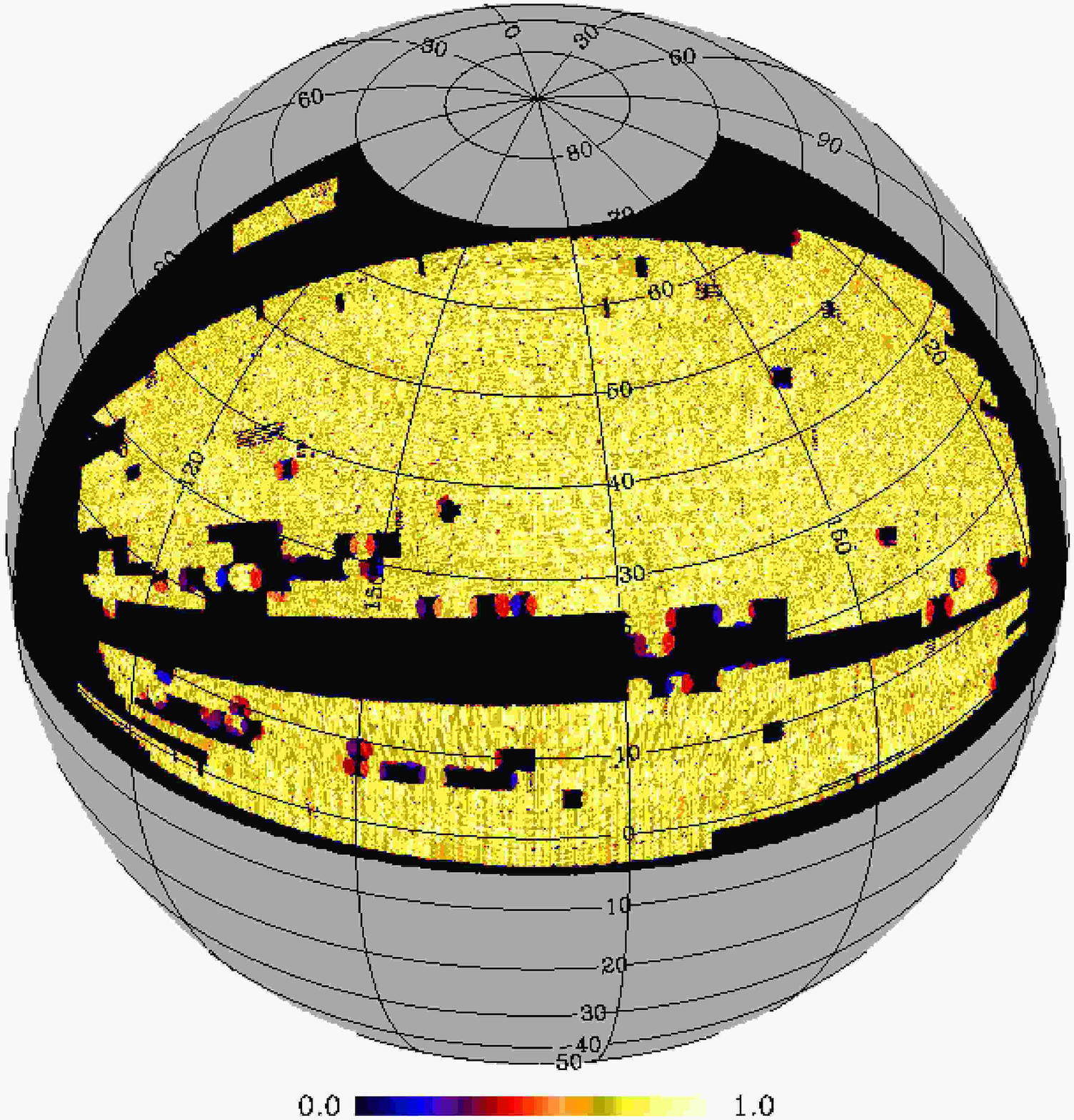}
\put(-240,0.5){{\huge (a)}}
\put(-55,3.){{$w_{\rm SKY}$}}
\includegraphics[width=9.0cm]{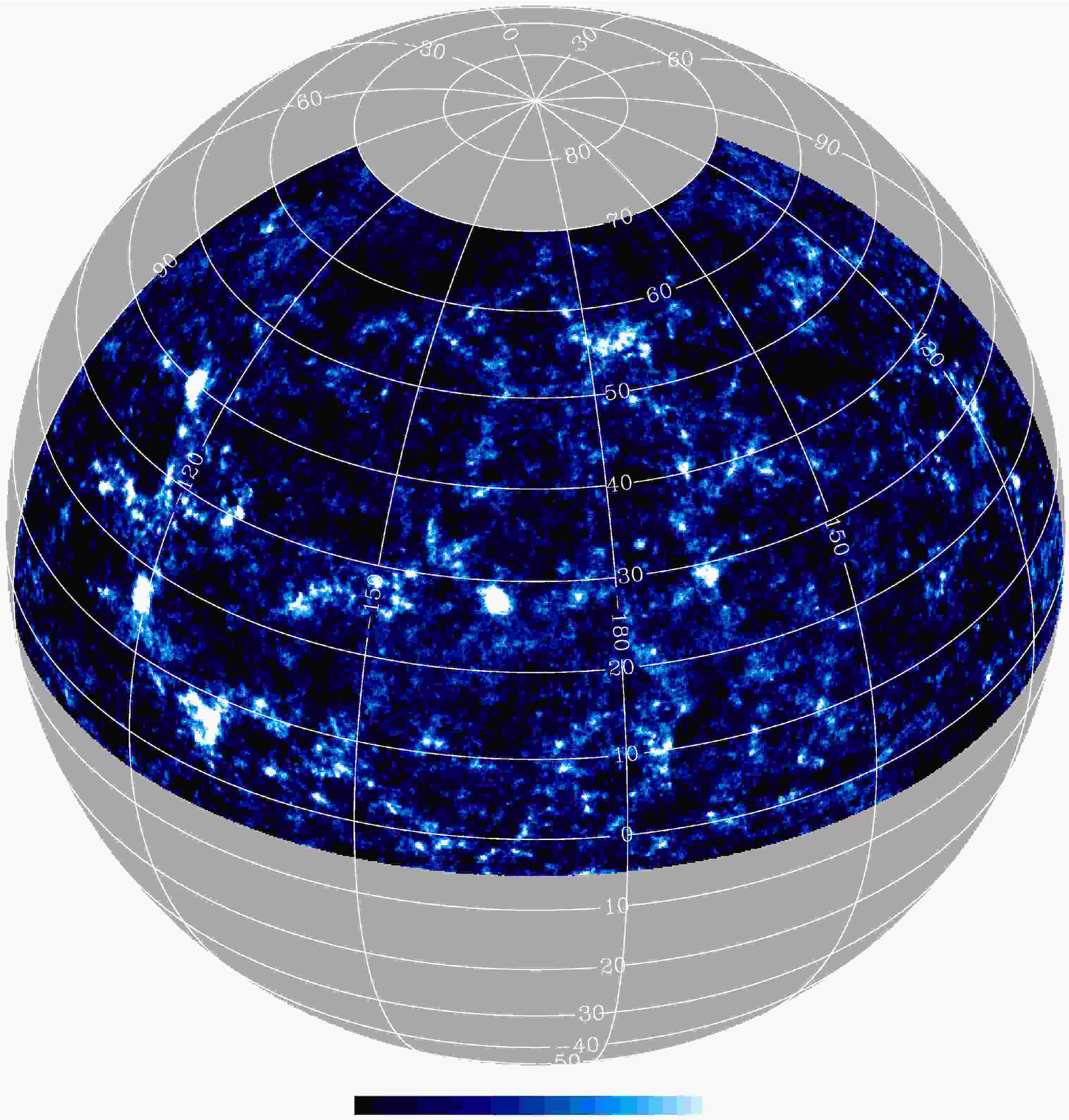}
\put(-240,0.5){{\huge (b)}}
\put(-185,3.){{0.5}}
\put(-80,3.){{2.5}}
\put(-55,3.){{$1+\delta$}}
\\
\includegraphics[width=9.0cm]{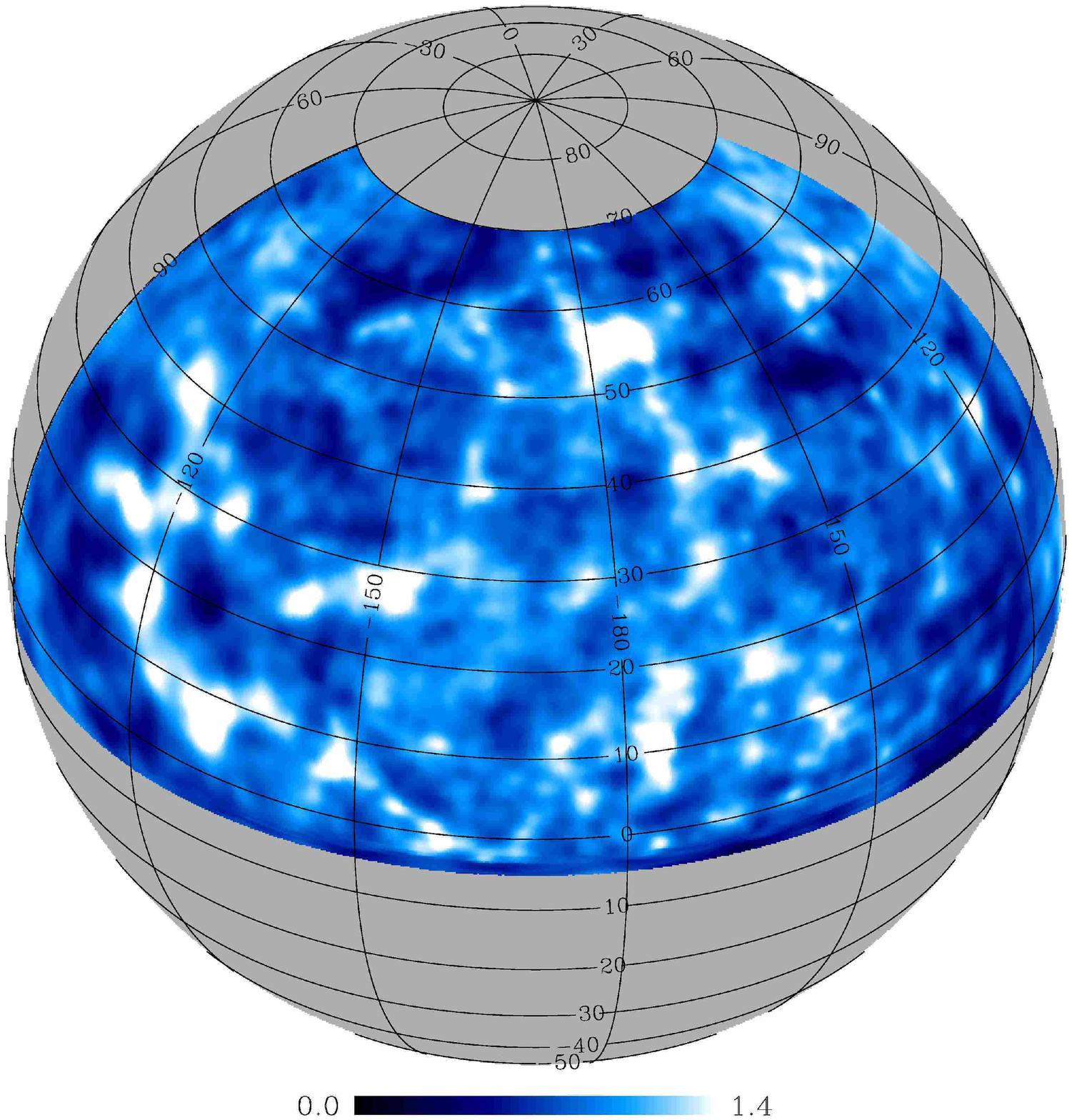}
\put(-240,0.5){{\huge (c)}}
\put(-55,3.){{$1+\delta$}}
\includegraphics[width=9.0cm]{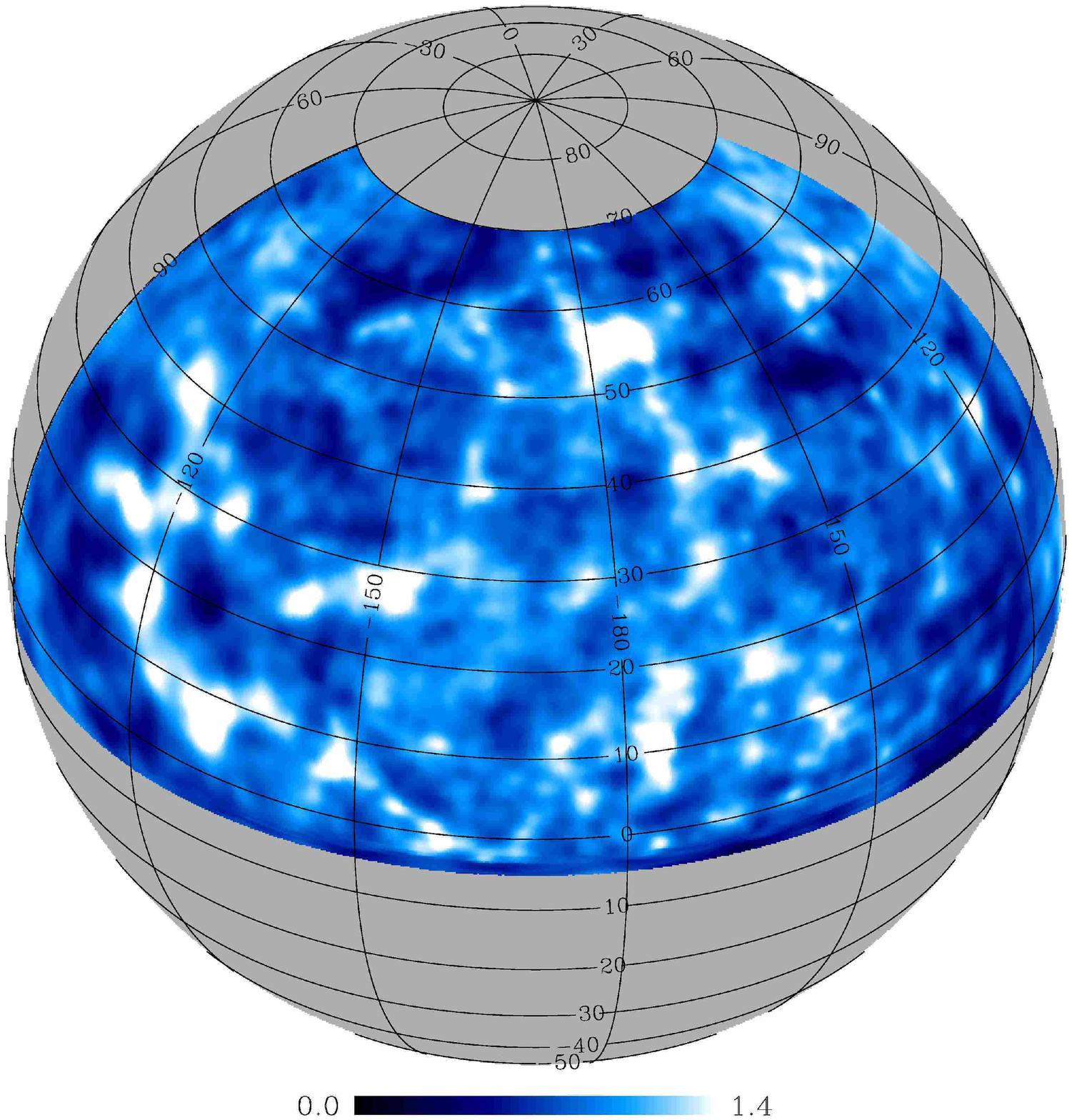}
\put(-240,0.5){{\huge (d)}}
\put(-55,3.){{$1+\delta$}}
\vspace{1cm}
\end{tabular}
\caption{Panel {(a)}: completeness of the observed patches on the sky. Shown are projections  on the sky of the three dimensional matter field  reconstruction, including the deconvolution with a redshift distortions operator and divided by the number of line-of sight grid-points used for the calculation to obtain a mean density field on the sky: without smoothing (panel {(b)}),  after a convolution with a Gaussian kernel with a smoothing radius of $r_{\rm S}=$5 Mpc (panel {(c)}) and   $r_{\rm S}=$10 Mpc (panel {(d)}). Note, that the longitude  angles -90$^\circ$, -120$^\circ$, -150$^\circ$ and -180$^\circ$ correspond to 270$^\circ$, 240$^\circ$, 210$^\circ$ and 180$^\circ$ right ascension angles, respectively, with the positive angles being equal. For a general right ascension angle $\alpha$ the longitude is calculated as: $\alpha$-360$^\circ$ for $\alpha\geq180^\circ$. The latitude angles are identical to the declination angles. }
\label{fig:SKY}
\end{figure*}

\section{Matter field reconstructions of the SDSS DR6}
\label{sec:results}

This work presents the first application of the
\textsc{argo}-code to observational data. This yields the matter field
reconstruction of the SDSS DR6 in the main area of the survey which is
located in the northern Galactic cap on a comoving cube of side 500 Mpc and $512^3$ cells.

In this section we describe  a few remarkable features in the
reconstructed matter field, demonstrating the quality of the
reconstruction and the scientific potential for future applications.
First, we discuss  the mask and the projected
three dimensional reconstruction without {smoothing} and after {smoothing} 
with a Gaussian kernel with a smoothing radius of $r_{\rm S}=$5 Mpc and $r_{\rm S}=$10
Mpc as displayed in Fig.~\ref{fig:SKY}.
We then describe the largest structures in the nearby Universe, in
particular the Sloan and CfA2 Great Walls \citep[see][]{2005ApJ...624..463G,1989Sci...246..897G}. Later, we analyze void or cluster detections which can be made
with this kind of work.  Finally, we
 analyze the statistical distribution of matter.

\subsection{Mask and completeness}
\label{sec:maskres}

The sky mask for the region  is shown in panel {(a)} of
Fig.~\ref{fig:SKY}. The high resolution ($36''$
in both $\alpha$ and $\delta$) permits us to visualize the plates of
the SDSS with the intersection of several plates leading to higher
completeness.  The mask is divided into three patches: one
small beam at high declination and right ascension angles and two wide
regions. All the patches together cover almost a quarter of the
sky. Between the two wider regions there is a large gap and there are
several additional smaller gaps inside the patches. Such a complex
mask is an interesting problem for the \textsc{argo}-code. 
It allows us to test, 
whether it can properly handle unobserved regions with zero
completeness.  Slices of the three dimensional mask calculated as the
product of the completeness on the sky and the selection function (see
section \ref{sec:mask}) are presented in panel {(a)} of
Figs.~\ref{fig:GW}, \ref{fig:VOID}, \ref{fig:GAP} and panel (e) of Fig.~\ref{fig:VOID}. In these plots one can see how
the selection function leads to a decrease of the completeness in
the radial direction. 
Note, that the observer is located at (0,0,0) in our
Cartesian coordinate system. We can see in panel {(a)} of
Fig.~\ref{fig:GW} that the completeness rapidly reaches  its maximum at
around 110 Mpc distance from the observer and decreases at larger
radii to values below 10\%.  
In the next section we show how
remarkably homogeneous structures are recovered in our
reconstruction, independent of the distance from the observer and despite
the low completeness values at large distances. We confirmed with 
additional reconstructions with larger volumes the same behavior for
boxes up to side lengths of around 750 Mpc. For even larger volumes
of 1 Gpc size, not shown here, however, the main sample becomes too sparse
and only the  large-scale structures are recovered. 
Including the three dimensional completeness for the SDSS DR6 data  (see section \ref{sec:mask}) in Eq.~\ref{eq:nmean} we obtain a mean galaxy density of about 0.05. 



\begin{table*}
\begin{tabular}{|c|c|c|c|c|c|} 
  \hspace{3cm} \\\hline
supercluster  &cluster & Abell number   & $\sim$ right ascension $\alpha$ [degrees] & $\sim$ declination $\delta$ [degrees] & $\sim$ redshift \\\hline\hline
Coma &Coma &A1656    & 195$^\circ$ (-165$^\circ$)& 28$^\circ$ & 0.0231 \\\hline\hline
Coma &Leo & A1367&    176$^\circ$ (176$^\circ$)&  20$^\circ$ & 0.0220\\\hline\hline
Hercules  && A2040 &   228$^\circ$ (-132$^\circ$)& 7$^\circ$ &     0.0448   \\\hline
Hercules && A2052 &   229$^\circ$ (-131$^\circ$)& 7$^\circ$ &     0.0338   \\\hline
Hercules && A2063  &  231$^\circ$ (-129$^\circ$)& 9$^\circ$ &     0.0341   \\\hline\hline
Hercules &Hercules&  A2151  &   241$^\circ$ (-119$^\circ$) & 18$^\circ$ &    0.0354   \\\hline
Hercules & & A2147   &  241$^\circ$ (-119$^\circ$) & 16$^\circ$ &   0.0338   \\\hline
Hercules & & A2152   &  241$^\circ$ (-119$^\circ$) & 16$^\circ$ &   0.0398    \\\hline\hline        
Hercules & & A2148  &   241$^\circ$ (-119$^\circ$) & 25$^\circ$ &   0.0418   \\\hline
Hercules & & A2162  &   243$^\circ$ (-117$^\circ$) & 29$^\circ$ &   0.0310   \\\hline\hline
Hercules & & A2197  &   247$^\circ$ (-113$^\circ$) & 41$^\circ$ &   0.0296   \\\hline           
Hercules & & A2199  &   247$^\circ$ (-113$^\circ$) & 40$^\circ$ &   0.0287   \\\hline
\end{tabular}
\caption{\label{tab:obscluster} Some of the most prominent clusters in the reconstruction with their corresponding right ascension and declination in degrees and redshift. Note, that the right ascension angle in Fig.~\ref{fig:SKY} is indicated in parenthesis and can be calculated as: $\alpha$-360$^\circ$ for $\alpha\geq180^\circ$.  }
\end{table*}

\subsection{Mapping the Sloan and the CfA2 Great Wall}
\label{sec:GW}

\begin{figure*}
\begin{tabular}{cc}
\includegraphics[width=9.0cm]{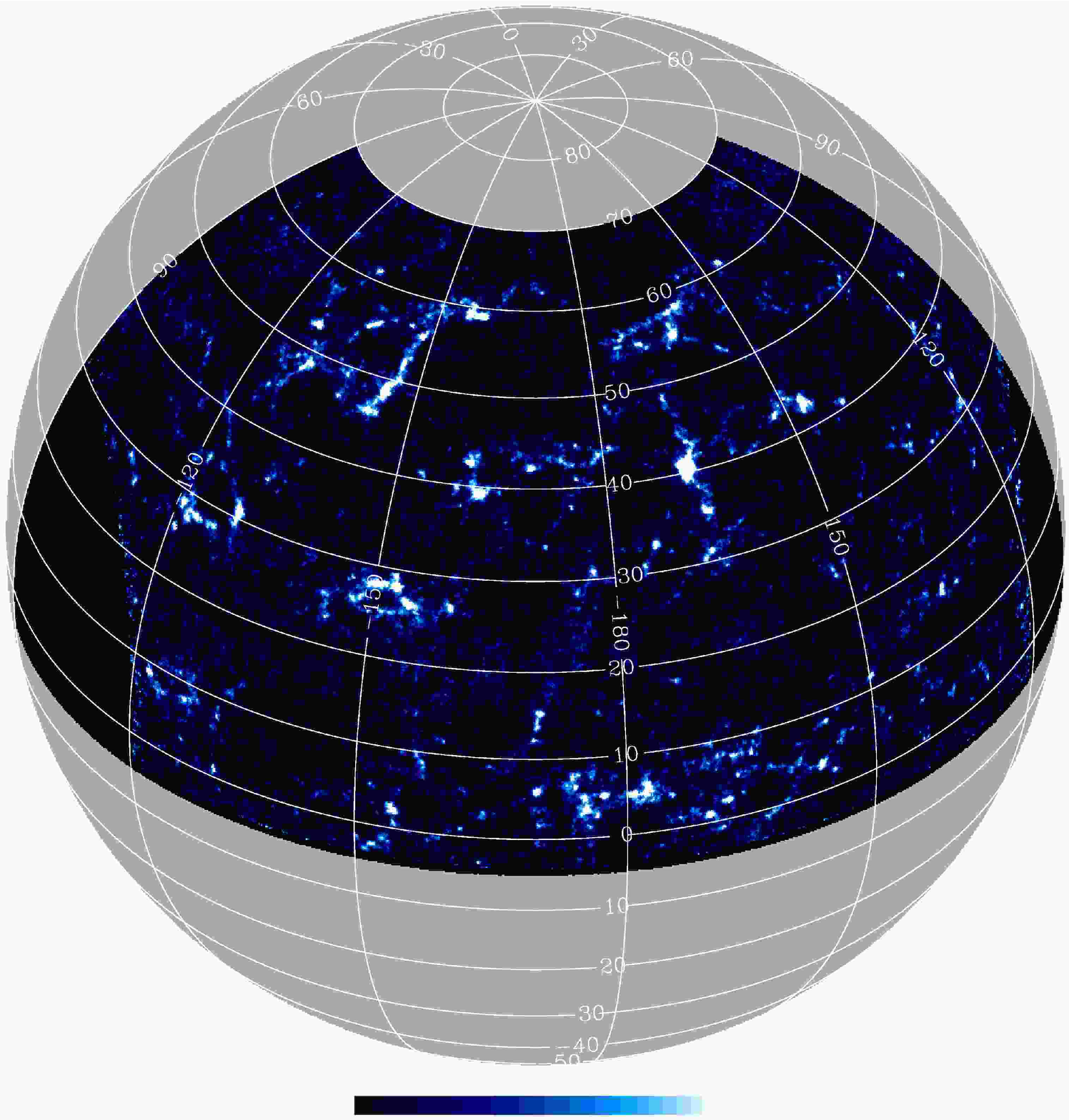}
\put(-240,0.5){{\huge (a)}}
\put(-185,3.){0.5}
\put(-80,3.){10.0}
\put(-55,3.){{1+$\delta$}}
\includegraphics[width=9.0cm]{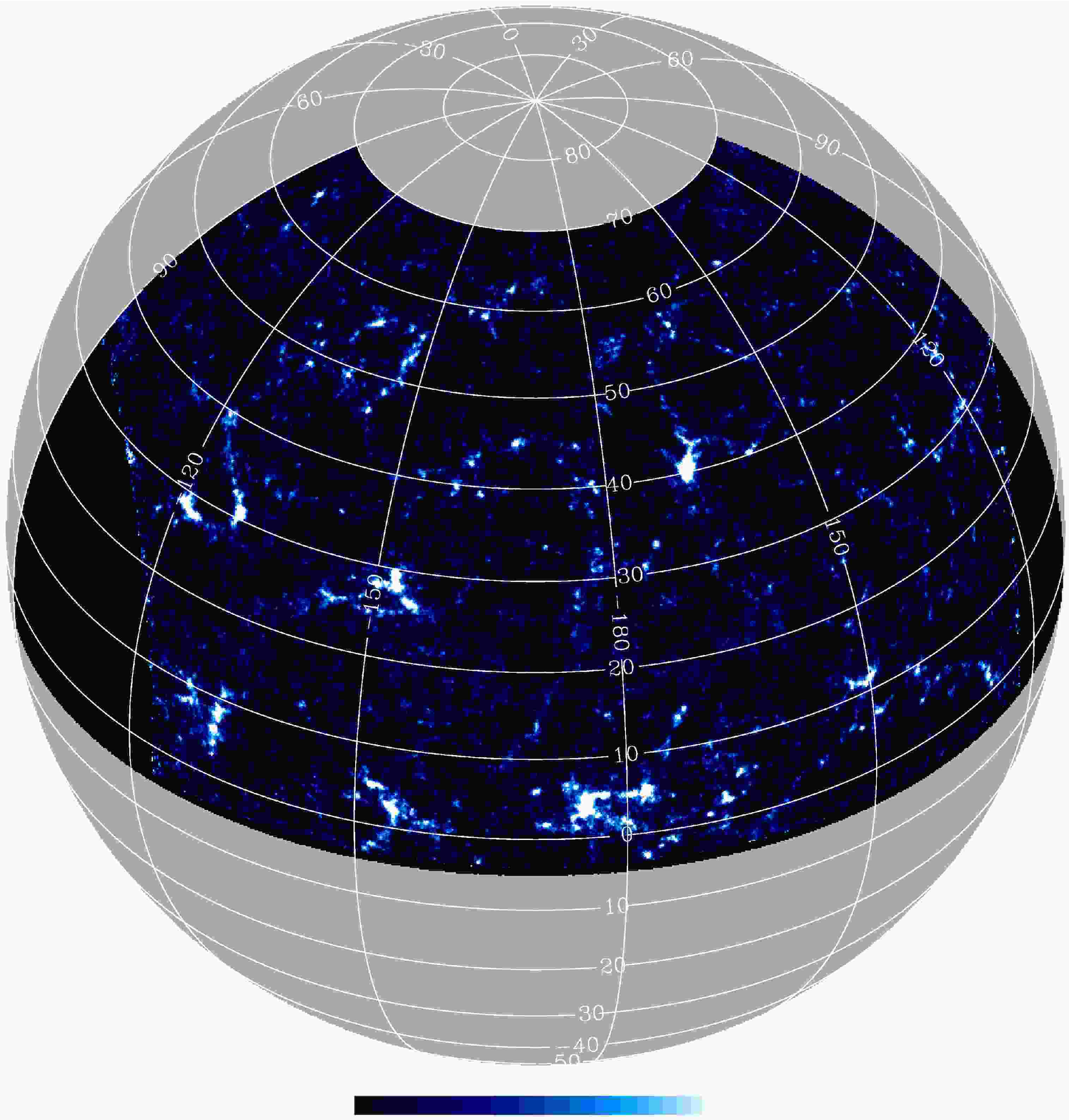}
\put(-240,0.5){{\huge (b)}}
\put(-185,3.){0.5}
\put(-80,3.){10.0}
\put(-55,3.){{1+$\delta$}}
\\
\includegraphics[width=9.0cm]{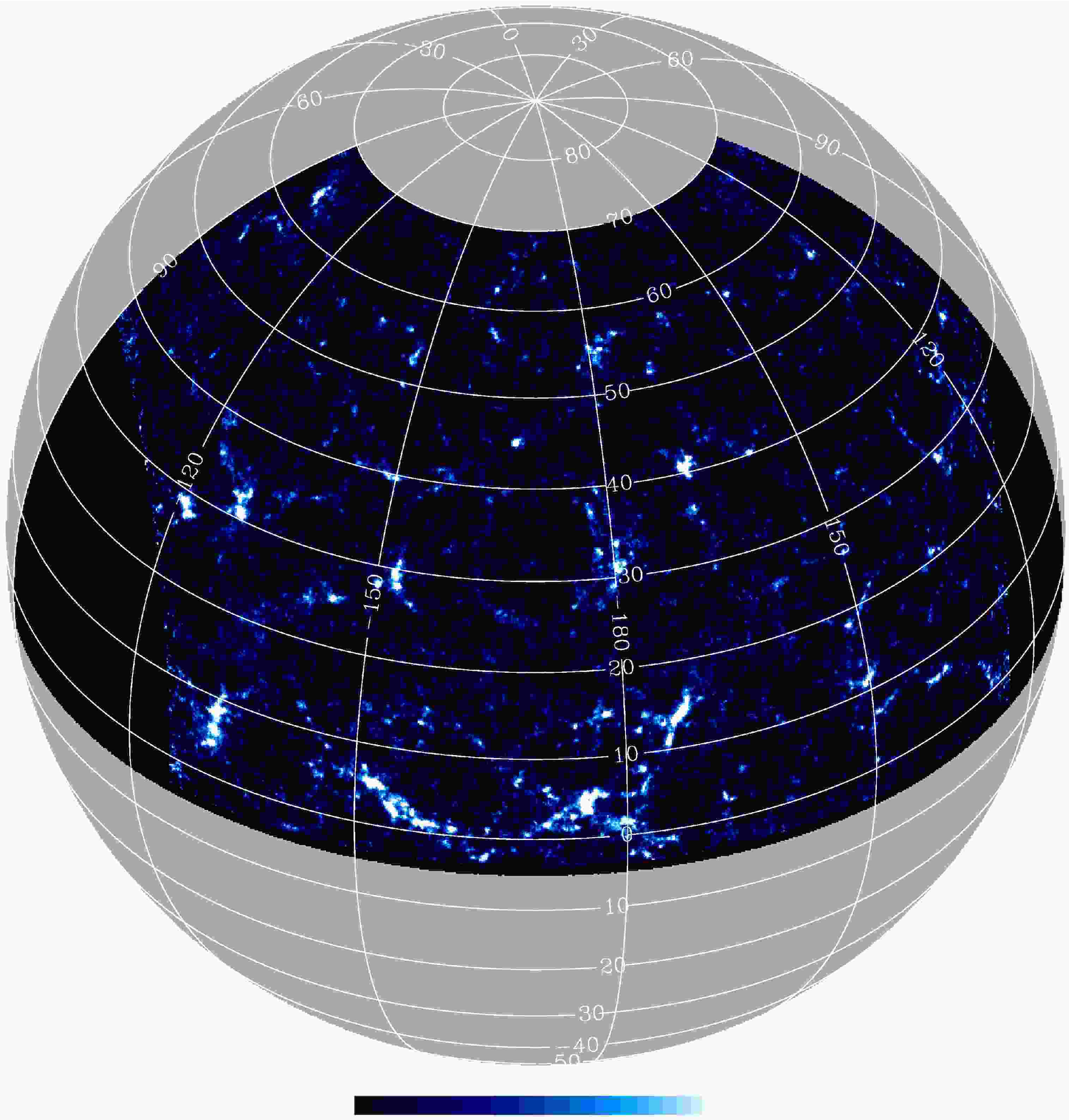}
\put(-240,0.5){{\huge (c)}}
\put(-185,3.){0.5}
\put(-80,3.){10.0}
\put(-55,3.){{1+$\delta$}}
\includegraphics[width=9.0cm]{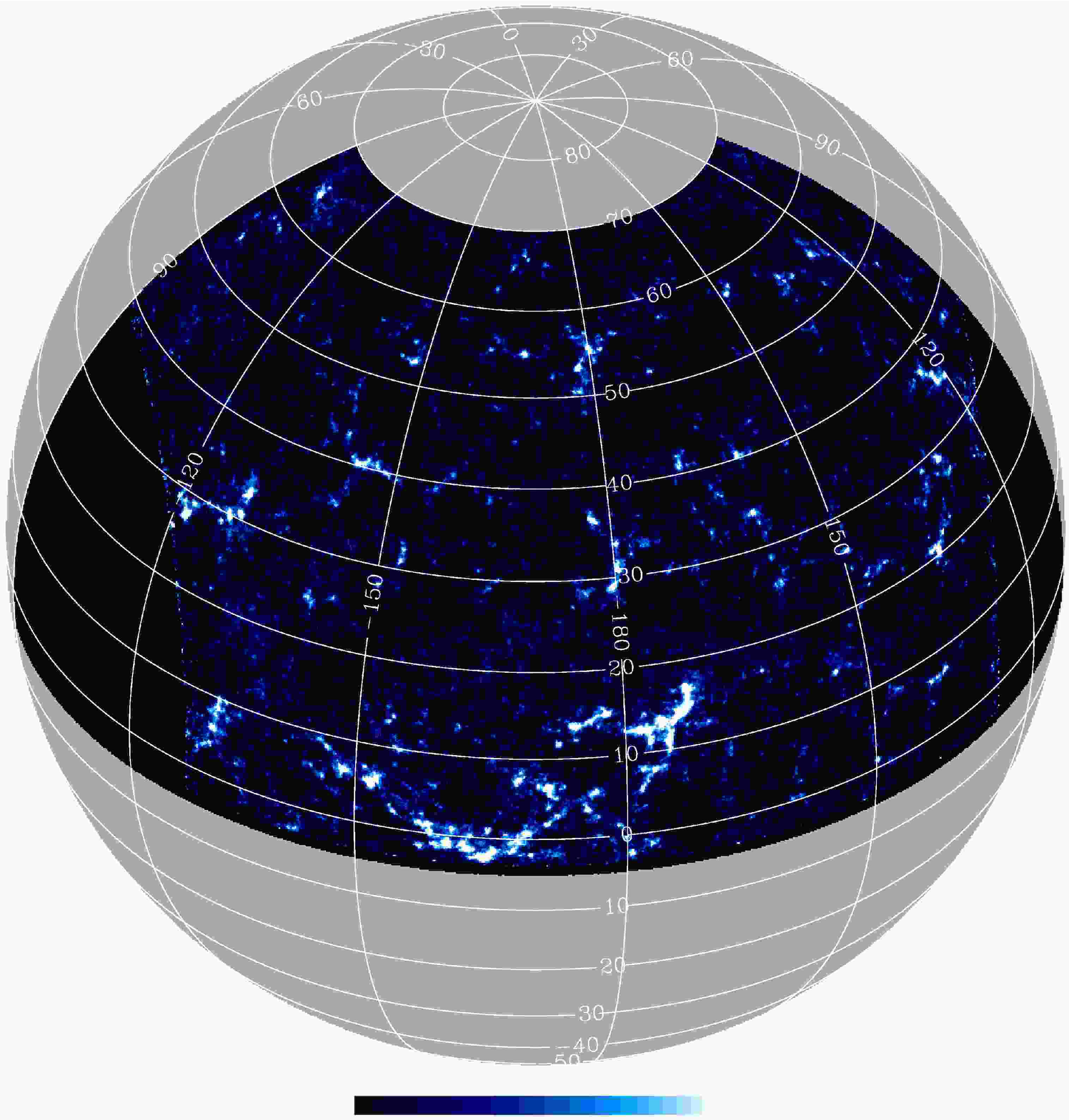}
\put(-240,0.5){{\huge (d)}}
\put(-185,3.){0.5}
\put(-80,3.){10.0}
\put(-55,3.){{1+$\delta$}}
\vspace{1cm}
\end{tabular}
\caption{Different radial slices around the Sloan Great Wall. Shown are projections of the three dimensional matter field  reconstruction on the sky  considering only cells with a comoving distance between 290 Mpc and 310 Mpc (panel {(a)}), 300 Mpc and 320 Mpc (panel {(b)}), 310 Mpc and 330 Mpc (panel {(c)}), and 320 Mpc and 340 Mpc (panel {(d)}). Note, that the longitude  angles -90$^\circ$, -120$^\circ$, -150$^\circ$ and -180$^\circ$ correspond to 270$^\circ$, 240$^\circ$, 210$^\circ$ and 180$^\circ$ right ascension angles, respectively, with the positive angles being equal. For a general right ascension angle $\alpha$ the longitude is calculated as: $\alpha$-360$^\circ$ for $\alpha\geq180^\circ$. The latitude angles are identical to the declination angles. }
\label{fig:RECSKYGW}
\end{figure*}

\begin{figure*}
\begin{tabular}{cc}
\includegraphics[width=8.5cm]{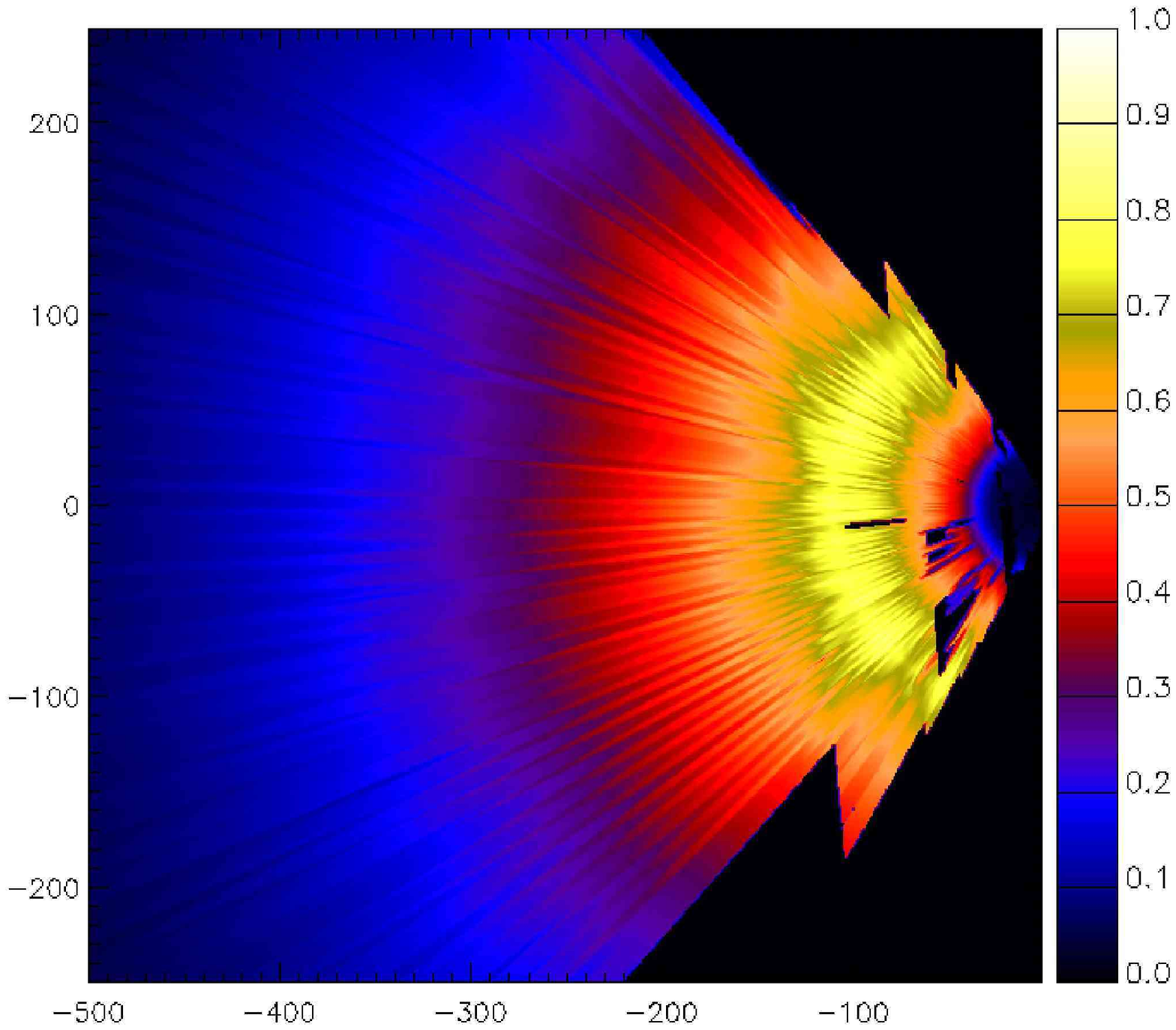}
\put(-250,0.5){{\huge (a)}}
\put(-138,-10){{X [Mpc]}}
\put(-245,93){\rotatebox[]{90}{Y [Mpc]}}
\hspace{0.5cm}
\includegraphics[width=8.5cm]{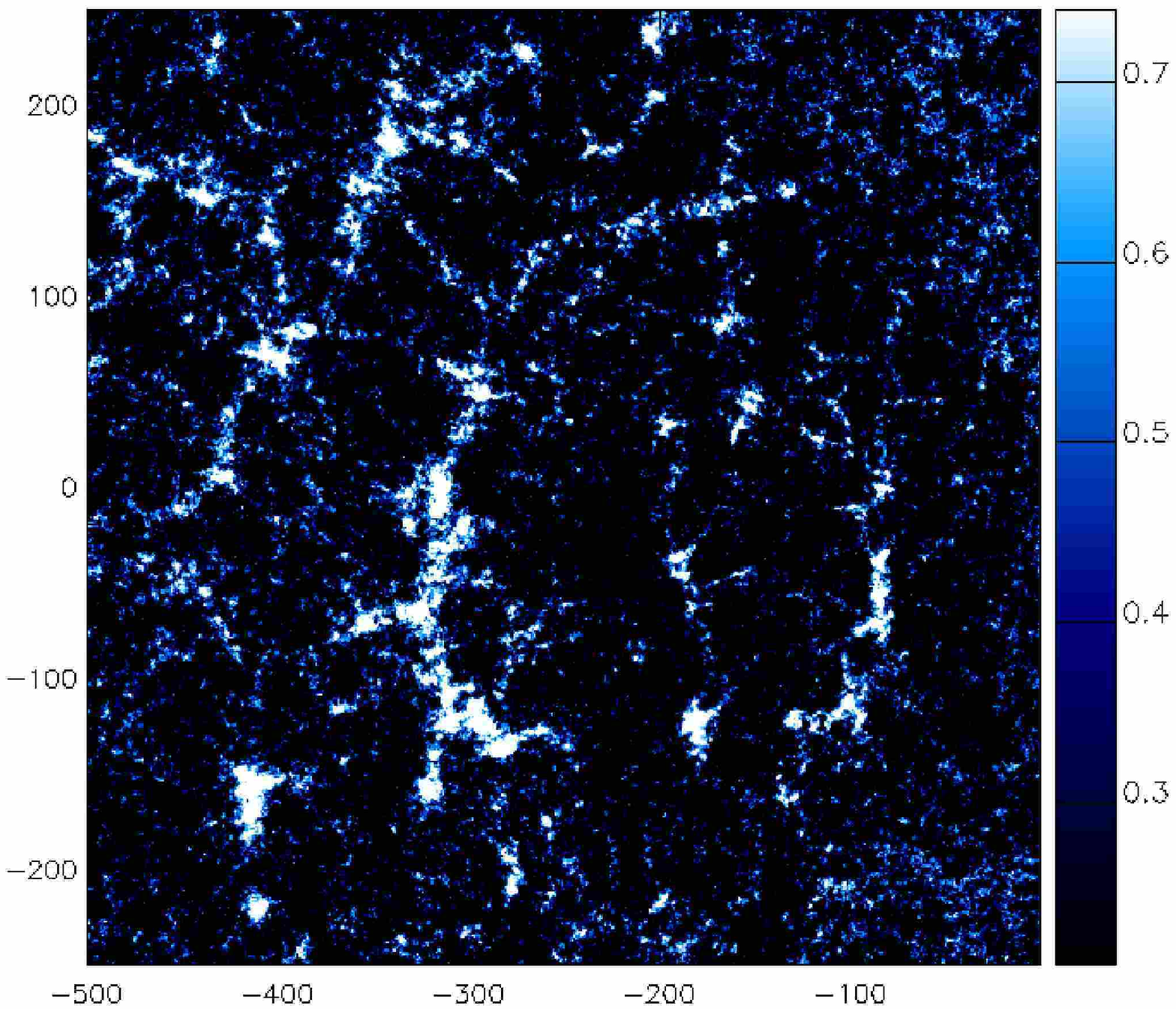}
\put(-250,1.5){{\huge (b)}}
\put(-138,-10){{X [Mpc]}}
\put(-245,93){\rotatebox[]{90}{Y [Mpc]}}
\\\\
\includegraphics[width=8.5cm]{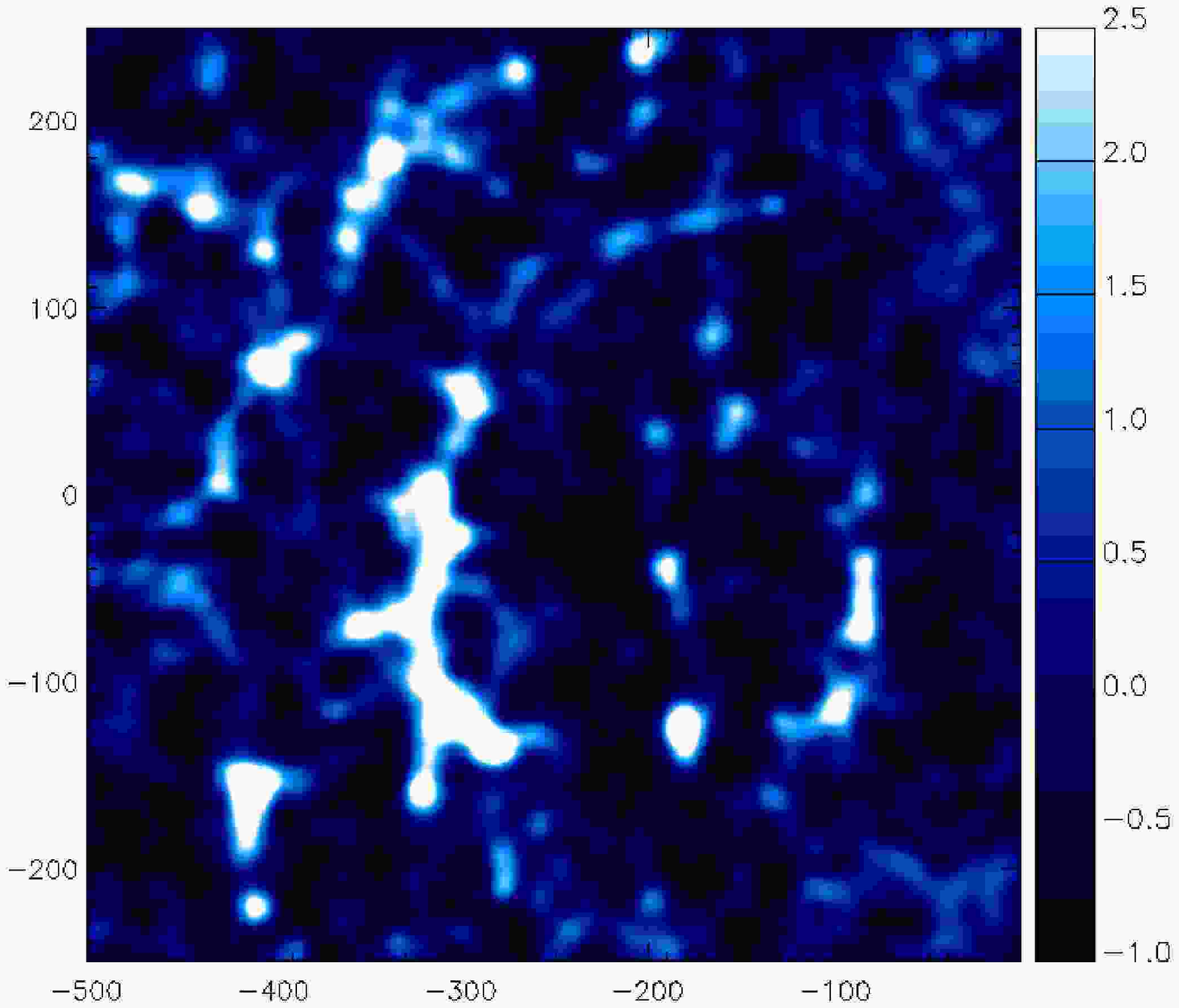}
\put(-250,0.5){{\huge (c)}}
\put(-138,-10){{X [Mpc]}}
\put(-245,93){\rotatebox[]{90}{Y [Mpc]}}
\hspace{0.5cm}
\includegraphics[width=8.5cm]{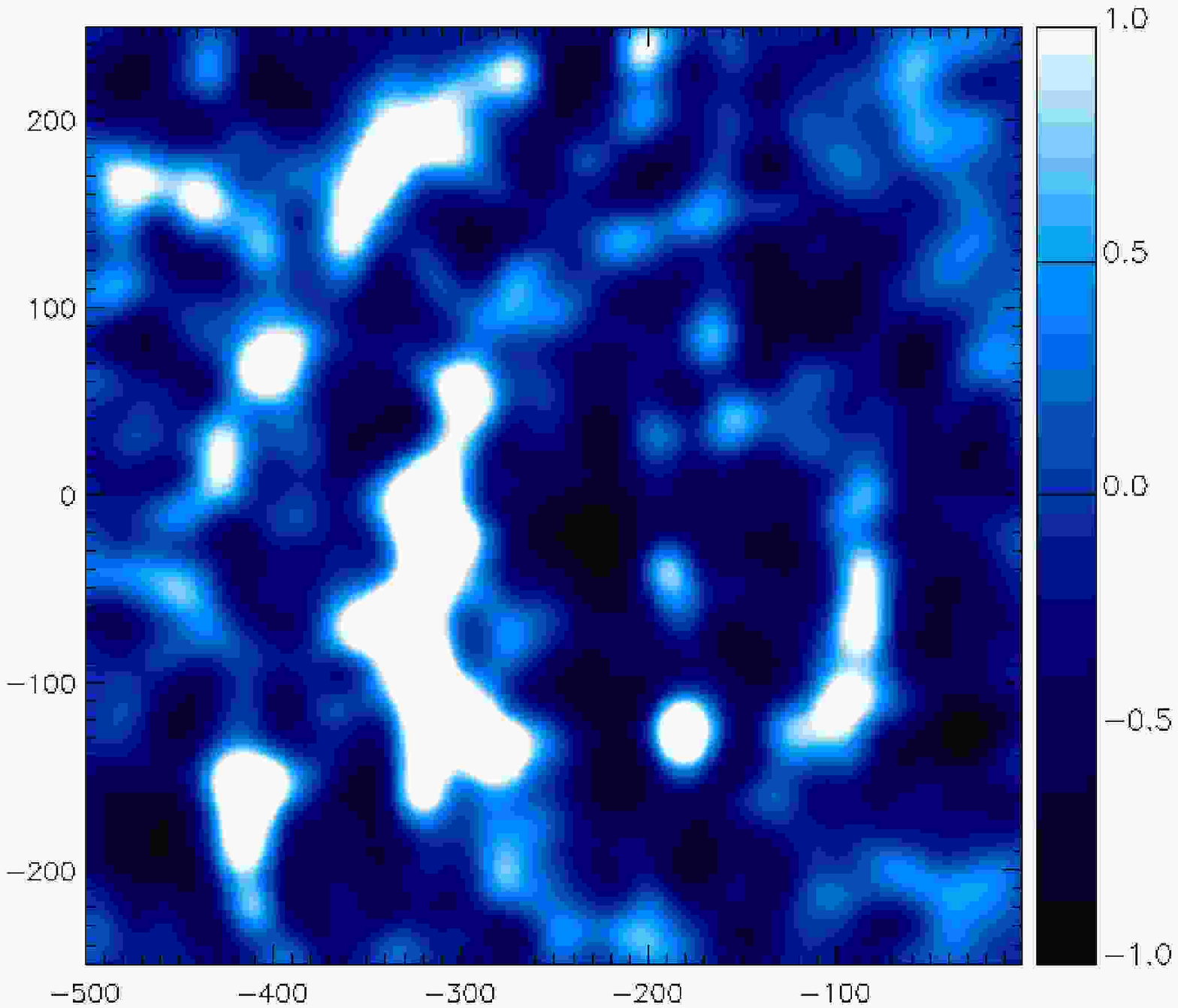}
\put(-250,1.5){{\huge (d)}}
\put(-138,-10){{X [Mpc]}}
\put(-245,93){\rotatebox[]{90}{Y [Mpc]}}
\\\\
\end{tabular}
\caption{ Slices around the Sloan and the CfA2 Great Wall. Panel {(a)}: slice through the three dimensional mask multiplied with the selection function at $\sim$7 Mpc in the Z-axis. Panels {(b)}, {(c)}, and {(d)} show slices through the reconstruction after taking the mean over 20 neighboring slices around the slice at $\sim$7 Mpc in the Z-axis, without smoothing, convolved with a Gaussian kernel with a smoothing radius of $r_{\rm S}=$5 Mpc and $r_{\rm S}=$10 Mpc, respectively.  Note, that panel {(b)} represents log($1+\delta_{\rm}$), whereas panels {(c)}, and {(d)} show $\delta_{\rm}$.} 
\label{fig:GW}
\end{figure*}

\begin{figure*}
\begin{tabular}{cc}
\includegraphics[width=7.5cm]{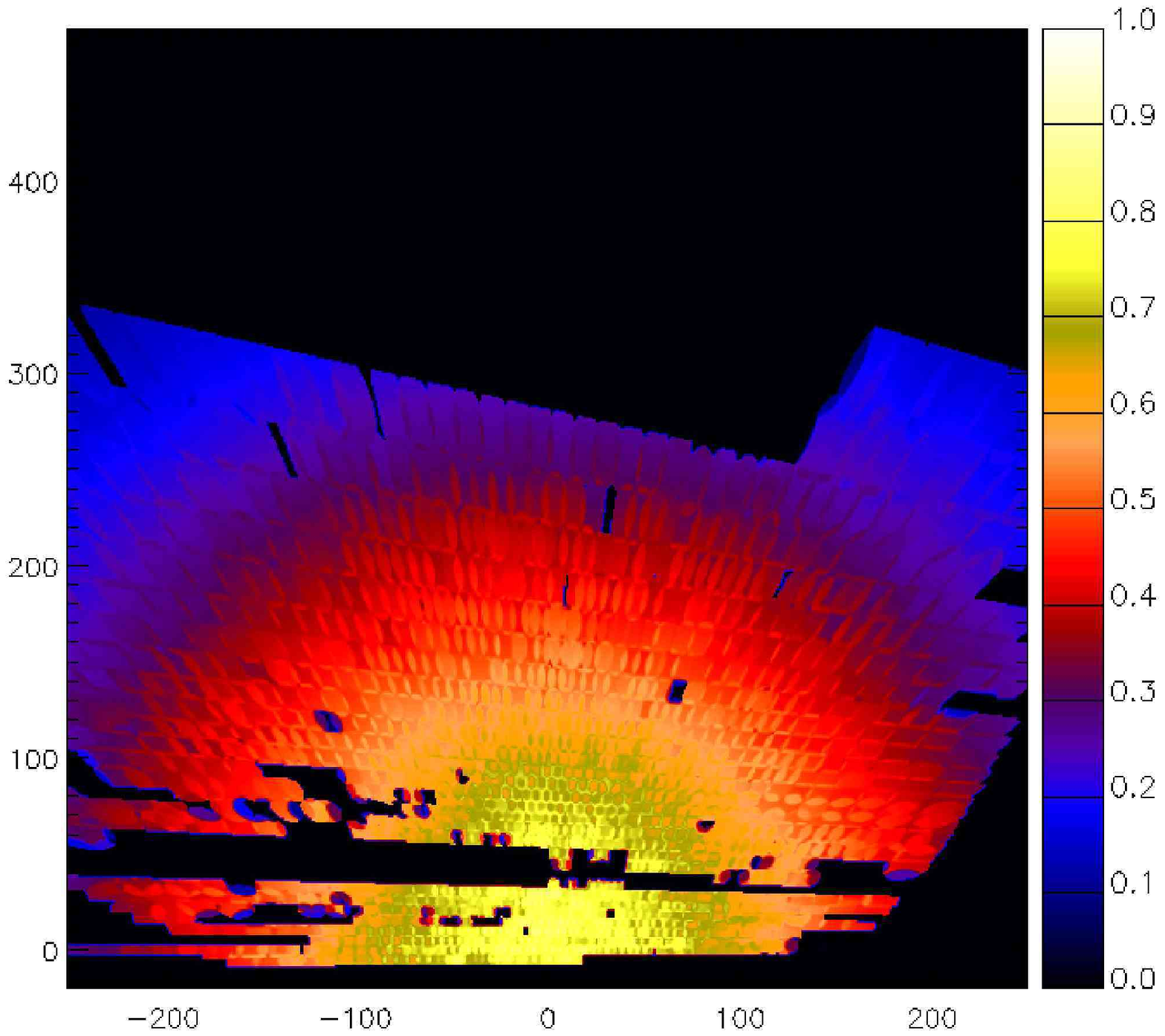}
\put(-225,0.5){{\huge (a)}}
\put(-123,-10){{Y [Mpc]}}
\put(-224,93){\rotatebox[]{90}{Z [Mpc]}}
\hspace{0.5cm}
\includegraphics[width=7.5cm]{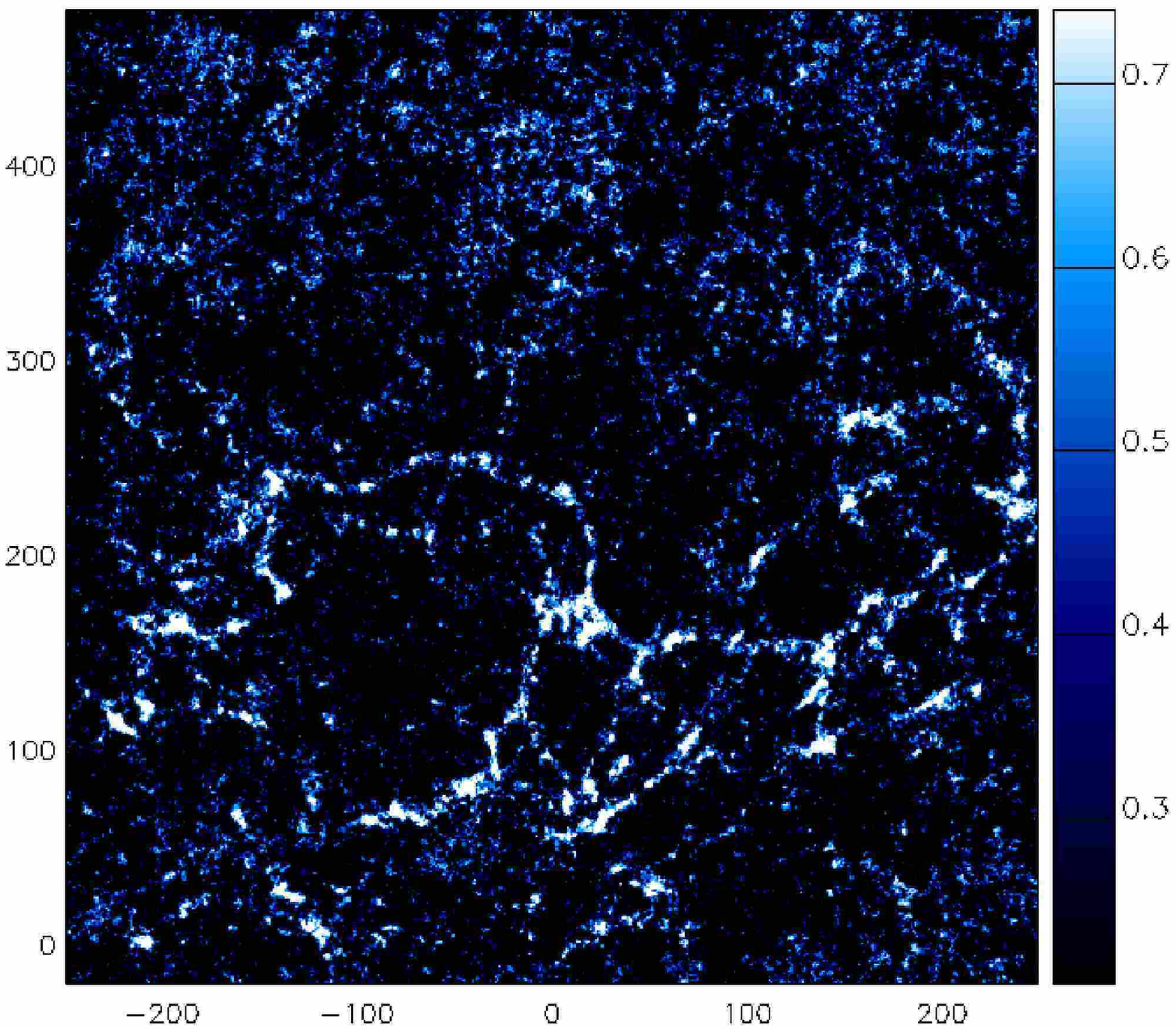}
\put(-225,1.5){{\huge (b)}}
\put(-123,-10){{Y [Mpc]}}
\put(-224,93){\rotatebox[]{90}{Z [Mpc]}}
\\\\
\includegraphics[width=7.5cm]{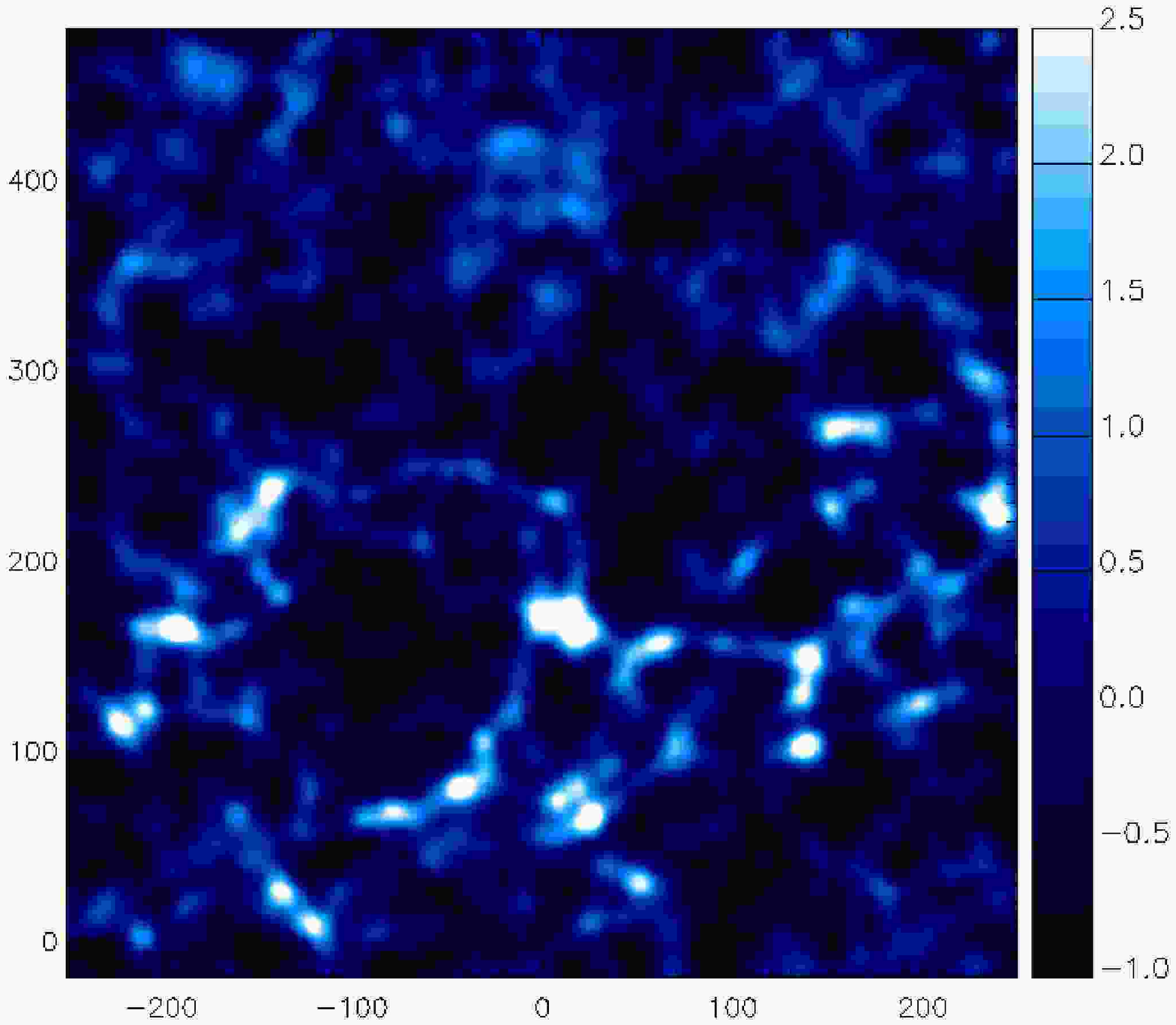}
\put(-225,0.5){{\huge (c)}}
\put(-123,-10){{Y [Mpc]}}
\put(-224,93){\rotatebox[]{90}{Z [Mpc]}}
\hspace{0.5cm}
\includegraphics[width=7.5cm]{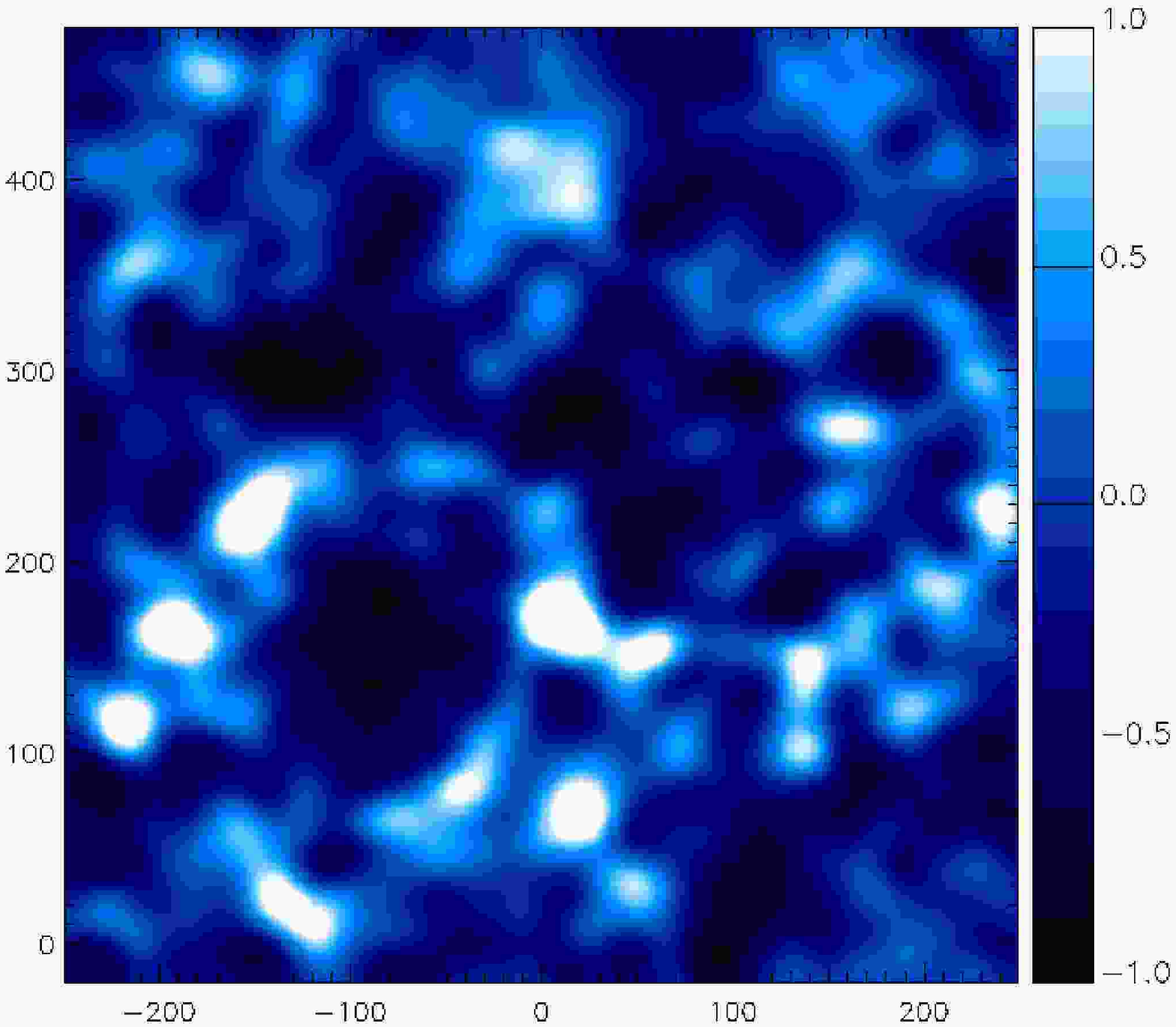}
\put(-225,1.5){{\huge (d)}}
\put(-123,-10){{Y [Mpc]}}
\put(-224,93){\rotatebox[]{90}{Z [Mpc]}}
\\\\
\includegraphics[width=7.5cm]{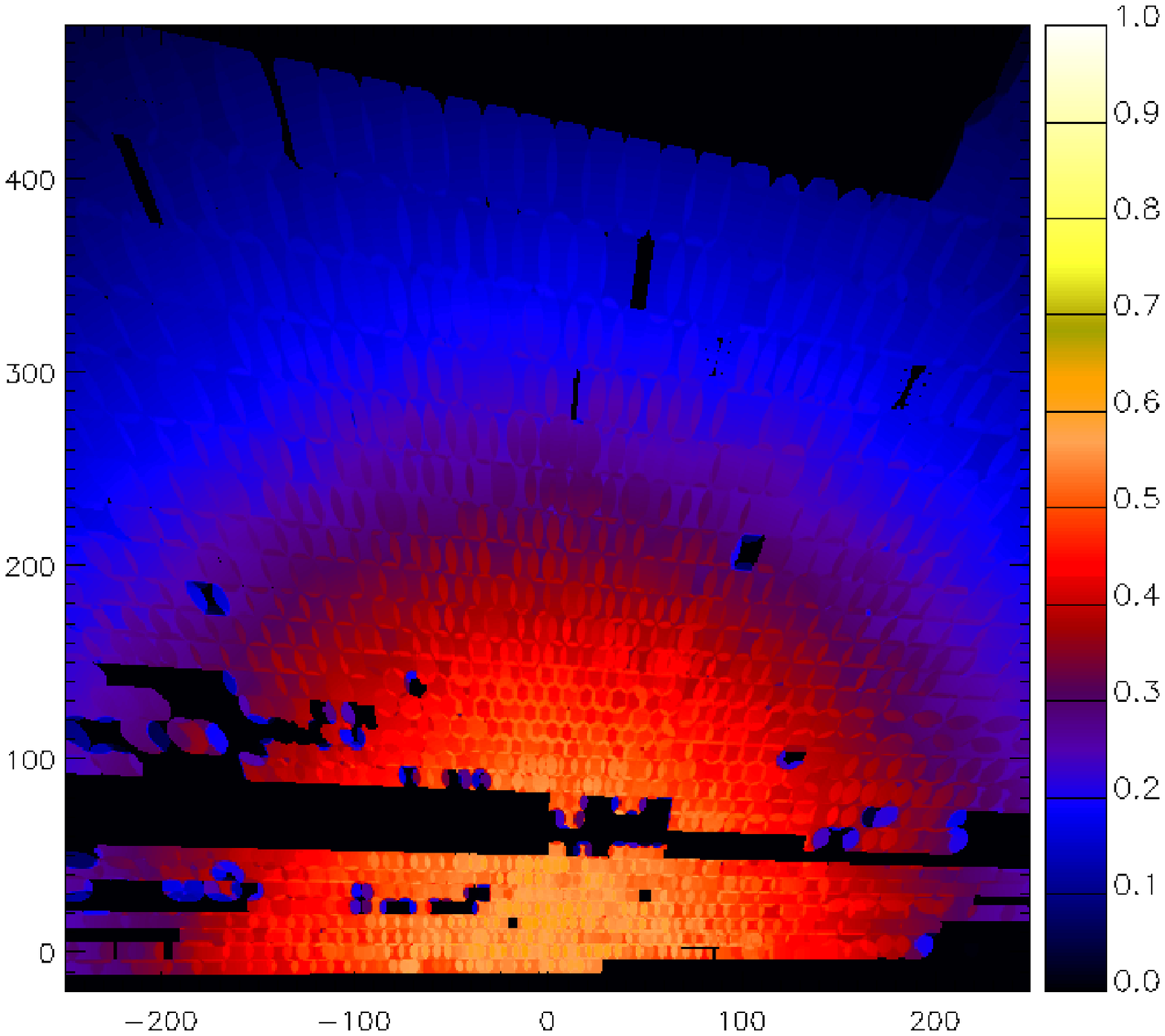}
\put(-225,0.5){{\huge (e)}}
\put(-123,-10){{Y [Mpc]}}
\put(-224,93){\rotatebox[]{90}{Z [Mpc]}}
\hspace{0.5cm}
\includegraphics[width=7.5cm]{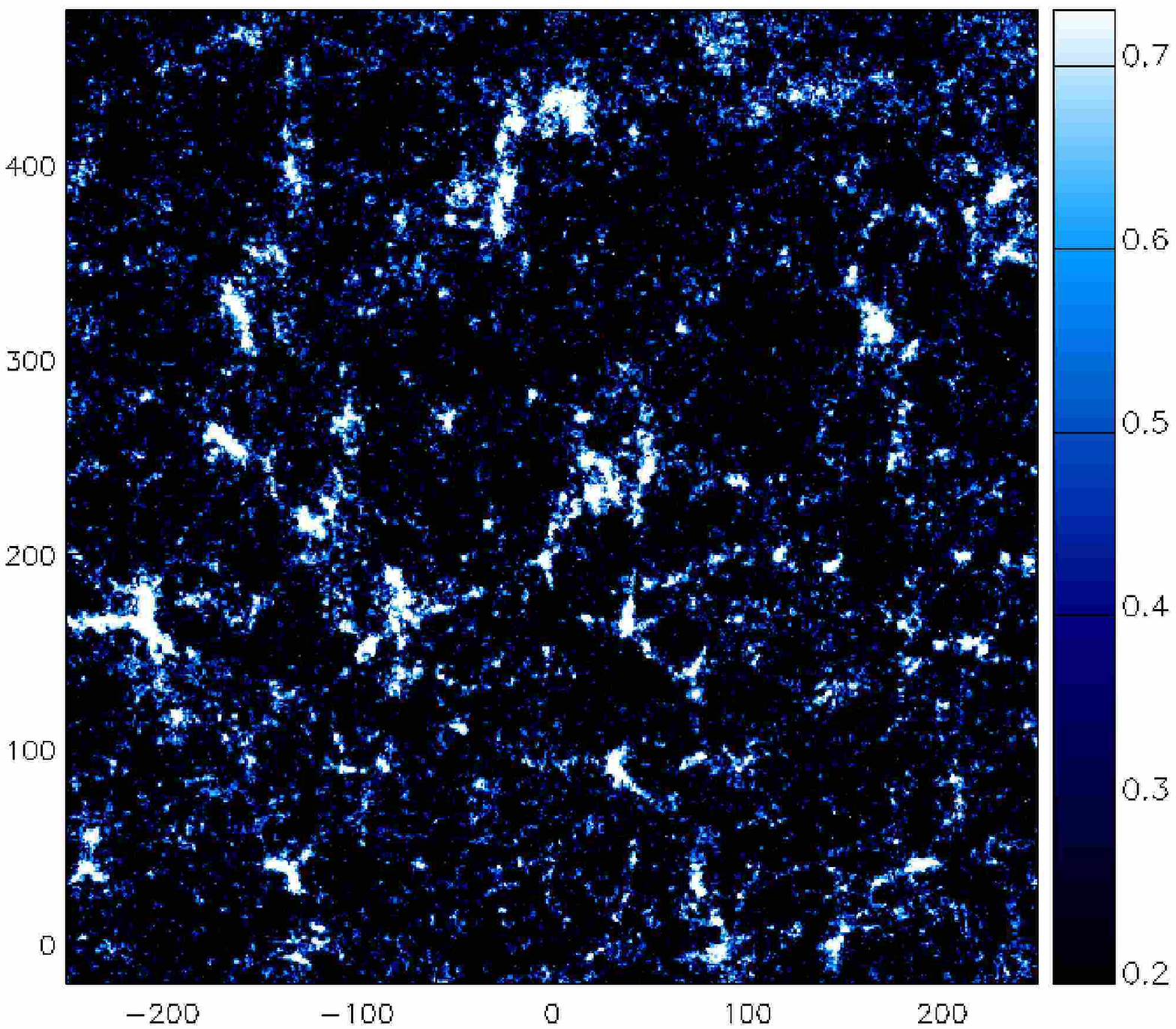}
\put(-225,1.5){{\huge (f)}}
\put(-123,-10){{Y [Mpc]}}
\put(-224,93){\rotatebox[]{90}{Z [Mpc]}}
\end{tabular}
\caption{Panel {(a)}: slice through the three dimensional mask multiplied with the selection function at $\sim$-109 Mpc in the X-axis. Panels {(b)}, {(c)}, and {(d)} show slices through the reconstruction after taking the mean over 20 neighboring slices around the slice at $\sim$-109 Mpc in the X-axis, without smoothing, convolved with a Gaussian kernel with a smoothing radius of $r_{\rm S}=$5 Mpc and $r_{\rm S}=$10 Mpc, respectively. Panel {(e)}: slice through the three dimensional mask multiplied with the selection function at $\sim$-168 Mpc in the X-axis. Panel {(f)} shows a slice through the reconstruction after taking the mean over 20 neighboring slices around the slice at $\sim$-168 Mpc in the X-axis without smoothing. Note, that panels {(b)} and (f) represent log($1+\delta_{\rm}$), whereas panels {(c)}, and {(d)} show $\delta_{\rm}$.} 
\label{fig:VOID}
\end{figure*}

The Sloan Great Wall is one of the largest structure known  in our local Universe
although it is not a gravitationally bound object
\citep[see][]{2005ApJ...624..463G}. It extends for about\footnote{Note, that the extension of the Sloan Great Wall is
  usually given in luminosity distance, which can be around 40 Mpc
  larger than in comoving distance as we represent it here.} 400
Mpc
\citep[for a detailed study see][]{2006ChJAA...6...35D} and  is
located around 300 Mpc distant from Earth. In Fig.~\ref{fig:RECSKYGW} we represent
different radial shells, picking out the
structures of the Sloan Great Wall, which extends from about
140$^\circ$ to 210$^\circ$ (-150$^\circ$ in Fig.~\ref{fig:SKY}) in right ascension and extends within a few
degrees around declination $\delta\approx 0^\circ$. In these shells
other complex structures can be observed at higher declinations,
showing filaments, voids and clusters of galaxies. Moreover, the
region which has not been observed, lying outside the mask (see panel
{(a)} in Fig.~\ref{fig:SKY}) is predicted to be filled with structures by the reconstruction method according to our assumed
correlation function (see section \ref{sec:RDO}).  The Sloan Great
Wall can also be seen in Fig.~\ref{fig:GW} almost in its full
extent. We can see, how \textsc{argo} recovers the matter
field, balancing the structures with low signal to noise ratio
against those with a higher signal,  leading to a homogeneously
distributed field, meaning that clusters close to and far from the observer are both well represented. Only where the signal to noise drops below unity, do structures tend to {\it blur}, as can be observed in the upper parts of the reconstruction shown in Fig.~\ref{fig:VOID}. 

The CfA2 Great Wall is also one of the largest structure known in our local
Universe and contains the Coma Cluster (Abell 1656) at its center
\citep[see][]{1989Sci...246..897G}. 
We can clearly see the Coma Cluster in the projected
reconstruction without smoothing, being the big spot at right ascension
$\alpha\approx 195^\circ$ (-165$^\circ$ in Fig.~\ref{fig:SKY}) and declination $\delta\approx 28^\circ$ in
panel {(b)} of Fig.~\ref{fig:SKY}, located at a distance of $\sim100$ Mpc from the observer \citep[see][]{1997ApJ...483L..37T,2008ApJS..176..424C}. 
  The
CfA2 Great Wall cannot be seen in its full extent in
Fig.~\ref{fig:GW} because it reaches higher declination angles than
selected in the plot. However, it can be partially seen as an
elongated matter structure at about 100 Mpc distance to the
observer, i.e.~at around -100 Mpc in the X-axis in
Fig.~\ref{fig:GW}. 
 Large filamentary structures are present even after smoothing
with a Gaussian kernel with a smoothing radius of $r_{\rm S}=$10 Mpc (see panel
{(d)} in Fig.~\ref{fig:GW}).
The second major cluster of the Coma super-cluster is the Leo Cluster
(Abell 1367) at a distance $\sim94$ Mpc
($z\approx0.022$), with galactic coordinates $\alpha\approx176^\circ$
and $\delta\approx20^\circ$.  It is weakly detected in our
reconstruction as can be seen in panel {(b)} of Fig.~\ref{fig:SKY},
since it is partially located in the major gap of DR6 and should be
therefore better detected with DR7.

%
%
%
%
%
%
%
%

 The Hercules supercluster also belongs to the CfA2 Great Wall.
Most of the clusters which  belong this supercluster can be identified in the
reconstructed area. Since the spatial range of these clusters is large, we have listed in table \ref{tab:obscluster} the groups of clusters with their respective localisation in the sky which appear as especially prominent overdensity regions in the projected reconstruction \citep[for references see][]{1989ApJS...70....1A,1999ApJS..125...35S}.
Note that close-by structures such as the Virgo Cluster, which is at a distance of only 
about 18 Mpc distance to us, cannot be detected in our
reconstruction, because the lower limit of our sample is set at $z=0.01$.

\subsection{Detection of a great void region}
\label{sec:VOID}

The {\it scorpion}-like form of the matter distribution spanning the whole observed region in
Fig.~\ref{fig:VOID} (see mask in panel {(a)}) shows large connected filamentary structures
with many clusters.  Interestingly, an extremely large void is spanned
in the region with -150 Mpc $<$ Y $<$ 30 Mpc and 70 Mpc $<$ Z
$<$ 220 Mpc (see panels {(a)}, {(b)} and {(c)} in
Fig.~\ref{fig:VOID}).  In order to evaluate the confidence of the
detection one should check how deeply this region has
been scanned by SDSS. By inspection of the three dimensional mask we
confirm a fairly high completeness ranging from about 30\% to about
65\% (see panel {(a)} of Fig.~\ref{fig:VOID}). The extension in the
X-axis is still unclear, since the gap in the mask grows in the void region to
larger distances to the observer.  \textsc{argo} predicts an extension
of about -250 Mpc $<$ X $<$ -450 Mpc.  From our results, we can
tell that it is one of the largest voids in
the reconstructed volume, having a diameter of about 150 Mpc.
Conclusive results can only be obtained after investigating DR7, which
fills the main gaps. Since, in this case, a proper treatment of the DR7
mask is required and this mask was not public at the time 
this project started, we postpone this study for later work.
The large overdensity region found in the unobserved region at about: -30 Mpc $<$ Y $<$ 30 Mpc and 370 Mpc $<$ Z
$<$ 430 Mpc results from the correlation with a huge cluster region which extends in the range:  -30 Mpc $<$ Y $<$ 30 Mpc and 350 Mpc $<$ Z
$<$ 450 Mpc and which can be best seen at about  X $\sim$ -170 Mpc (see panels (e) anf (f) in Fig.~\ref{fig:VOID}).

\begin{table*}
\begin{tabular}{|c|c|c|} 
  \hspace{3cm} \\\hline
c$_1$    & -220 Mpc $<$ Y $<$ -200 Mpc & 140 Mpc $<$ Z $<$ 180 Mpc\\\hline
c$_2$    & -140 Mpc $<$ Y $<$ -100 Mpc & 120 Mpc $<$ Z $<$ 160 Mpc\\\hline
c$_3$    & 10 Mpc $<$ Y $<$ 20 Mpc & 120 Mpc $<$ Z $<$ 160 Mpc\\\hline
c$_4$    & 10 Mpc $<$ Y $<$ 30 Mpc & 70 Mpc $<$ Z $<$ 90 Mpc\\\hline
c$_5$    & 60 Mpc $<$ Y $<$ 70 Mpc & 70 Mpc $<$ Z $<$ 90 Mpc\\\hline
c$_6$    & 150 Mpc $<$ Y $<$ 160 Mpc & 60 Mpc $<$ Z $<$ 70 Mpc\\\hline 
c$_7$    & 220 Mpc $<$ Y $<$ 240 Mpc & 90 Mpc $<$ Z $<$ 110 Mpc\\\hline
c$_8$    & -210 Mpc $<$ Y $<$ -200 Mpc & 70 Mpc $<$ Z $<$ 90 Mpc\\\hline
c$_9$    & -110 Mpc $<$ Y $<$ -90 Mpc & 70 Mpc $<$ Z $<$ 90 Mpc\\\hline 
c$_{10}$ & -40 Mpc $<$ Y $<$ -60 Mpc & 70 Mpc $<$ Z $<$ 90 Mpc\\\hline
\end{tabular}
\caption{\label{tab:cluster} Approximate positions of cluster candidates c$_i$  (with $i$ ranging from 1 to 10) at a slice around -265 Mpc $<$ X $<$ -245 Mpc in the reconstructed box which are located close to gaps (see Fig~\ref{fig:GAP}). }
\end{table*}

\begin{figure*}
\begin{tabular}{cc}
\includegraphics[width=8.5cm]{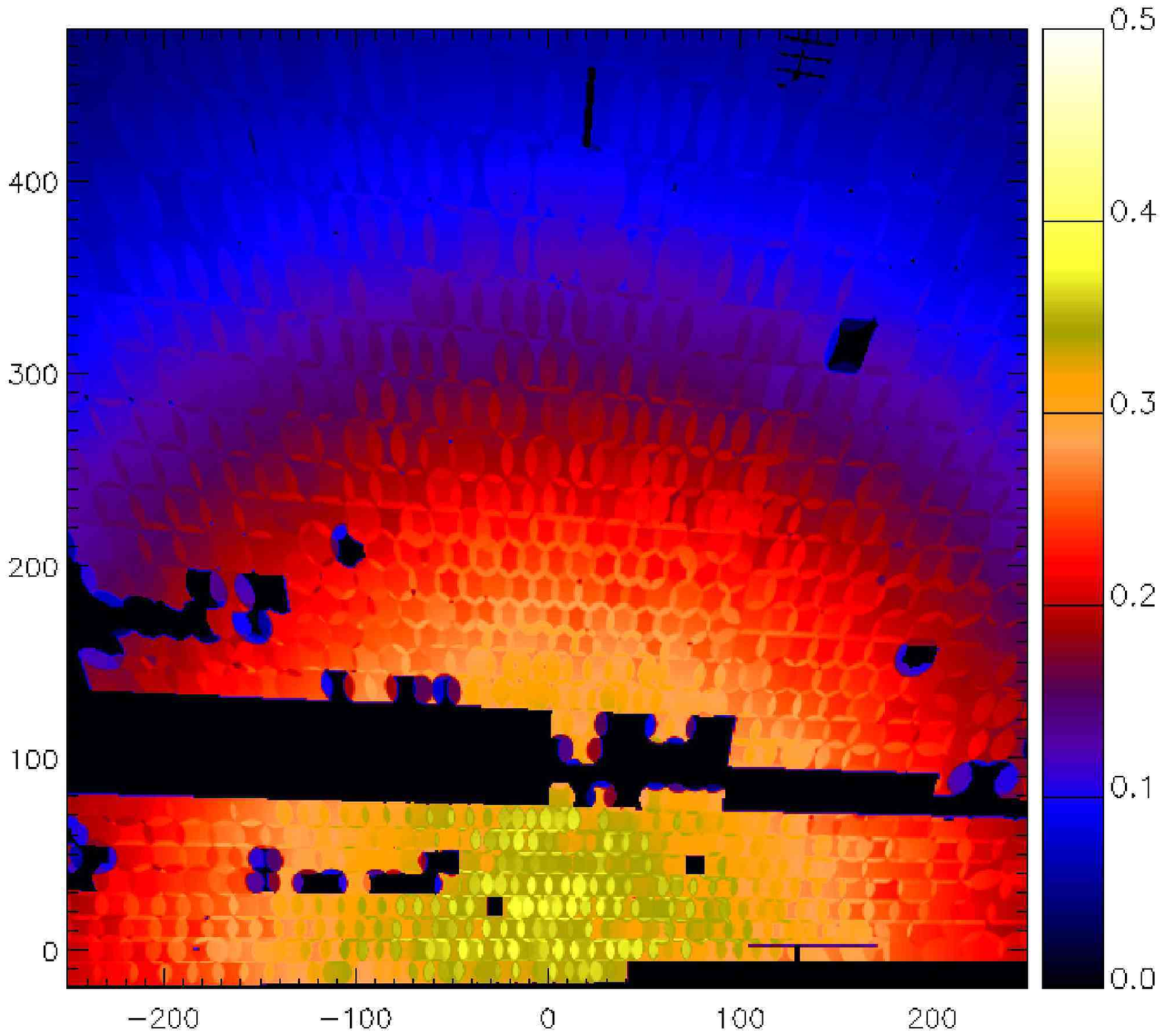}
\put(-250,0.5){{\huge (a)}}
\put(-138,-10){{Y [Mpc]}}
\put(-249,93){\rotatebox[]{90}{Z [Mpc]}}
\hspace{.5cm}
\includegraphics[width=8.5cm]{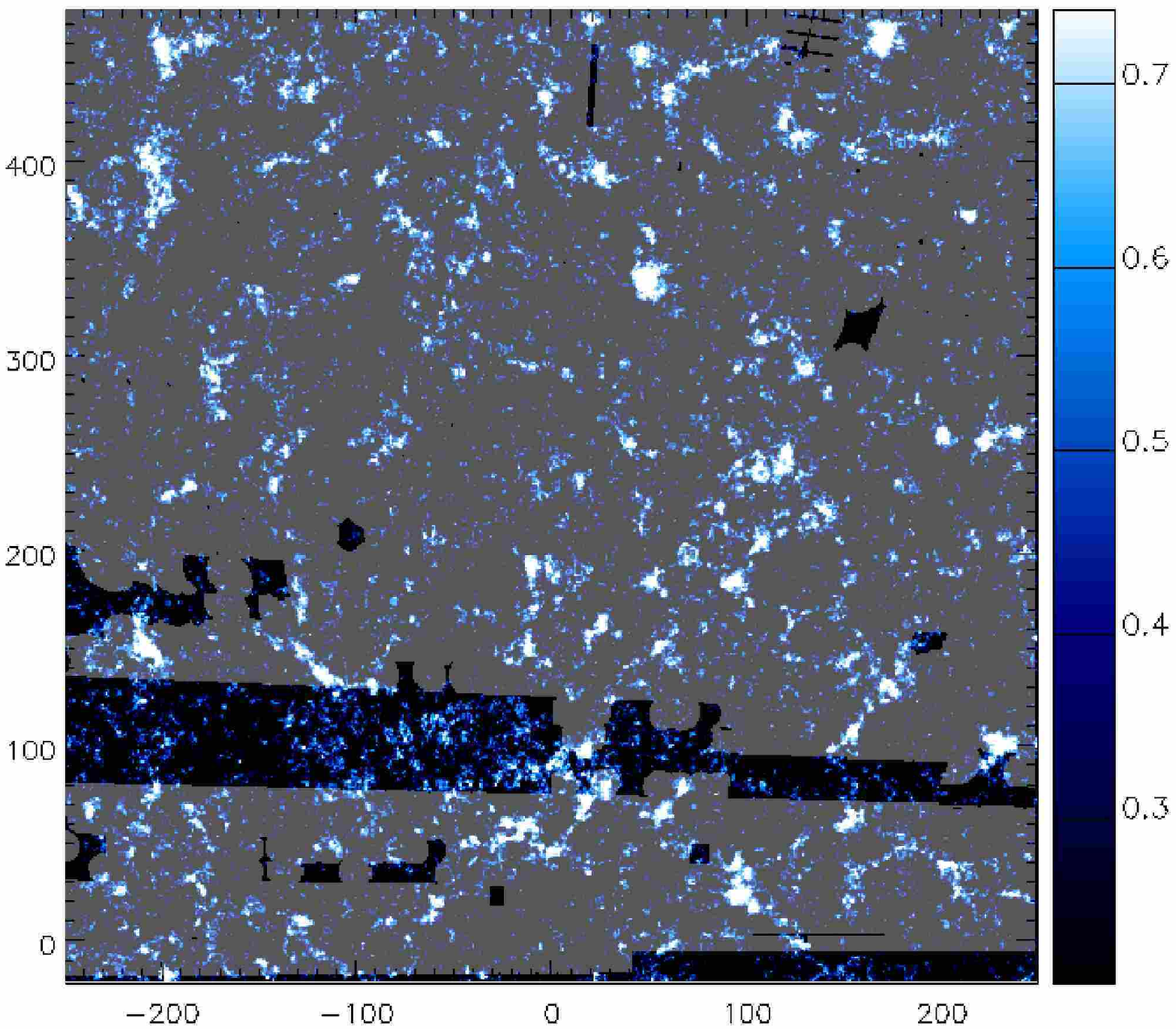}
\put(-250,1.5){{\huge (b)}}
\put(-138,-10){{Y [Mpc]}}
\put(-249,93){\rotatebox[]{90}{Z [Mpc]}}
\\\\
\includegraphics[width=8.5cm]{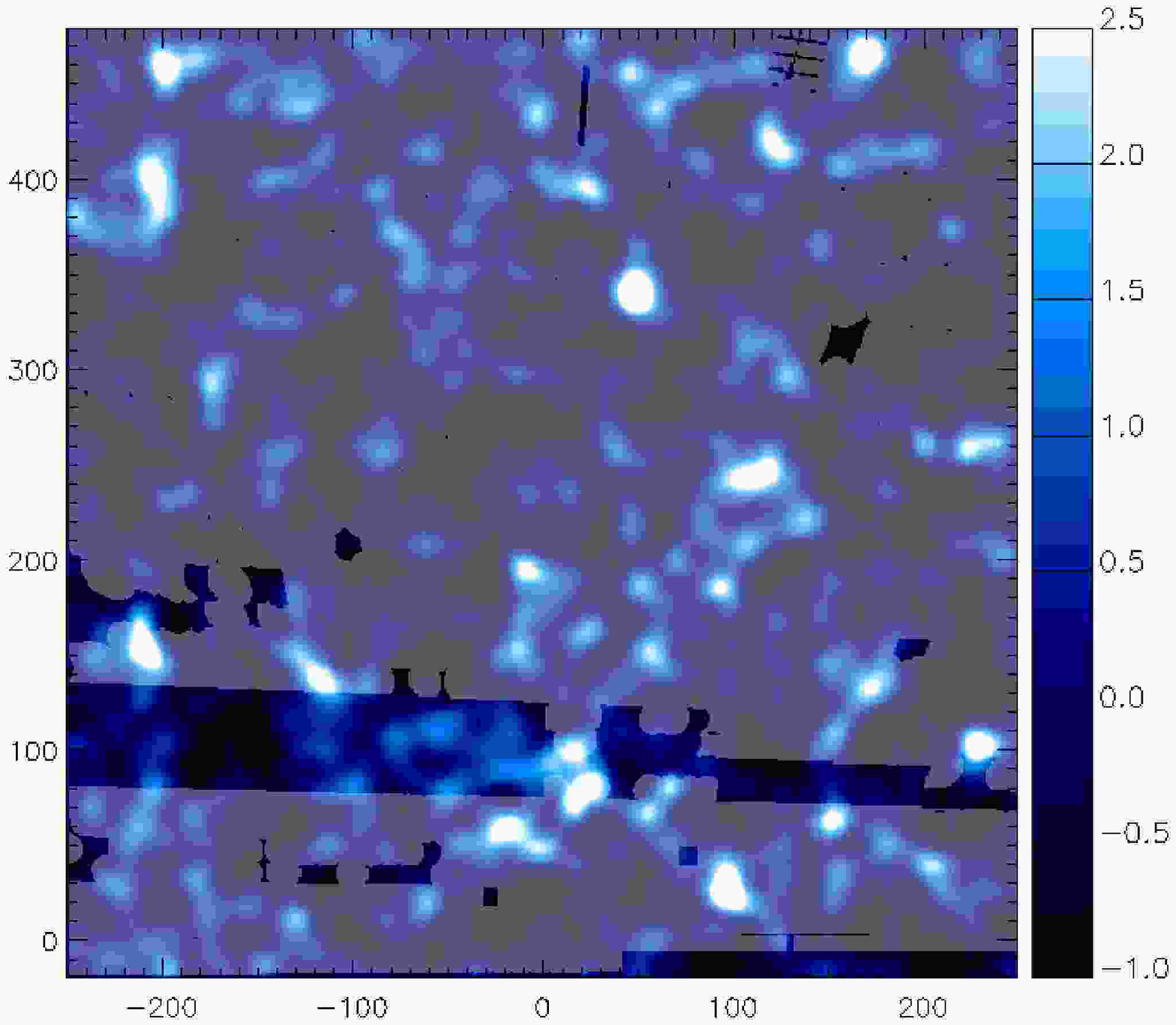}
\put(-250,0.5){{\huge (c)}}
\put(-138,-10){{Y [Mpc]}}
\put(-249,93){\rotatebox[]{90}{Z [Mpc]}}
\hspace{.5cm}
\includegraphics[width=8.5cm]{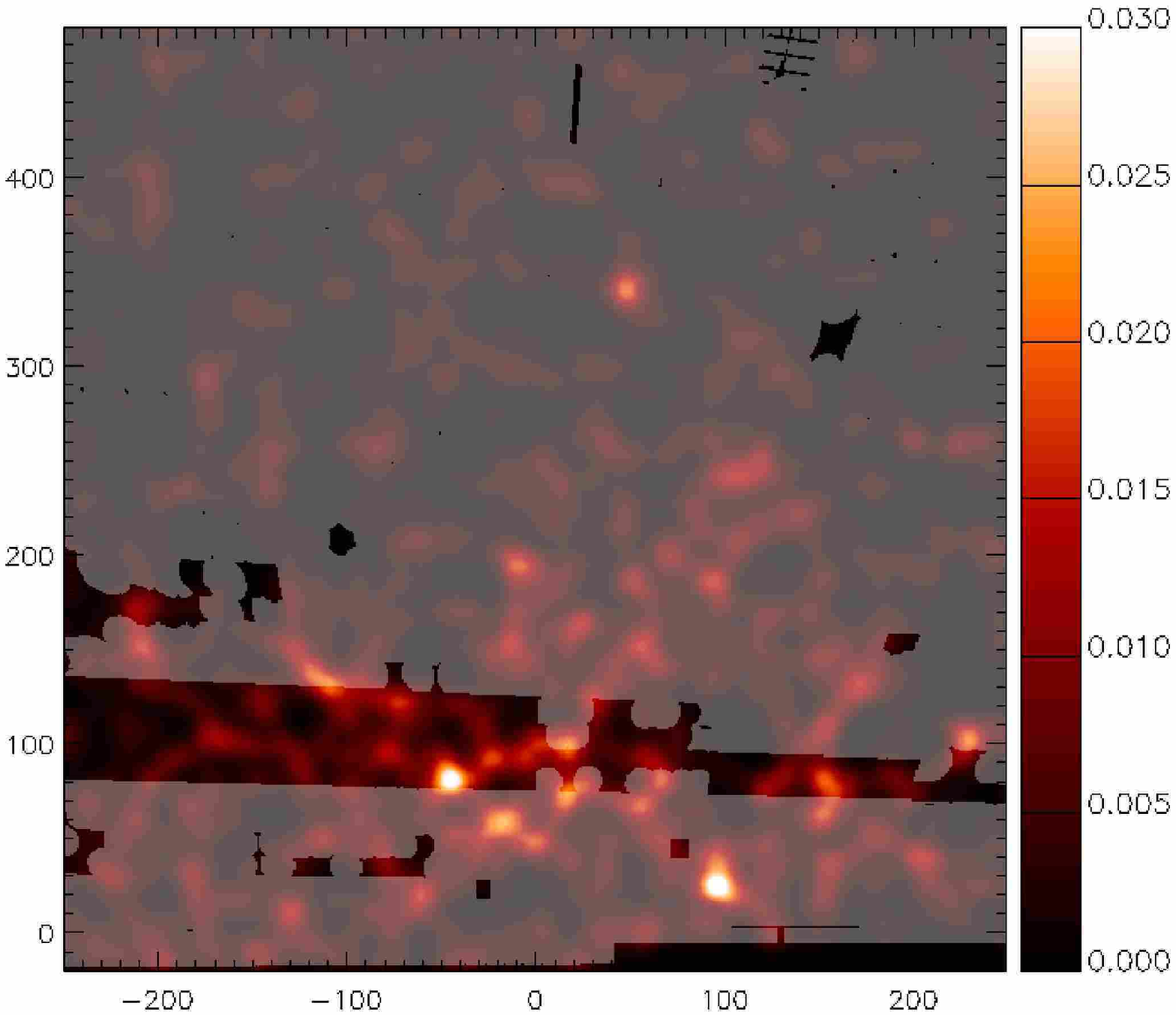}
\put(-250,1.5){{\huge (d)}}
\put(-138,-10){{Y [Mpc]}}
\put(-249,93){\rotatebox[]{90}{Z [Mpc]}}
\\\\
\end{tabular}
\caption{Panel {(a)}: slice through the three dimensional mask multiplied with the selection function at $\sim$-256 Mpc in the X-axis. Panels {(b)} and {(c)} show slices through the reconstruction after taking the mean over 20 neighboring slices around the slice at $\sim$-256 Mpc in the X-axis, without smoothing and convolved with a Gaussian kernel with a smoothing radius of $r_{\rm S}=$5 Mpc, respectively. Panel {(d)}: DR7 sample gridded with NGP and convolved with a Gaussian kernel with a smoothing radius of $r_{\rm S}=$5 Mpc. In panels {(c)} and {(d)} the DR6 mask is over-plotted. Note, that there is some correspondance between the structures predicted in the gap from the {\tt Sample dr6fix} and the observed galaxy distribution there in DR7. Note, that panel {(b)} represents log($1+\delta_{\rm}$), whereas panels {(c)}, and {(d)} show $\delta_{\rm}$.} 
\label{fig:GAP}
\end{figure*}

\subsection{Cluster prediction}
\label{sec:GAP}

The signal-space representation of the Wiener-filter (see section
\ref{sec:WF}) enables us to deal with unobserved regions, i.e.~cells
with zero completeness. Note, that for those cells the noise term
vanishes in the Wiener-filter expression (Eq.~\ref{eq:WF}). The
filter can then be regarded as a convolution with the non-diagonal
autocorrelation matrix of the underlying signal propagating the
information from the windowed region into the unobserved cells. This
gives a prediction for the Large-Scale Structure in these regions. 
Such an {\it extrapolation} can be clearly seen in panels {(b)},
{(c)} and {(d)} of Fig.~\ref{fig:SKY}. These show the projected
three dimensional reconstruction on the sky without smoothing and
after a convolution with a Gaussian with a smoothing radius $r_{\rm S}$ of 5
and 10 Mpc, respectively. In these plots the gaps are hardly
distinguishable, due to the signal prediction given by the
Wiener-filter.
We have chosen a slice, in which the propagation of the information
through gaps can be analyzed. 
In panel {(a)} of Fig.~\ref{fig:GAP}
we can see the three-dimensional mask through our selected slice.  The
main gap crosses the entire box through the Y-axis and reaches about
50 Mpc width in the Z-axis. Several other smaller gaps are
distributed in the slice.  In the reconstruction in panel {(b)} we
can see how the main gap is partially filled with some {\it diffuse}
overdensity structures which are produced precisely  as described above.
Panel {(c)} shows the same reconstruction smoothed with a Gaussian kernel
with a smoothing radius of $r_{\rm S}=$5 Mpc.  Overplotted is the mask showing the regions in which it was observed.   
We identify seven clusters close to gaps extending into  unobserved regions at a slice around -265 Mpc $<$ X $<$ -245 Mpc (see clusters c$_1$-c$_{7}$ in Tab.~\ref{tab:cluster}).
In addition, there are some weaker detections (see clusters c$_8$-c$_{10}$ in Tab.~\ref{tab:cluster}).
The gap which cluster c$_1$ extends into, and the largest gap, are the ones in which more information propagation occurs.
There is an especially interesting region in the main gap around -140 Mpc $<$ Y $<$ 30 Mpc in which the algorithm predicts a high chance to find overdense structures.
The rest of the gaps remains with low density values, since no prominent structures are in their vicinity.
We investigate the public DR7 archive (see Section \ref{sec:IGS}) to check for overdense regions in the gap.
Note, that without a full angular  and radial selection function treatment a quantitative comparison is not possible.
We restrict our study by gridding the galaxy sample with NGP, ignoring mask or selection
function effects, and convolving it with a Gaussian kernel with a
smoothing radius of $r_{\rm S}=$10 Mpc (see panel {(d)} in Fig.~\ref{fig:GAP}). 
Though, faint features like the filaments lying at around -230 Mpc
$<$ Y $<$ -130 Mpc cannot be recovered, stronger features like the
clusters located at -100 Mpc $<$ Y $<$ 0 Mpc show that there is
indeed an overdense region in the gap confirming our prediction based
on DR6. In particular the extension of the clusters c$_1$ and c$_2$ are very well predicted by our algorithm. Cluster c$_{10}$ is weakly predicted. 
The filament connecting clusters c$_3$ and c$_{10}$ is predicted by  \textsc{argo}, perhaps by chance, but the resemblance in the gap of the reconstruction to the real underlying distribution shows that use of the correlation function of the LSS allows for  plausible predictions.

\begin{figure*}
\includegraphics[width=17cm]{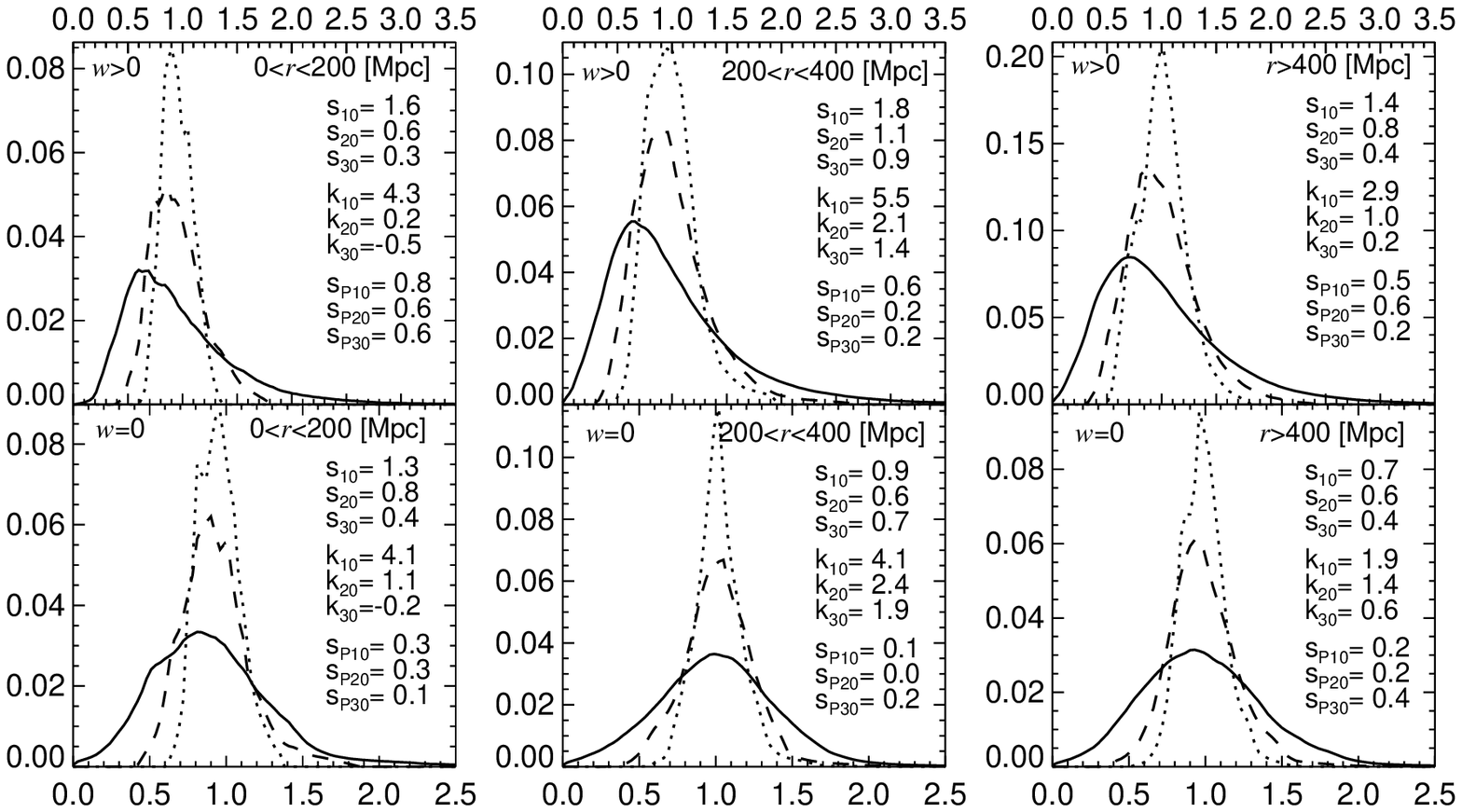}
\put(-245,-13.){\Large \bf $1+\delta^{\rm B}_{{\Delta r},i}$}
\put(-490,130.){\rotatebox[]{90}{\Large\bf $f^{\rm B}_{{\Delta r},i}/N^{\Delta r}_{\rm cells}$}}\\
\caption{\label{fig:stats_shells} Statistical distribution of cells at different  densities with a density binning of 0.03 in $(1+\delta_{\rm m})$. The curves represent the distribution for the reconstructed matter field at different scales ($r_{\rm S}$: continuous: 10 Mpc, dashed: 20 Mpc, dotted: 30 Mpc). The upper panels show the statistics at different radial shells in the observed region ($w>0$), and the lower panels show the same in the unobserved region ($w=0$). The corresponding skewness: s$_{10}$, s$_{20}$, s$_{30}$, kurtosis: k$_{10}$, k$_{20}$, k$_{30}$, and Pearson's skewness: s$_{\rm P 10}$, s$_{\rm P 20}$, s$_{\rm P 30}$ are also given.   }  
\end{figure*}

\begin{figure*}
\includegraphics[width=17cm]{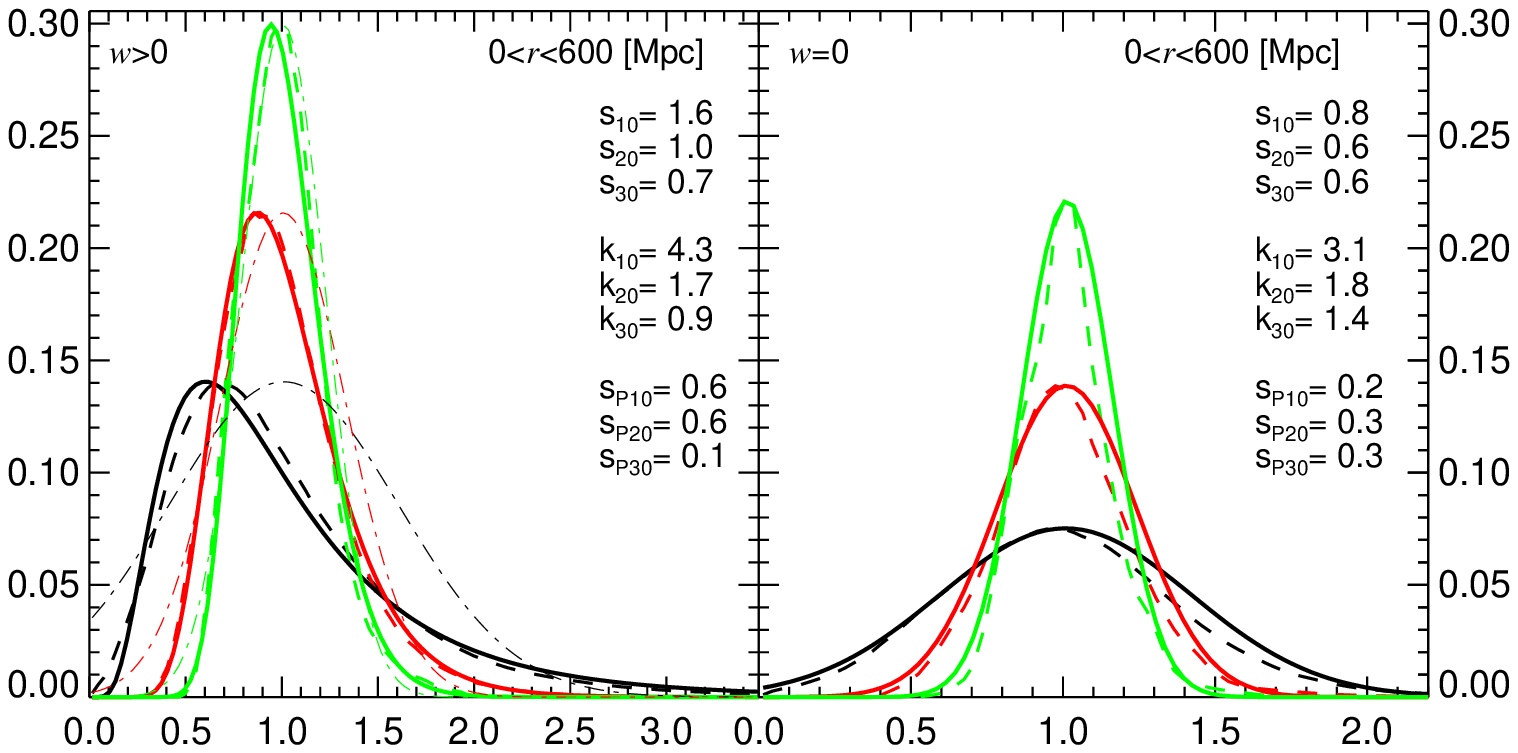}
\put(-255,-5.){\Large \bf $1+\delta^{\rm B}_{{\Delta r},i}$}
\put(-495,130.){\rotatebox[]{90}{\Large\bf $f^{\rm B}_{{\Delta r},i}/N^{\Delta r}_{\rm cells}$}}\\
\caption{\label{fig:stats_tot} Statistical distribution of cells at different  densities with a density binning of 0.03 in $(1+\delta_{\rm m})$. The dashed curves represent the distribution for the reconstructed matter field at different scales ($r_{\rm S}$: black: 10 Mpc, red: 20 Mpc, green: 30 Mpc).  The corresponding skewness: s$_{10}$, s$_{20}$, s$_{30}$, kurtosis: k$_{10}$, k$_{20}$, k$_{30}$, and Pearson's skewness: s$_{\rm P 10}$, s$_{\rm P 20}$, s$_{\rm P 30}$ are also given.  
On the left: (observed region: $w>0$) continuous lines: best fit lognormal distributions using a nonlinear least squares fit based on a gradient-expansion algorithm, dashed-dotted curves: Gaussian distributions for the measured means and variances. 
On the right: (unobserved region: $w=0$) continuous lines: Gaussian distributions for the measured means and variances with the corresponding statistical correlation coefficients $r_{20}$, $r_{40}$, and $r_{60}$.    }  
\end{figure*}

\subsection{Statistics of the density field}
\label{sec:statistics}

From a physical point of view, one would expect  a log-normal distribution  of smoothed density for a certain
                  range of smoothing scales, if one assumes an initial Gaussian velocity field and  extrapolates the continuity equation for the matter flow into the nonlinear regime with linear velocity fluctuations \citep[see][]{1991MNRAS.248....1C}.
Since the log-normal field is not able to describe caustics, we expect this distribution to fail below a threshold smoothing scale. There should also be a  transition at a certain scale between  this quasilinear regime and the linear regime where the matter field is still Gaussian distributed. 
Due to use of the Wiener-filter which considers only the correlation function to reconstruct the density field and the Gaussian smoothing, we expect the density field to be closely Gaussian distributed in the unobserved regions.
 Here, we analyse the statistical distribution of the density field by counting
 the number of cells at different  densities with a density binning of 0.03 in $(1+\delta_{\rm m})$ at different scales, defined by convolving the reconstruction with a Gaussian kernel with smoothing radii $r_{\rm S}$ of: 10, 20, and 30 Mpc. We performed the analysis for  different radial shells in the $\Delta r$\footnote{Note, that we considered the density at the center of the bins.} ranges: $0<{r}<200$ Mpc, $200<r<400$ Mpc, $r>400$ Mpc, and $0<r<600$ Mpc, separating observed ($w>0$) and unobserved  ($w=0$) regions (see Figs.~\ref{fig:stats_shells} and \ref{fig:stats_tot}).
Note, that due to shot noise, we are missing power in the filtered reconstruction on small scales. Moreover, the discrete Fourier representation of the signal implies negative densities \citep[see][]{dsp}.
  This obliges us to perform this statistical analysis on scales larger  than the smallest grid scales. We can see this in the excess of low density cells for the dashed black curve ($r_{\rm S}=$5 Mpc).  
In addition to that, we are also limited by the size of the box, having less  information as we go to larger and larger scales. This effect can be appreciated in the stronger deviation from the log-normal fit around the peak for the green line ($r_{\rm S}=$30 Mpc). For this reason, we restrict this analysis to the  range of scales given above.
The plots in Figs.~\ref{fig:stats_shells} and \ref{fig:stats_tot} show how the distribution tends towards Gaussianity as we go to larger and larger scales.

We calculated the skewness and kurtosis to quantify the deviation from Gaussianity.
Let us define here the statistical quantities required for our analysis.
The number of cells contained in a shell of radial range $\Delta r$ is given by
the sum of the number counts in each density bin $f^{\rm B}_{{\Delta r},i}$:
\begin{equation}
{N^{\Delta r}_{\rm cells}} \equiv \sum^{N^{\Delta r}_{\rm bins}}_i\,f^{\rm B}_{{\Delta r},i}{.}
\end{equation}
The mean overdensity in ${\Delta r}$ which is very close to zero, is calculated as:
\begin{equation}
\overline{\delta^{\rm B}_{{\Delta r}}}\equiv{\frac{1}{N^{\Delta r}_{\rm cells}} \sum^{N^{\Delta r}_{\rm bins}}_i\,f^{\rm B}_{{\Delta r},i}\delta^{\rm B}_{{\Delta r},i}}{,}
\end{equation}
with the superscript $^{\rm B}$ standing for bin.
These two previously defined quantities permitted us to calculate the central $n$-moments $\mu_n$ of the distribution with:
\begin{equation}
{\mu_{n}}({\Delta r})\equiv{\frac{1}{N^{\Delta r}_{\rm cells}} \sum^{N^{\Delta r}_{\rm bins}}_i\,f^{\rm B}_{{\Delta r},i}\left(\delta^{\rm B}_{{\Delta r},i}-\overline{\delta^{\rm B}_{{\Delta r}}}\right)^n}{.}
\end{equation}
Note, that the variance is just the second moment: $\sigma^2\equiv{\mu_2}$.
Now, we can define the skewness\footnote{Note, that for a Gaussian distribution: $s=0$.}:
\begin{equation}
s\equiv \frac{\mu_3}{\sigma^3}{,}
\end{equation}
and the kurtosis\footnote{Note, that for a Gaussian distribution: ${\mu_4}/{\sigma^4}$=3 and thereby: $k=0$.}:
\begin{equation}
k\equiv \frac{\mu_4}{\sigma^4}-3{.}
\end{equation}
Let us also introduce Pearson's skewness defined as
the  mean $\overline{\delta^{\rm B}}$ minus the  mode $\delta^{\rm B}_{{\rm max}(f)}$ (overdensity bin with the maximum number of counts ${\rm max}(f)$) normalized by the square root of the variance: 
\begin{equation}
s_{\rm P}({\Delta r})\equiv \frac{\overline{\delta^{\rm B}_{\Delta r}}-\delta^{\rm B}_{{\rm max}(f^{\rm B}({\Delta r}))}}{\sigma({\Delta r})}{.}
\end{equation}
The results are shown in Figs.~\ref{fig:stats_shells} and \ref{fig:stats_tot} demonstrating large deviations from Gaussianity in the observed regions and negligible deviations for the unobserved regions. Since the Wiener filter uses only the first two moments of the matter distribution, we do not expect large deviations from Gaussianity in the  unobserved regions where there is almost no data constraining the result. 
Note, that in Figs.~\ref{fig:stats_shells} and \ref{fig:stats_tot} the skewness and kurtosis are also given (skewness: s$_{10}$, s$_{20}$, s$_{30}$, kurtosis: k$_{10}$, k$_{20}$, k$_{30}$, Pearson's skewness: s$_{\rm P 10}$, s$_{\rm P 20}$, s$_{\rm P 30}$, with the subscript denoting the smoothing radius in Mpc). Pearson's skewness is always larger for the observed regions than for the unobserved regions after smoothing with $r_{\rm S}=$10 and $r_{\rm S}=$20 Mpc and all distributions show a positive skewness. 
The skewness and kurtosis values show that the matter distribution starts to be closely Gaussian distributed after smoothing with a radius $r_{\rm S}$ of 30 Mpc. Nevertheless, for the region $200<r<400$ Mpc we find a large deviation from Gaussianty even at that scale.
Large scale structures like the Sloan Great Wall can be responsible for this.
Furthermore, we analyzed in great detail the matter distribution in the region $0<r<600$ Mpc which has better statistics. 
On the right panel of Fig.~\ref{fig:stats_tot} we can see the statistics for the unobserved region. 
The dashed curves show the measured distributions at different scales  (black: $r_{\rm S}=$10 Mpc, red: $r_{\rm S}=$20 Mpc, green: $r_{\rm S}=$30 Mpc). 
We calculated the means and the variances for each distribution and plotted the corresponding Gaussian distributions with light dashed-dotted lines.

 On the left panel of Fig.~\ref{fig:stats_tot} we can see the statistics for the observed region with the dashed curves showing again the measured distributions at different scales  (black: $r_{\rm S}=$10 Mpc, red: $r_{\rm S}=$20 Mpc, green: $r_{\rm S}=$30 Mpc). 
We modelled the distribution by a log-normal \citep[see][]{1991MNRAS.248....1C} and calculated the  best fit using a nonlinear least squares fit based on a gradient-expansion algorithm\footnote{\textsc{CURVEFIT} from \textsc{IDL}}.
For that, we parameterized the log-normal distribution as:
\begin{equation}
P(\delta_{\rm m}|\mbi p)=\frac{a}{\log(1+\delta_{\rm m})}\exp\left[b\,(\log(1+\delta_{\rm m})-c)^2)\right]{,}
\end{equation}
with $\mbi p=[a,b,c]$ being a set of parameters. 
The results of the best fits normalized with the number of cells are shown as the continuous lines on the left panel in Fig.~\ref{fig:stats_tot}.
One can appreciate in all curves for $w>0$ small tails towards low densities and long tails towards high densities showing a clear deviation from Gaussianity. The measured distributions are well fitted by the log-normal distribution of smoothed density  for smoothing radii $r_{\rm S}$ of 10, 20, and 30 Mpc. 
We also calculated the mean and the variance and plotted the corresponding Gaussian distributions with light dashed-dotted lines. 
 We conclude therefore, that the distribution of the matter field is in good agreement with the log-normal distribution at least in the scale range from about $10\,{\rm Mpc}\lsim1{r_{\rm S}}\lsim130\,{\rm Mpc}$.
This result is especially strong, since we did not assume a log-normal prior distribution in the reconstruction method. 
From a frequentist approach the Wiener-filter just gives the least squares estimator without imposing any statistical distribution to the matter distribution. The picture from a Bayesian perspective is more precise: a Gaussian prior distribution for the underlying density field is assumed. The posterior distribution, however, is conditioned on the data, which finally imposes its statistical behavior onto  the reconstruction, as can be seen in our results.

\section{Conclusions}
\label{sec:conc}

We have presented the first application of the \textsc{argo} computer code to
observational data.  In particular, we have performed a reconstruction
of the density field based on data from {\tt Sample dr6fix} of the New
York University Value Added Catalogue (NYU-VAGC) (see section
\ref{sec:IGS}). This yielded the largest Wiener-reconstruction of
the Large-Scale Structure made to date requiring the effective inversion of a matrix with about
$10^8\times 10^8$ entries. 
The use of optimized iterative
inversion schemes within an operator formalism \citep[see][]{kitaura},
together with a careful treatment of aliasing effects
\citep[see][]{dsp} permitted us to recover the field on a  Mpc mesh with an effective  resolution of the order of $\sim$10 Mpc.
Furthermore, we have investigated in detail the statistical problem 
 in particular  the noise covariance employed for performing Wiener-reconstructions.  


  We have demonstrated that Wiener-filtering leads to different
results than those obtained by the commonly used method of inverse
weighting the galaxies with the selection function. Both methods are
comparable when the galaxy number counts per cell is high. However, in regions with
sparse observed galaxy densities inverse weighting delivers very noisy
reconstructions.  This finding could have important consequences in
power-spectrum estimation and galaxy biasing estimation on large scales.  

As part of the results the Sloan Great Wall has been presented in
detail (see section \ref{sec:GW}) and some other prominent structures
like the Coma, the Leo, and the Hercules Cluster, have been discussed,
as well as the detection of a large void region (see section
\ref{sec:VOID}).  Our results also show the detection of
overdensity regions close to edges of the mask and predictions for structures in within gaps in the mask which compare well with the DR7 data in which the gaps are filled (see section \ref{sec:GAP}).
Finally, we have analyzed the statistical distribution of the density
field finding a good agreement with the log-normal distribution for Gaussian smoothing with radii in the range $10\,{\rm Mpc}\lsim1{r_{\rm S}}\lsim130\,{\rm Mpc}$.
We hope that this work highlights the potential of Bayesian large-scale structure reconstructions for cosmology and is helpful in establishing them as a widely used technique.

\section*{Acknowledgements}
We thank Andreas Faltenbacher and Jeremy Blaizot for very useful
discussions about the distribution of density fields and galaxy
formation. Some of the plots in this paper have been done using
the \textsc{HEALPix} \citep[][]{gorski} package. We also thank Samuel Leach
for encouraging conversations and helping substantially with the sky
plots using \textsc{HEALPix}.  Moreover, we thank Rainer Moll, Martin
Reinecke, Hans-Werner Paulsen and Heinz-Ado Arnolds for their
inestimable computational support.

The authors  thank the Intra-European Marie Curie fellowship and
the Transregio TR33 Dark Universe, as well as the Munich cluster {\it
  Universe} for supporting this project and both the Max Planck
Institute for Astrophysics in Munich and the Scuola Internazionale
Superiore di Studi Avanzati in Trieste for generously providing the
authors with all the necessary facilities. 

Funding for the SDSS and SDSS-II has been provided by the Alfred P.
Sloan Foundation, the Participating Institutions, the National Science
Foundation, the U.S.  Department of Energy, the National Aeronautics
and Space Administration, the Japanese Monbukagakusho, the Max Planck
Society, and the Higher Education Funding Council for England. 

The SDSS is  managed by the Astrophysical Research  Consortium for the
Participating  Institutions. The  Participating  Institutions are  the
American Museum  of Natural History,  Astrophysical Institute Potsdam,
University  of Basel,  University of  Cambridge, Case  Western Reserve
University,  University of Chicago,  Drexel University,  Fermilab, the
Institute  for Advanced  Study, the  Japan Participation  Group, Johns
Hopkins University, the Joint  Institute for Nuclear Astrophysics, the
Kavli Institute  for Particle  Astrophysics and Cosmology,  the Korean
Scientist Group, the Chinese  Academy of Sciences (LAMOST), Los Alamos
National  Laboratory, the  Max-Planck-Institute for  Astronomy (MPIA),
the  Max-Planck-Institute  for Astrophysics  (MPA),  New Mexico  State
University,   Ohio  State   University,   University  of   Pittsburgh,
University  of  Portsmouth, Princeton  University,  the United  States
Naval Observatory, and the University of Washington.

We finally thank the German Astrophysical Virtual Observatory (GAVO),
which is supported by a grant from the German Federal Ministry of
Education and Research (BMBF) under contract 05 AC6VHA, for providing
us with mock data.

{\small
\bibliographystyle{mn2e}
\bibliography{lit}
}

\clearpage

\end{document}